%% file: main.tex
\documentclass[twocolumn]{aastex631}

\usepackage{amsmath}

\newcommand{\teff}{$T_{\text{eff}}$}

\newcommand{\vsini}{$v\sin{(i)}$}

\newcommand{\rprs}{$R_{\text{p}}/R_{\text{*}}$}

\newcommand{\kms}{$\text{km }\text{s}^{-1}$}
\newcommand{\ms}{$\text{m }\text{s}^{-1}$}
\newcommand{\cms}{$\text{cm }\text{s}^{-1}$}
\newcommand{\degree}{$^\circ$}

\newcommand{\rearth}{\mbox{R$_{\Earth}$}}
\shorttitle{TOI-1759A b is Aligned}
\shortauthors{Polanski et al. 2024}
\graphicspath{{./}{}}
\begin{document}

\title{An Aligned Sub-Neptune Revealed with MAROON-X and a Tendency Towards Alignment for Small Planets }

\correspondingauthor{Alex S.\ Polanski}
\email{aspolanski@lowell.edu}

\author[0000-0001-7047-8681]{Alex S. Polanski} 
\altaffiliation{Percival Lowell Postdoctoral Fellow}
\affil{Lowell Observatory, 1400 W Mars Hill Road, Flagstaff, AZ, 86001, USA}
\affil{Department of Physics and Astronomy, University of Kansas, Lawrence, KS 66045, USA}

\author{Ian J.~M.\ Crossfield}
\affil{Department of Physics and Astronomy, University of Kansas, Lawrence, KS 66045, USA}

\author[0000-0003-4526-3747]{Andreas Seifahrt}
\affiliation{Gemini Observatory/NSF NOIRLab, 670 N. A'ohoku Place, Hilo, HI 96720, USA}

\author[0000-0003-4733-6532]{Jacob L.\ Bean}
\affiliation{Department of Astronomy \& Astrophysics, University of Chicago, Chicago, IL 60637, USA}

\author[0000-0002-2072-6541]{Jonathan Brande}
\affiliation{Department of Physics and Astronomy, University of Kansas, Lawrence, KS 66045, USA}

\author[0000-0001-6588-9574]{Karen A.\ Collins}
\affiliation{Center for Astrophysics \textbar \ Harvard \& Smithsonian, 60 Garden Street, Cambridge, MA 02138, USA}

\author[0000-0002-1221-5346]{David R.\ Coria}
\affiliation{Department of Physics and Astronomy, University of Kansas, Lawrence, KS 66045, USA}

\author[0000-0002-4909-5763]{Akihiko Fukui}
\affiliation{Komaba Institute for Science, The University of Tokyo, 3-8-1 Komaba,
Meguro, Tokyo 153-8902, Japan}
\affiliation{Instituto de Astrof\'{i}sica de Canarias (IAC), 38205 La Laguna, Tenerife, Spain}

\author[0000-0001-8511-2981]{Norio Narita}
\affiliation{Komaba Institute for Science, The University of Tokyo, 3-8-1 Komaba,
Meguro, Tokyo 153-8902, Japan}
\affiliation{Astrobiology Center, 2-21-1 Osawa, Mitaka, Tokyo 181-8588, Japan}
\affiliation{Instituto de Astrof\'{i}sica de Canarias (IAC), 38205 La Laguna, Tenerife, Spain}

\author[0000-0002-4410-4712]{Julian St{\"u}rmer}
\affiliation{Landessternwarte, Zentrum f{\"u}r Astronomie der Universität Heidelberg, K{\"o}nigstuhl 12, D-69117 Heidelberg, Germany}

% \author[0000-0003-2404-2427]{Madison Brady}
% \affiliation{Department of Astronomy \& Astrophysics, University of Chicago, Chicago, IL 60637, USA}
%No response yet

\author[0000-0002-8965-3969]{Steven Giacalone}
\altaffiliation{NSF Astronomy and Astrophysics Postdoctoral Fellow}
\affiliation{Department of Astronomy, California Institute of Technology, Pasadena, CA 91125, USA}

\author[0000-0003-0534-6388]{David Kasper}
\affiliation{Department of Astronomy \& Astrophysics, University of Chicago, Chicago, IL 60637, USA}

\submitjournal{AJ}
\received{April 23, 2025}
\revised{June 16, 2025}

\begin{abstract}

We present the Rossiter-McLaughlin measurement of the sub-Neptune TOI-1759A b with MAROON-X. A joint analysis with MuSCAT3 photometry and nine additional TESS transits produces a sky-projected obliquity of $|\lambda|$= $4^\circ\pm18^{\circ}$. We also derive a true obliquity of $\psi$=24$\pm12^{\circ}$ making this planet consistent with full alignment albeit to $<1\sigma$. With a period of 18.85 days and an $a/R_{*}$ of 40, TOI-1759A b is the longest period single sub-Neptune to have a measured obliquity. It joins a growing number of smaller planets which have had this measurement made and, along with K2-25 b, is the only single, aligned sub-Neptune known to date. We also provide an overview of the emerging distribution of obliquity measurements for planets with R$<8$ \rearth. We find that these types of planets tend toward alignment, especially the sub-Neptunes and super-Earths implying a dynamically cool formation history. The majority of misaligned planets in this category have 4$<$R$\leq$8 \rearth~and are more likely to be isolated than planets rather than in compact systems. We find this result to be significant at the $3\sigma$ level, consistent with previous studies. In addition, we conduct injection and recovery testing on available archival radial velocity data to put limits on the presence of massive companions in these systems. Current archival data is insufficient for most systems to have detected a giant planet.

\end{abstract}

%\keywords{Abundance ratios (11), Chemical abundances (224), Metallicity (1031), Stellar abundances (1577), Exoplanets (498)}


\section{Introduction} \label{sec:intro}

The projected angle between a planet's orbital angular momentum vector and the rotation vector of the host star (obliquity, $\lambda$) is an important tracer of a planetary system's evolution. The majority of these measurements have been made for short-period, giant planets (hot-Jupiters) and have enabled a deeper understanding of their origins \citep{Winn2010,Rice2022a}. However hot-Jupiters are rare population compared to planets between the size of Earth and Neptune, which are ubiquitous in our galaxy \citep{Petigura2018}. Unfortunately, the obliquity distribution of these planets, and even those of sub-Jovians (R$<$8 \rearth), has been left fairly unexplored.

The lack of small planet obliquities is largely due to the inherent limitations of the Rossiter-McLaughlin (RM) effect, which is the primary method of making these measurements \citep{Rossiter1924, McLaughlin1924}. The amplitude of a radial velocity (RV) anomaly due to the RM effect is dependent on both the transit depth and the \vsini~of the star. A typical hot-Jupiter, owing to its large cross-section, can induce an RM amplitude of tens of \ms~even if its host star has a low \vsini. Measuring the obliquity for smaller planets, on the other hand, is a balancing act. These planets typically have shallower transit depths, especially for host stars of earlier spectral types, and so one might target planets around smaller stars. However these late-type host stars are cooler and therefore fainter in the optical requiring longer integration times to reach the required RV precision. If the integration time is too long, one risks undersampling the RM effect since the transit duration can be short. In addition, if one chooses to target systems with higher \vsini~, increasing the RM amplitude, there is a trade-off for lower RV precision due to rotational broadening.

Spectrographs capable of sub-\ms~precision can overcome some of these difficulties and we are now measuring obliquities for a growing number of sub-Neptune sized planets and smaller \citep{Lubin2023,Zak2024,Zhao2023}. Due to their lower mass and larger average orbital separation, tidal interactions likely do not play as large of a role in shaping the the obliquity distribution of small planets as they do for hot-Jupiters. Indeed, massive planets on longer orbits (warm-Jupiters) tend to be aligned regardless of their host star temperature hinting at a quiescent formation history inside an aligned protoplanetary disk \citep{Rice2022b,Wang2024,EspinozaRetamal2024b}. Growing the number of small planet obliquities will reveal if this preference for alignment holds for more intrinsically common planets. In addition, \textit{intra-system} obliquities (i.e measurements for more than one planet in a system) will be vital for understanding the impact of multiplicity on planet formation.

This paper presents the obliquity measurement for TOI-1759A b, a 3.25 \rearth~sub-Neptune orbiting an early M-dwarf \citep{Martioli2022, Espinoza2022, Polanski2024} and is organized as follows: \S\ref{sec:observations} describes the collection of radial velocity and photometric data. \S\ref{sec:photmod} describes our joint analysis of TESS and MuSCAT3 photometry and subsequent search for transit timing variations (TTVs). In \S\ref{sec:rmmod} we combine our photometric model with MAROON-X radial velocities to constrain the obliquity. In this section we also assess the false alarm probability of our detection and examine the impact of stellar activity. In \S\ref{sec:small_planets} we discuss the obliquities currently measured for planets with R $<$8 \rearth~. Specifically, we examine the dependence on \teff~and orbital separation, compare systems of compact and isolated planets, and set limits on the existence of giant companions for single planet systems. Finally, we discuss TOI-1759A in this context.

\section{Observations}{\label{sec:observations}}

\begin{figure}
    \centering
    \includegraphics[width=0.43\textwidth]{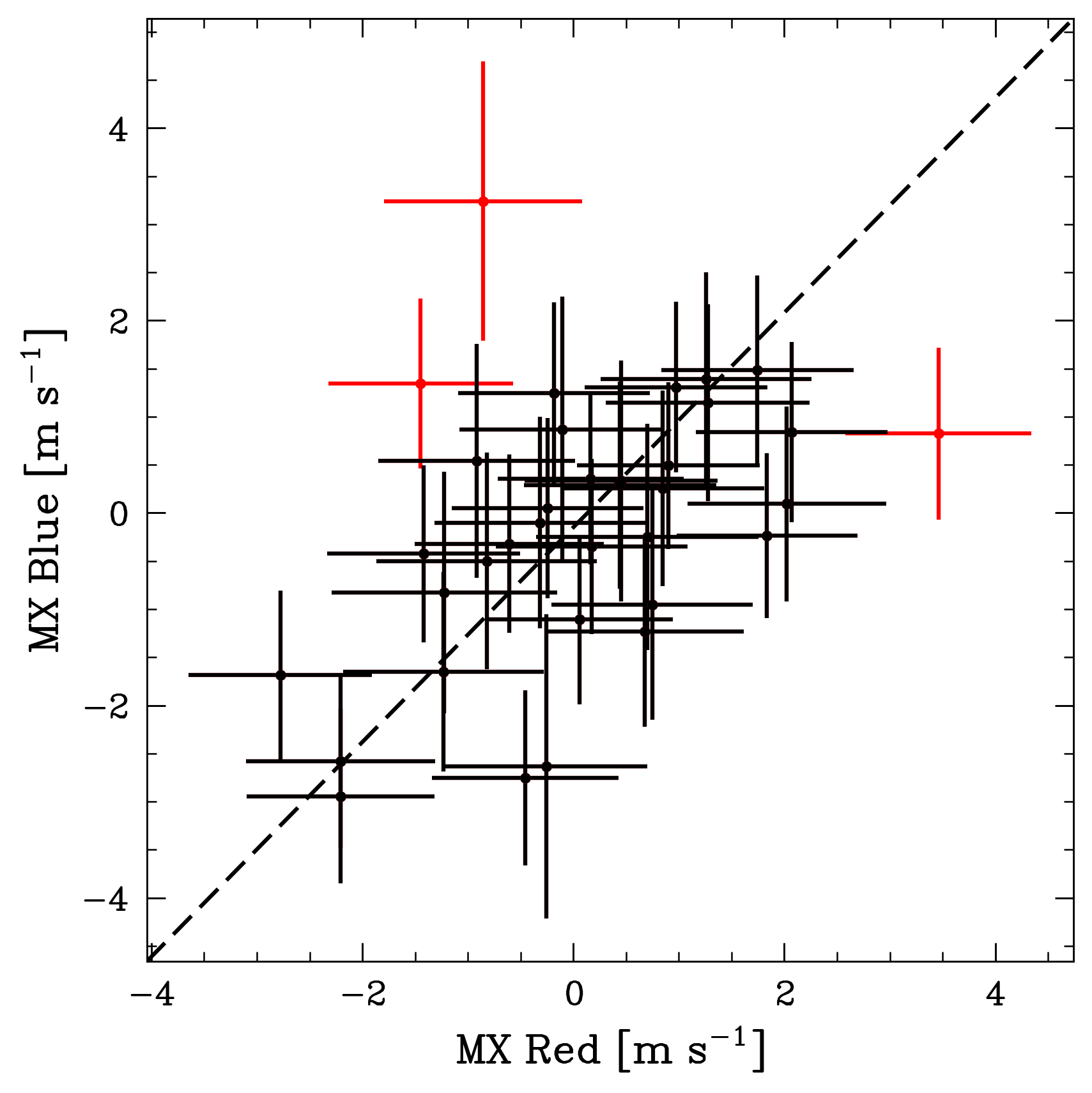}
    \caption{Comparison of radial velocities derived from the red and blue arms of MAROON-X. Points in red are discrepant at the 2$\sqrt{\sigma_{\text{Red}}^2 + \sigma_{\text{Blue}}^2}$ level however we found no additional reason to exclude these points from the analysis.}
    \label{fig:mx_comp}
\end{figure}

\begin{figure*}
    \centering
    \includegraphics[width=0.98\linewidth]{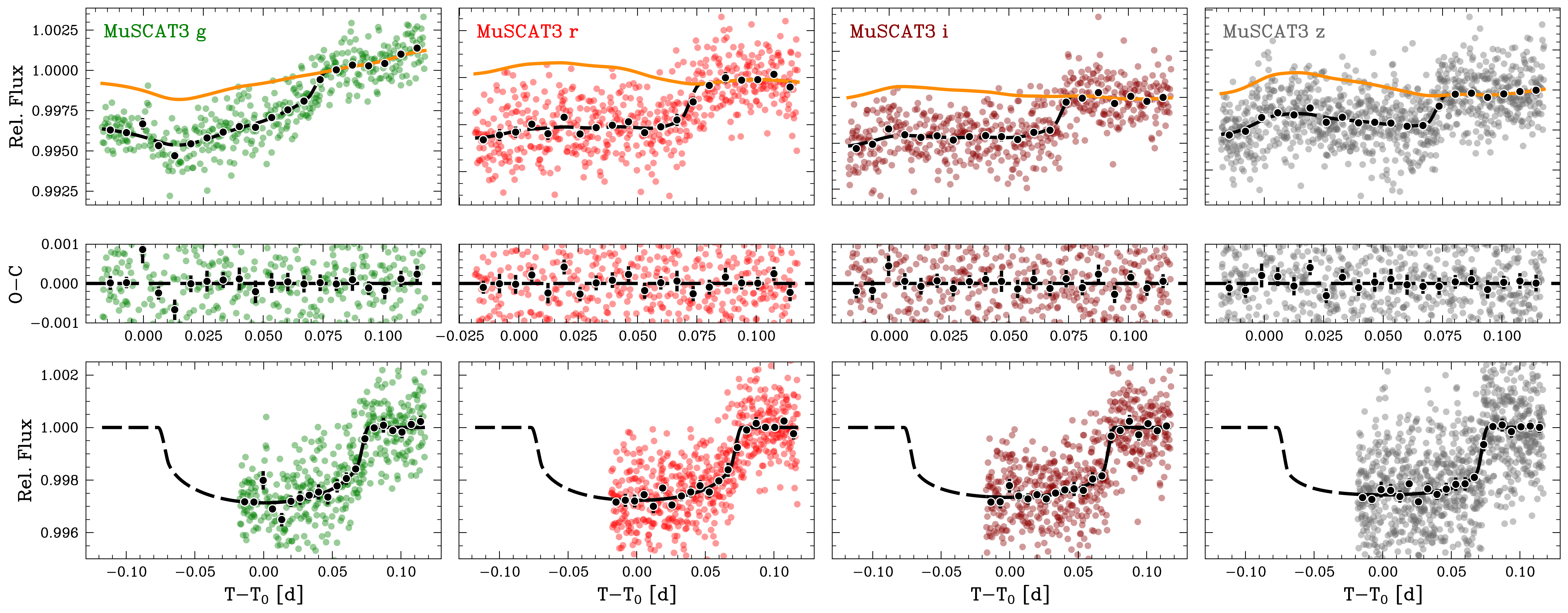}
    \caption{Normalized transit lightcurves from MuSCAT3 in the $g$, $r$, $i$, and $z$ passbands. The top row shows both the lightcurve model (black, dashed lines) and the Gaussian Process used to account for the systematics (orange line). }
    \label{fig:muscat_transits}
\end{figure*}

\subsection{MAROON-X Spectroscopy}{\label{sec:maroonx}}

MAROON-X is a high-resolution spectrograph ($\lambda / \Delta \lambda \sim 85000 $) operating in the optical range (500-920 nm) and installed on the 8.1-m Gemini North telescope atop Mauna Kea, Hawaii \citep{Seifahrt2018}. The spectrograph is fiber-fed, highly stabilized, and bench-mounted. It was designed to achieve sub-\ms\ radial velocity (RV) precision.

We obtained 35 RVs of TOI-1759A on 2021-08-16 UTC (PI: Crossfield, PID:  GN-2021B-Q-203), 20 of which were in-transit. Conditions during the transit were ideal, with seeing estimated at $\sim$0\farcs6. The observation sequence started with the target at an airmass of 2.0, however the entirety of the transit was observed below airmass 1.5 and down to the target's minimum airmass of 1.3. An exposure time of 600 seconds was chosen which resulted in a signal to noise ratio per spectral element of 215$\pm11$ and 123$\pm9$ in the red and blue arms, respectively. The MAROON-X raw data was been reduced using the standard procedure implemented in the instrument Python3 pipeline \citep{Seifahrt2020}. This procedure involved bias and background subtraction, order tracing and the extraction of one-dimensional wavelength-calibrated spectra. Wavelength solutions and instrumental drift corrections were based on the simultaneous calibration data of a stabilized Fabry–Pérot etalon \citep{Sturmer2017}, which allows order-by-order drift corrections at the sub–\ms\ level. The flux-weighted midpoint of each observation was used to calculate the barycentric corrections.

We examine the agreement of the RVs derived from the red and blue arms, respectively (Figure \ref{fig:mx_comp}) and find that the majority are consistent within the quadrature sum of their uncertainties. Three points are discrepant at greater than 2$\sqrt{\sigma_{\text{Red}}^2 + \sigma_{\text{Blue}}^2}$ but we found no additional reasons to omit these data from the analysis. 

\subsection{MuSCAT3 Photometry}{\label{sec:muscat}}

In order to pin down the mid-transit time we obtained simultaneous multiband photometric observation of the transit of TOI-1759A b using MuSCAT3 \citep{2020SPIE11447E..5KN} mounted on the 2m Faulkes Telescope North (FTN) at Haleakala Observatory in Hawaii, US. MuSCAT3 has four channels that enable simultaneous imaging in the $g$, $r$, $i$, and $z_s$ bands. Each channel is equipped with a 2k $\times$ 2k pixel CCD, providing a pixel scale of 0.27" pixel$^{-1}$ and a field of view of 9\farcm1 $\times$ 9\farcm1.  To increase the observing efficiency, we defocused the telescope and increased the exposure times while avoiding saturation. The exposure times were set to 20, 14, 15, and 8 s for the $g$, $r$, $i$, and $z_s$ bands, respectively. After applying dark and flat-field corrections, we performed aperture photometry on the acquired images using a custom-built pipeline \citep{2011PASJ...63..287F} with optimal aperture radii of 20, 22, 20, and 20 pixels for the $g$, $r$, $i$, and $z_s$ bands, respectively.

\subsection{TESS Photometry}{\label{sec:tess}}

Both \cite{Martioli2022} and \cite{Espinoza2022} used three sectors of TESS data (sectors 16, 17, and 24) which included three transit events. Since then, TESS has re-observed this system for an additional seven sectors: 56, 57, 58 (September-November 2022), 76, 77, 78 (March-May 2024), and 85 (October-November 2024) providing an additional nine transits. We downloaded the Presearch Data Conditioning (PDC) flux time series from the Mikulski Archive for Space Telescopes (MAST). The PDC flux is processed by the TESS Science Processing Operations Center (SPOC) pipeline \citep{Jenkins2016}. 

To prepare the flux time series for analysis, we detrend the data with \texttt{W{\={o}}tan} using the sum of cosines method \citep{Hippke}. A window size of 0.3 days was chosen, following the recommendation that the window size be $\gtrsim 2.2~T_{\text{14}}$ \citep{Hippke}. Each sector was detrended individually with a transit mask applied to minimize impact on the transit depth. 

\section{Photometric Modeling}{\label{sec:photmod}}

\subsection{Transit Model}

To model the photometry, we used the \texttt{exoplanet} package \cite{Foreman-Mackey2021} and simultaneously fit both the TESS and MuSCAT3 datasets. The free parameters in the lightcurve model were the scaled planet radius, \rprs, the impact parameter, $b$, the full transit duration, $T_{\text{14}}$, the period, $P$, the mid-transit time, $T_{\text{0}}$, and a white noise term and mean offset for each dataset, $\sigma$ and $\mu$. We used a quadratic limb darkening law and calculated the coefficients by interpolating the tables found in \cite{Claret2011} as described in \cite{Eastman2013}. Since we observed only a partial transit with MuSCAT3, we hold the limb darkening parameters constant whereas we impose a Gaussian prior of width 0.2 for the TESS limb darkening parameters. Additional priors on other parameters are given in Table \ref{tab:params}. In all of our analyses, eccentricity is held at 0. The MuSCAT3 lightcurves exhibit systematics which we modeled with a Gaussian Process using the \texttt{celerite2} implementation of the simple harmonic oscillator kernel \citep{celerite2}. A common  oscillator period was used between the four band passes. MuSCAT3 data and the best fit models are shown in Figure \ref{fig:muscat_transits}.

\subsection{Search for TTVs}{\label{sec:ttvs}}

\begin{figure}
    \centering
    \includegraphics[width=0.47\textwidth]{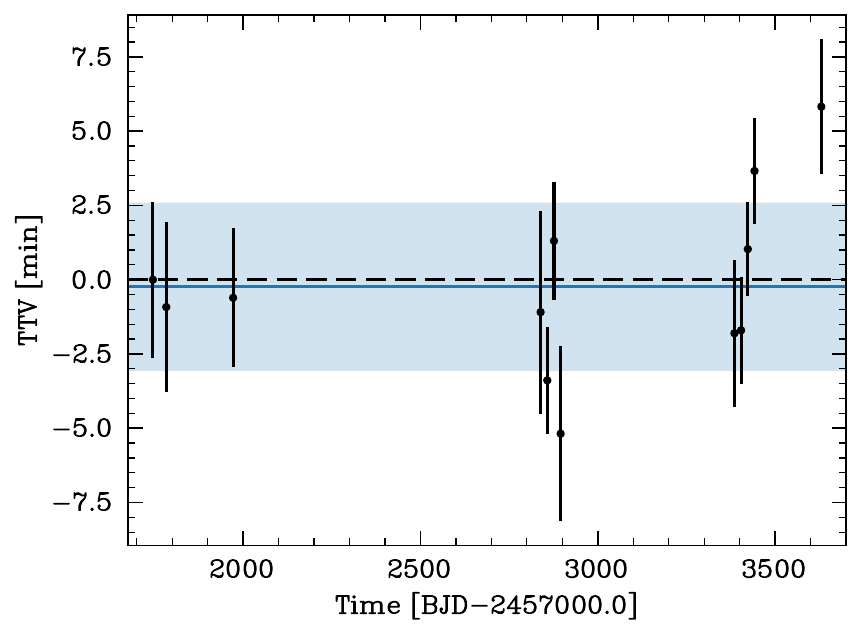}
    \caption{Transit timing variations for TOI-1759A b from 12 TESS transits. The shaded blue area represents the 1$\sigma$ spread (2.8 minutes) which is comparable to the cadence of the TESS photometry (2 minutes).  }
    \label{fig:ttv}
\end{figure}

With the addition of nine transits, we also conduct a search for transit timing variations (TTVs) that could be caused by a non-transiting companion. Using the \texttt{Batman} \citep{batman} package, we model each individual TESS transit holding the period, $a/R_*$, and $i$ constant at the values found in \S\ref{sec:photmod} while fitting for the transit midpoint and \rprs. We use \texttt{emcee} \citep{emcee} to sample the posterior distribution and then compare the individual transit times with those expected from a linear ephemeris. Our results are shown in Figure \ref{fig:ttv}. We do not observe any significant TTVs, with the standard deviation being comparable to the photometric exposure time (2 minutes).

\section{Rossiter-McLaughlin Modeling}{\label{sec:rmmod}}

\begin{figure*}[t!]
    \centering
    \includegraphics[width=0.9\linewidth]{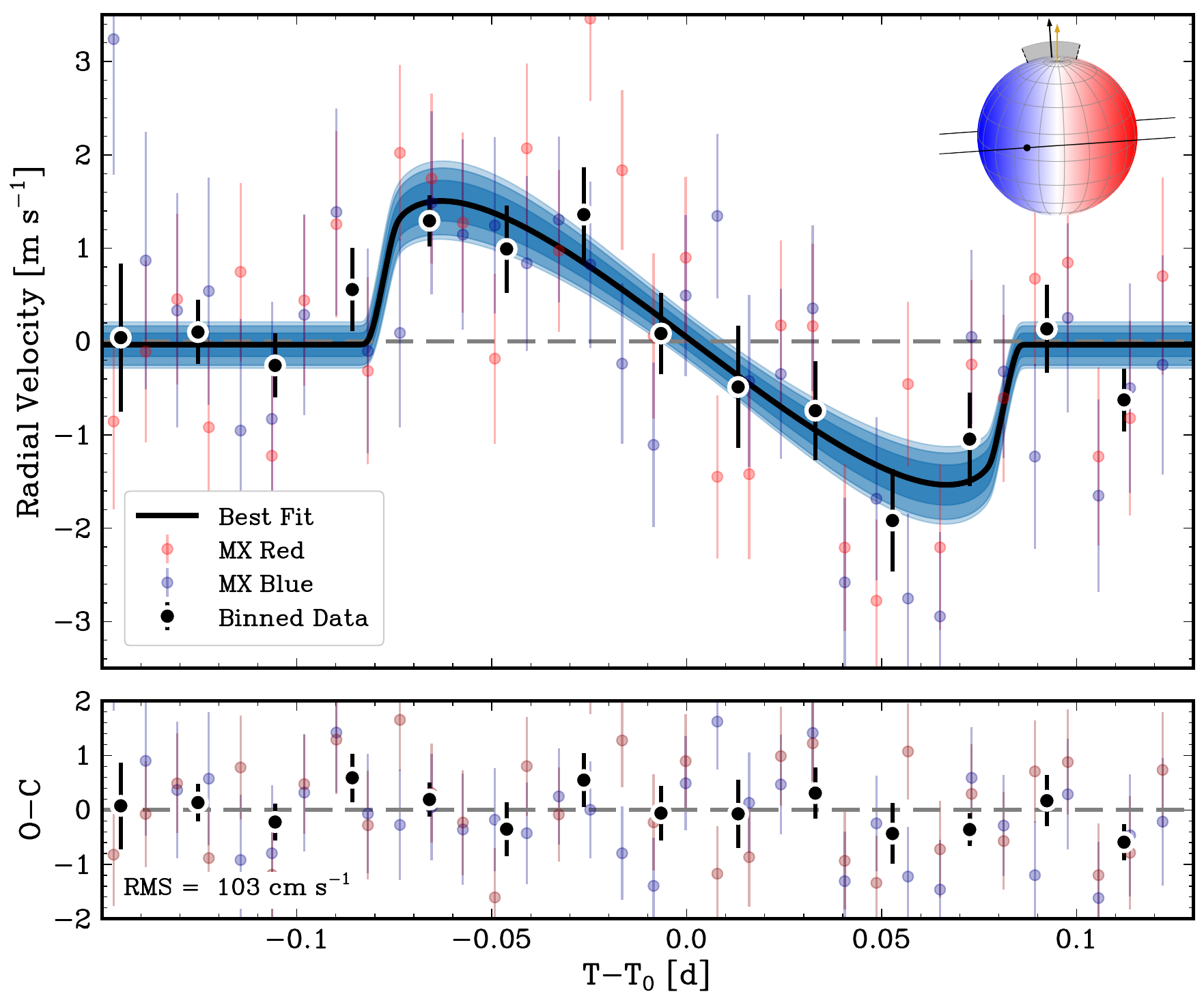}
    \caption{Radial velocity timeseries from MAROON-X with the median Rossiter-McLaughlin and 1$\sigma$ confidence shown as the black line and blue shaded region. Black points are the binned radial velocities. The inset shows the geometry of the system with the black arrow indicating the orbital angular momentum vector and the gray shaded area representing the uncertainty. }The bottom panel shows the residuals with an RMS of $\sim1$ \ms.
    \label{fig:rm}
\end{figure*}

\subsection{RM Model}{\label{sec:rmmod_desciption}}

To model the RM effect, we build on the \texttt{exoplanet} framework in \S\ref{sec:photmod} and model the MAROON-X radial velocities jointly with TESS and MuSCAT3 photometry. For the RM effect, we use the model presented by \cite{Hirano2010} which includes an intrinsic line-broadening parameter, $\beta$, and rotational broadening parameter $\sigma$. We set $\beta$ to the width of the MAROON-X resolution element with a normal prior $\mathcal{N}(3.5,0.5)$ \kms. The width of the prior is intended to account for macrotuburlence or other processes that may broaden the line profile. Following from \cite{Hirano2010,Hirano2011}, we define $\sigma=$\vsini/1.31. Coefficients for a quadratic limb darkening law were calculated for the MAROON-X bandpass using ExoCTK \citep{exoctk} and held fixed. Previous published works for TOI-1759A b produce differing K-amplitudes ranging from $\sim1$ to nearly 4 \ms~depending on which RV data is used. Assuming a maximum K-amplitude of 4 \ms~for TOI-1759A b would produce only a 50 \cms~shift across our observation window and we found that including out of transit RVs had a negligible affect on the obliquity.

Although it is common to treat RVs from the red and blue arms of MAROON-X as having come from separate instruments \citep{Brinkman2023,Tyler2025}, given the low-amplitude signal, we adopt a common RM model for both datasets. The main differences would be in the RV zero point and the limb darkening parameters. However, the ingress and the egress of the transit are not well enough samples (2-3 observations) to warrant the difference in limb darkening parameters. When fitting for separate zero points, we found both of them to be consistent with one another. The good agreement between the RVs from both arms of MAROON-X also motivates a common RM model (Figure \ref{fig:mx_comp}).

\input{posterior_vals}

\subsection{Constraints on \vsini~and the True Obliquity}

The low \vsini~of TOI-1759A makes it difficult to directly constrain it, especially for our low-amplitude signal. Poor constraints on \vsini~can result in similarly poor obliquity constraints (see Appendix \ref{sec:low-b-models}). Since \vsini~should be less than the equatorial velocity, $v_{eq}$, we can use the stellar radius and rotation period to put limits on the \vsini~through $2\pi R_{*}/P_{\text{rot}}$. \cite{Martioli2022} identified a rotation period of 35.7 days, as measured by both SPIRou Stokes V polarized spectra, that is also detected in Keck/HIRES RVs \citep{Polanski2024}. Using the stellar radius from \cite{Martioli2022} we find the equatorial velocity to be $v_{eq}$=0.89$\pm$0.03 \kms~, consistent with the estimated \vsini~of $<2$\kms. There is also a peak in the periodogram seen at $\sim$18 days. If this were the true rotation period, $v_{eq}$ would be 1.7 \kms. However, since stellar rotation periods are usually manifested in RV timeseries at either $P_{\text{rot}}$ or its harmonics \citep{Vanderburg2016}, the 18 day rotation period is unlikely and we take the true rotation period to be around 35 days.

Following \cite{Stefansson2022} (who follow from \cite{Masuda2020}), instead of sampling \vsini~directly, we sample the stellar radius and rotation period to obtain $v_{eq}$. We then sample the cosine of the stellar inclination to estimate \vsini~through $v_{eq}\sqrt{1-\cos^2{i_{*}}}$,~which accounts for the fact that \vsini~should always be smaller than $v_{eq}$. We placed normal priors on $R_{*}$ and $P_{\text{rot}}$. The latter was given a width of $\sim5$ days to account for the presence of additional peaks seen in periodograms around the 35 day period.

With this parameterization, we obtain a \vsini~of 0.95$\pm$0.12 \kms~and an obliquity of $|\lambda|$=$4^\circ\pm18^{\circ}$ (Figure \ref{fig:rm_corner}), making TOI-1759A b consistent with an aligned system and ruling out a polar orbit to 4.7$\sigma$ confidence. We also find a slightly shorter rotation period of 32$\pm$4 days, although it is still consistent with the 35 day reported period to 1$\sigma$. The best fit model is shown in Figure \ref{fig:rm}. To calculate the true obliquity, $\psi$, we use Equation 9 from \cite{Fabrycky2009} and find $\psi$=$24^\circ\pm12^{\circ}$ which is below the 30\degree~convention for alignment, but only satisfies this to less than 1$\sigma$.

\subsection{Modeling on Random Noise}{\label{sec:false_positive_test}}

For TOI-1759A b, we measure a peak RM amplitude of 1.5 \ms~which is larger than, but still comparable to, the RMS scatter in the RV residuals of $\sim1$ \ms. \cite{Albrecht2011b} demonstrated that the measurement of a misaligned orbit for WASP-2 b by \cite{Triaud2010} was likely obtained through chance by fitting an RM model to simulated random radial velocity measurements. To ensure we are not claiming a detection from noise, we followed a similar prescription by simulating MAROON-X RVs without an RM effect where the noise properties are sampled from a 2D Gaussian distribution with a covariance matrix of the form

\begin{align}
    \begin{bmatrix}
    \sigma_{Red,i}^2 & 0.82 \\
    0.82 & \sigma_{Blue,i}^2
\end{bmatrix}
\end{align}

\begin{figure}
    \centering
    \includegraphics[width=0.45\textwidth]{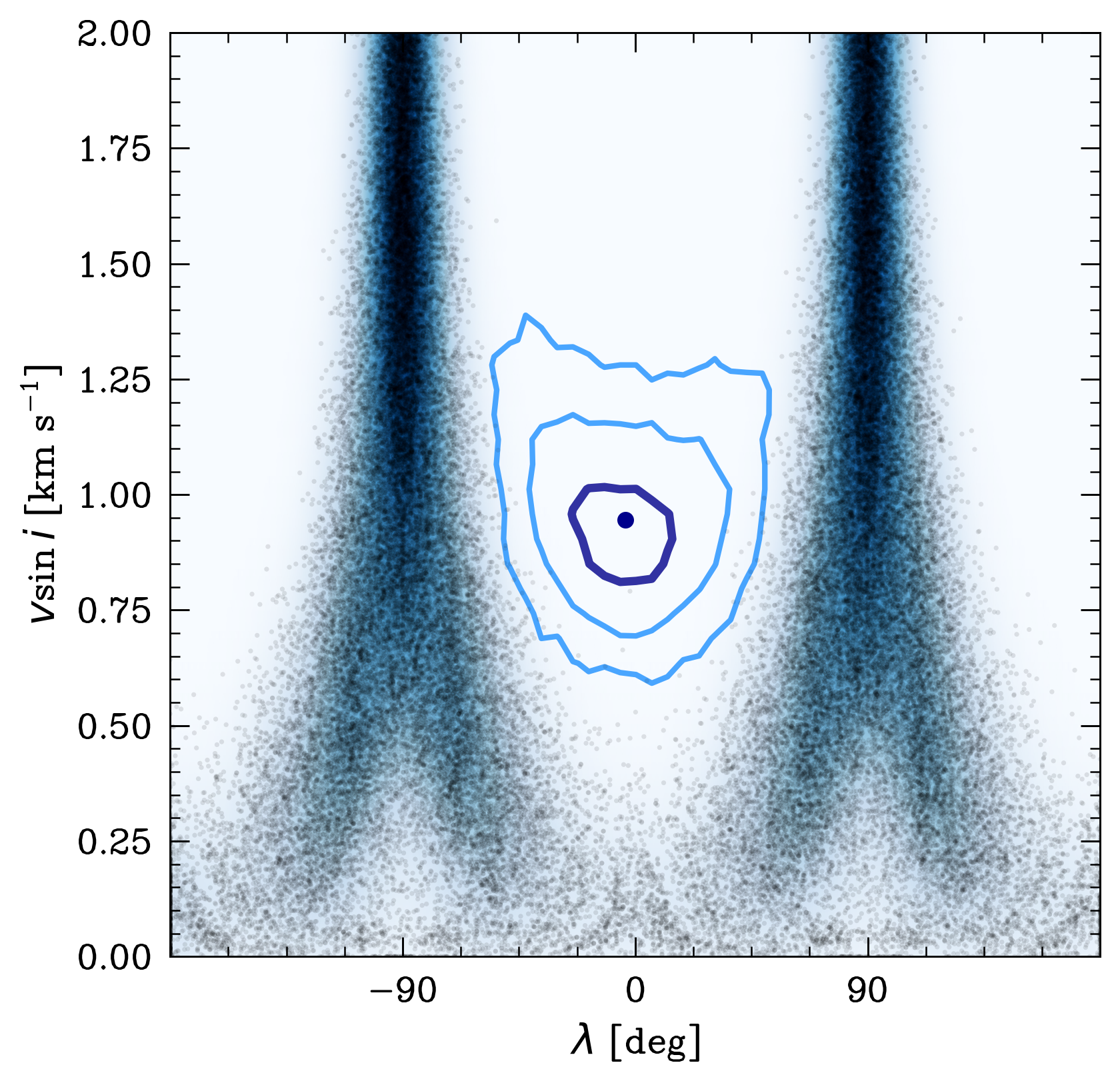}
    \caption{\vsini-$\lambda$ space for the maximum likelihood values of an RM model fit to 100,000 simulated datasets absent of an RM effect. A Gaussian kernel density estimate is overlaid in blue to emphasize density of points. The posterior distribution for the real MAROON-X dataset is shown as the blue contours representing the 1, 2, and 3 $\sigma$ confidence levels.}
    \label{fig:min_tests}
\end{figure}

\noindent The off-diagonal elements are included to capture the covariance between RVs measured with both the red and blue arms. We generated 100,000 datasets and, with the RM model described in \S\ref{sec:rmmod_desciption}, minimized a negative log-likelihood function to obtain the maximum \textit{a priori} estimate of $\lambda$ and \vsini. We impose normal priors on the photometric parameters informed from the joint RV-photometric fit in the previous section. Due to the computational cost, we do not perform a full MCMC exploration of the posterior for each simulation. Our results are given in Figure \ref{fig:min_tests} showing that our simulated RVs without an RM effect prefer polar, high-\vsini~ states and few of the simulations returned parameters consistent with those found in the real dataset to 3$\sigma$. From this we conclude that the false-alarm probability is likely less than 1\%; however, we note that we have ignored the possibility of correlated noise which could increase the false-alarm probability. 

\begin{figure*}
    \centering
    \includegraphics[width=0.95\linewidth]{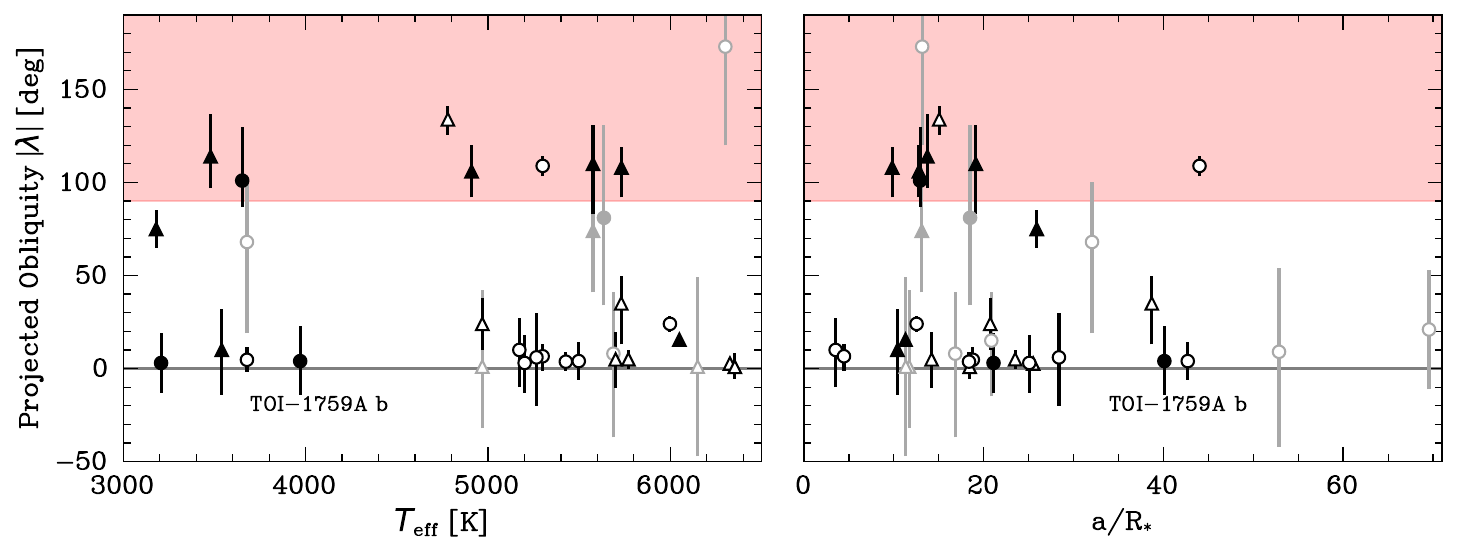}
    \caption{\textit{Left:} Projected obliquity $|\lambda|$ as a function of host star effective temperature for planets with R $\leq$ 8 \rearth~from a query of the TEPCAT catalog \citep{Southworth2011}. Triangles indicate planets with 4$<$ R $\leq$8 \rearth~ while circles indicate planets with R $\leq$ 4 \rearth. Open points indicate members of multiplanet systems while closed points are systems of a single detected planet. For planets with $\sigma_{\lambda}>30$\degree~the points are colored gray. The red shaded region indicates retrograde orbits. \textit{Right:} The same as left but for projected obliquity as a function of scaled orbital separation.}
    \label{fig:small_planets_alt}
\end{figure*}

\subsubsection{Correlations with Activity \& Instrumental Indices}{\label{sec:rv_activity_corr}}

TOI-1759A is a moderately active M dwarf ($\log{R'_{HK}}=-4.69$) with a known activity signal caused by rotational modulation with a period of $\sim$35 days. While this likely does not contribute significantly to our in-transit measurements, we still investigate whether our RVs are correlated with common activity indicators: the Ca II infrared triplet (Ca IRT), H$\alpha$, and the Na D doublet in addition to the chromatic index (CRX) and differential line width (dLW). For the CaIRT and H$\alpha$ features, we find marginal evidence of correlation with our RVs (Pearson R $\sim$0.1-0.3, Figure \ref{fig:activity}) but the two-sided p-values are all greater than 0.05 suggesting a low probability of correlation. The CRX and dLW indices show no correlation with RVs for both arms of MAROON-X. For the Na D doublet, we find a moderate positive correlation with RVs derived from the blue arm of MAROON-X with two-sided p-values of 0.02 and 0.03 for the D$_{1}$ and D$_{2}$ lines, respectively. While this suggests an activity signal that we may need to account for in our analysis, we note that the significance of the correlations are dependent on a single observation. Omitting this point we find the Pearson R values are reduced and the p-values surpass 0.05. This point also corresponds to the observation with the highest discrepancy between RVs derived from the two arms of MAROON-X. The lack of clear correlations suggests that stellar activity, at least as measured with these indicators, does not impact our analysis.

\begin{figure*}
    \centering
    \includegraphics[width=\linewidth]{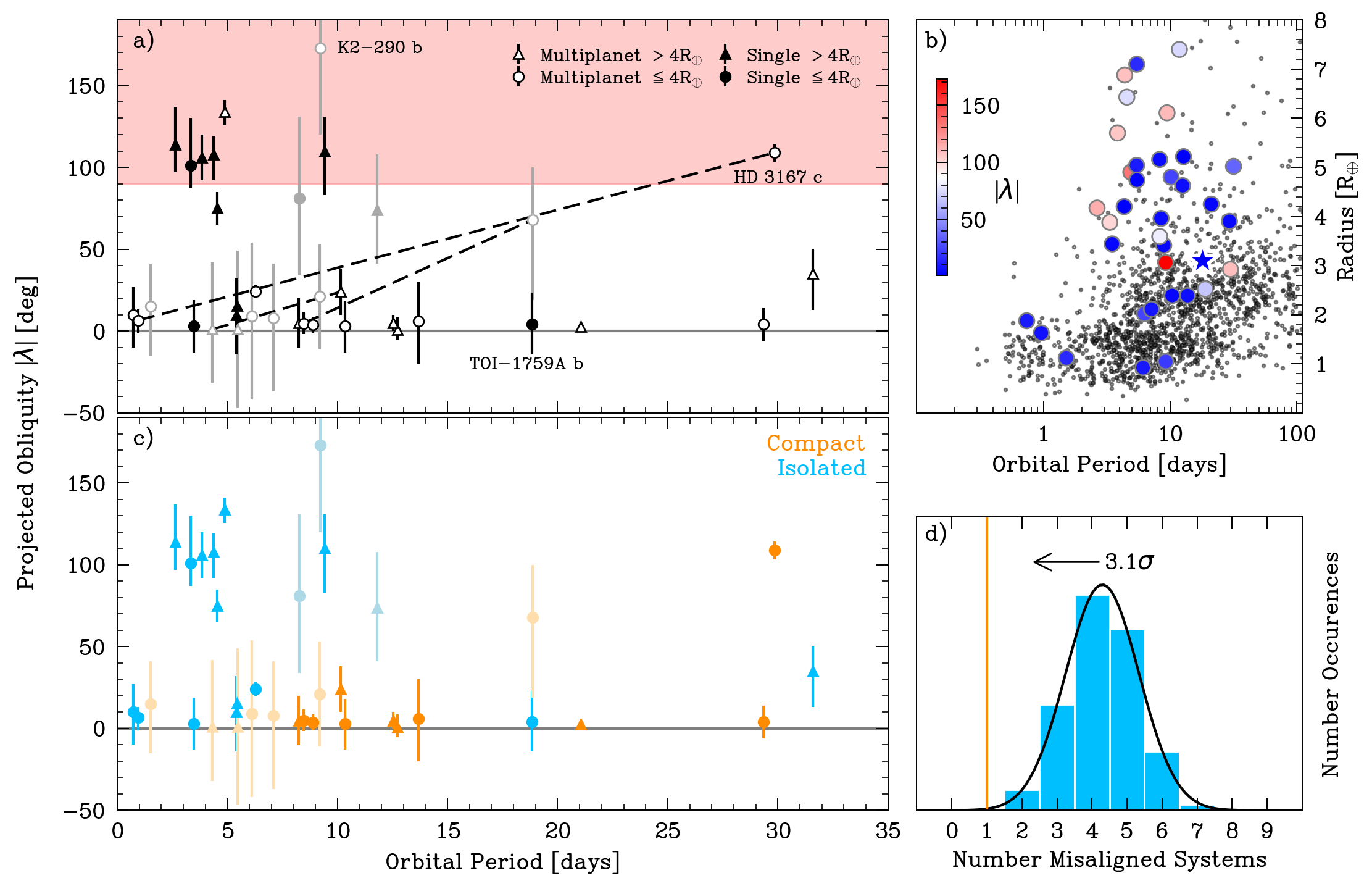}
    \caption{\textbf{a)} The same data as in Figure \ref{fig:small_planets_alt} but with projected obliquity as a function of orbital period. Dashed lines connect planets within the same system. For planets with $\sigma_{\lambda}>30$\degree~the points are colored gray. The red shaded region indicates retrograde orbits. \textbf{b)} Planet radius versus orbital period for the same sample. Points are color-coded according to their obliquity $|\lambda|$. The star denotes TOI-1759A b and black points are the sample of \textit{Kepler} planets from the California-Kepler Survey \citep{Petigura2017}. \textbf{c)} Similar to a) but planets are color-coded according to their classification as compact or isolated. The definition of compact is taken from \cite{Radzom2024} and defined as having an adjacent planet within a period ratio of 4. \textbf{d)} Distribution of isolated, misaligned planets after 100,000 iterations of sampling with replacement and a Gaussian fit given in black. The orange line indicates the one compact, misaligned planet in our sample showing that misaligned systems are more likely to be isolated to 3.1$\sigma$. }
    %\caption{\textit{Left:} Projected obliquity $|\lambda|$ as a function of orbital period for planets with R$<$8 \rearth from a query of the TEPCAT catalog \citep{Southworth2011}. Open circles indicate members of multiplanet systems while closed circles are systems of a single detected planet. Dashed lines connect planets within the same system. For planets with $\sigma_{\lambda}>30$\degree~the points are colored gray. The red shaded region indicates retrograde orbits. \textit{Right:} Planet radius versus orbital period for the same sample. Points are color-coded according to their obliquity $|\lambda|$. The star denotes TOI-1759A b. Black points are the sample of \textit{Kepler} planets from the California-Kepler Survey \citep{Petigura2017} }
    \label{fig:lambda_period}
\end{figure*}

\section{Obliquities of Small Planets}{\label{sec:small_planets}}

The current inventory of obliquity measurements for planets smaller than 8 \rearth~is limited with a query of the TEPCAT catalog \citep{Southworth2011} returning just 36 planets with such a measurement. While the field is at a similar point with these planets as \cite{Winn2010} was with hot-Jupiters, the current sample size makes it difficult to make quality cuts when searching for trends in the obliquity distribution of smaller planets. In this section we consider planets with R$<8$ \rearth, whose obliquities have been constrained through various measurements. Out of the 31 total systems, 6 have a known stellar companion. Obliquity measurements are taken from the TEPCAT catalog while additional parameters (\teff, P, a/R$_*$) are queried from the NASA Exoplanet Archive \citep[][as of 25 February 2025,]{NEA,nea_new}. To reduce the heterogeneity of the sample, we pull parameters from as few sources as possible. We provide these values in Table \ref{tab:literature} and also provide them in an online format.

Both the measurement method and stellar multiplicity are factors that have been controlled for in the much larger samples of hot- and warm-Jupiters \citep{Wang2024}, but we leave these considerations to future studies that will benefit from the steadily growing sample of small planet obliquities.

\subsection{Trends with \teff~and $a/R_{*}$}

The obliquity distribution of hot-Jupiters is marked by a strong dependence on stellar effective temperature. Below the Kraft Break \citep[$\sim6250$ K,][]{Kraft1967,Schatzman1962}, the majority of hot-Jupiters are well-aligned while above the Kraft break, a broader obliquity distribution is observed. Since this temperature is associated with a transition from a convective outer envelope in cooler stars to a mostly radiative envelope in hotter stars, this suggests that tides play an important role in the observed hot-Jupiter obliquity distribution. %This is supported by the apparent lack of a \teff-$\lambda$ dependence in longer period warm-Jupiters, perhaps owing to the strong dependence of the tidal realignment timescale on orbital separation ($\tau_{\lambda}\propto(a/R_{*})^6$).

Nearly all of the obliquity measurements for planets R$<$8 \rearth~have host star \teff~below the Kraft break, primarily due to the decreased transit depths for planets around larger stars which results in smaller RM amplitudes. Currently, our sample is not able to assess how the obliquity distribution changes across the Kraft break, but we do observe that planets R$<$8 \rearth~around cool stars do not show the same degree of preferential alignment as is seen in hot-Jupiters (Figure \ref{fig:small_planets_alt}). Instead, $\sim1/3$ of our sample is significantly misaligned. Since the tidal realignment timescale ($\tau_{\lambda}$) is inversely dependent on planet mass, realignment through tides is more inefficient for smaller planets. Indeed, \cite{Hebrard2011a} noted that Jovian planets 0.3$<$M$<$3 M$_{\text{Jup}}$ are more likely to be retrograde than their more massive super-Jupiter counterparts. The fraction of aligned and misaligned systems are similar to our sample ($\sim$ 2/3 aligned, 1/3 misaligned), but we note that most of the misaligned planets considered here have radii between 4 and 8 \rearth. The subset sub-Neptunes and super-Earths (R$<$4 \rearth) show a return to preferential alignment. 

%We also note that there are few systems in our sample that have intermediate obliquity states, rather planets tend to be well-aligned or cluster near 90\degree. A similar partitioning of obliquity states was also noted by \cite{Albrecht2021} for a subset of systems where the true obliquity could be derived. While this study was focused on giant planets (with some overlap with the sample we consider here), they found that systems tend to either be well-aligned or have obliquities near 90\degree~with few systems that have intermediate obliquities.

Even stronger than the tidal realignment timescale's ($\tau_{\lambda}$) dependence on mass is its dependence on orbital separation, $\tau_{\lambda}\propto(a/R_{*})^6$. In this respect, we see another similarity between the Jovian and small planet populations. For $a/R_{*}>12$, the obliquity distribution of giant planets becomes preferentially aligned, regardless of host star temperature \citep{Wang2024,Rice2022b,EspinozaRetamal2024b}. When plotted as a function of $a/R_{*}$, the majority of misaligned planets in our sample have $a/R_{*}$ between 10 to 20. Beyond this we find that small planets become preferentially aligned, including TOI-1759A b. The combination of low mass and large orbital separation means that tidal realignment is inefficient for these systems, and these planets likely formed in a well-aligned disk then smoothly migrated inward to their current positions. Many of these planets are also accompanied by more massive companions (Figure \ref{fig:abacus}) which further supports a dynamically cool formation scenario, especially when considering that warm-Jupiters also tend towards alignment. 

These results are in broad agreement with studies that attempt to infer the obliquity distribution of planets through stellar rotational velocity. Although first attempted with photometric variation \citep{Mazeh2015}, subsequent studies have identified differences in spectroscopic \vsini~measurements between stars with and without known planets \citep{Winn2017}. In particular, \cite{Louden2021} found that hot \textit{Kepler} host stars had lower $\langle$\vsini$\rangle$ than a sample of control stars suggesting that planets around hotter stars had higher obliquity and confirming previous tentative results \citep{Winn2017,Mazeh2015}. \cite{Louden2024} extended this trend to small planets with a sample of combined \textit{Kepler} and TESS planets but found that the \teff~dependence was itself dependent on orbital period. The difference in $\langle$\vsini$\rangle$ becomes insignificant for planets on orbital periods less than 10 days. This is comparable to our finding that planets with larger $a/R_{*}$ tend to be aligned while the majority of misaligned planets in our sample are on short orbital periods. 

%``Cleaning'' our sample by selecting measurements with average uncertainties less than 30$^{\circ}$, we find a more bimodal distribution, with many measurements near the Solar obliquity and the misaligned systems clustering near 100$^{\circ}$.

\subsection{Isolated Systems Tend to be Misaligned}{\label{sec:multis}}

In this sample, just over half of the measured obliquities are for planets that reside in multiplanet systems. From Figure \ref{fig:lambda_period}, we find planets closer to alignment are more likely to be in multiplanet systems whereas many of the misaligned planets are often the only one in the system. By ``cleaning'' our sample and selecting measurements with average uncertainties less than 30$^{\circ}$, this dichotomy becomes more apparent. Recently,  \cite{Radzom2024} found that sub-Saturn misalignment is more likely in systems that are isolated, defined as lacking an adjacent planet within a period ratio of 4. We expand on this result by classifying our sample into isolated and compact systems using similar criteria. Our choice to use planet radius to define our sample, in contrast to \cite{Radzom2024} who use mass, results in some overlap with roughly 1/3 of the systems considered here also appearing in the previous study. 

For small planets (R$\leq4$ \rearth), 11/19 are considered compact systems while only 7/17 of the larger planets are compact. All compact systems tend towards alignment, with the exception being HD 3167 c and AU Mic c. In contrast, isolated planets comprise the entirety of the short-period, misaligned systems seen in Figure \ref{fig:lambda_period} with a mix of planet sizes. We do find some isolated systems have low obliquities, but we note that the ultra-short period planets in this category, HD 3167 b and 55 Cnc e, reside in large multiplanet systems. In these cases, tidal dissipation may have pulled these planets from their compact systems \citep{MillholandSpalding2020,PuLai2019}. To quantify the significance of our result, we follow a similar prescription as \cite{Radzom2024} and randomly select 7 members of the isolated group and count the number of planets that are considered to be misaligned\footnote{$|\lambda|>10$ to $\sigma_{\lambda}$ and $|\lambda|>0$ to 2$\sigma_{\lambda}$ } (see Figure \ref{fig:lambda_period}d). Fitting a Gaussian function to the resulting distribution, we find that isolated planets are more often misaligned than those with nearby neighbors at a level of 3.1$\sigma$. Testing the robustness of this result, we exclude systems of multiple stars which slightly increases the significance (3.3$\sigma$) and by increasing the misalignment threshold from 10\degree~to 15\degree~which decreases the significance (2.8$\sigma$). Overall, our current sample shows that misaligned planets are more likely to be isolated to $\sim3\sigma$ confidence.

A preference towards alignment in compact systems is likely due in part to the capture of planetary systems into deep mean motion resonances (MMR) as it facilitates very slow disk migration \citep{Dai2023,Esteves2023,Xu2017}. This is especially pertinent for notably compact systems of six or more planets such as HD 191939 \citep{Lubin2022,OrellMiqual2023}, TOI-1136 \cite{Beard2024a}, HD 110067 \citep{Luque2023}, and TRAPPIST-1 \citep{Gillon2016}. TRAPPIST-1 b, d, and e all have had their obliquities measured, however \cite{Hirano2020a} stresses the false alarm probability for the latter two to be too high to claim a detection. The system-wide obliquity has since been measured by \cite{Brady2023} revealing a well-aligned system. Dynamical simulations suggest that the HD 191939 and TOI-1136 systems are likely coplanar \citep{Lubin2023,Dai2023}. 

However, MMR chains are thought to be quite fragile. \cite{Izidoro2021} suggests that more than 95\% of resonance chains become dynamically unstable after gas disk dispersal. The chaotic era that would ensue could lead to the large mutual inclinations seen in HD 3167 and AU Mic, although the obliquity measurement for AU Mic c is fairly unconstrained. These near-perpendicular systems stress the need for more intra-system obliquity measurements. 

\input{companion}

\subsection{Searching for Companions to Single Planets}

\begin{figure*}
    \centering
    \includegraphics[width=\linewidth]{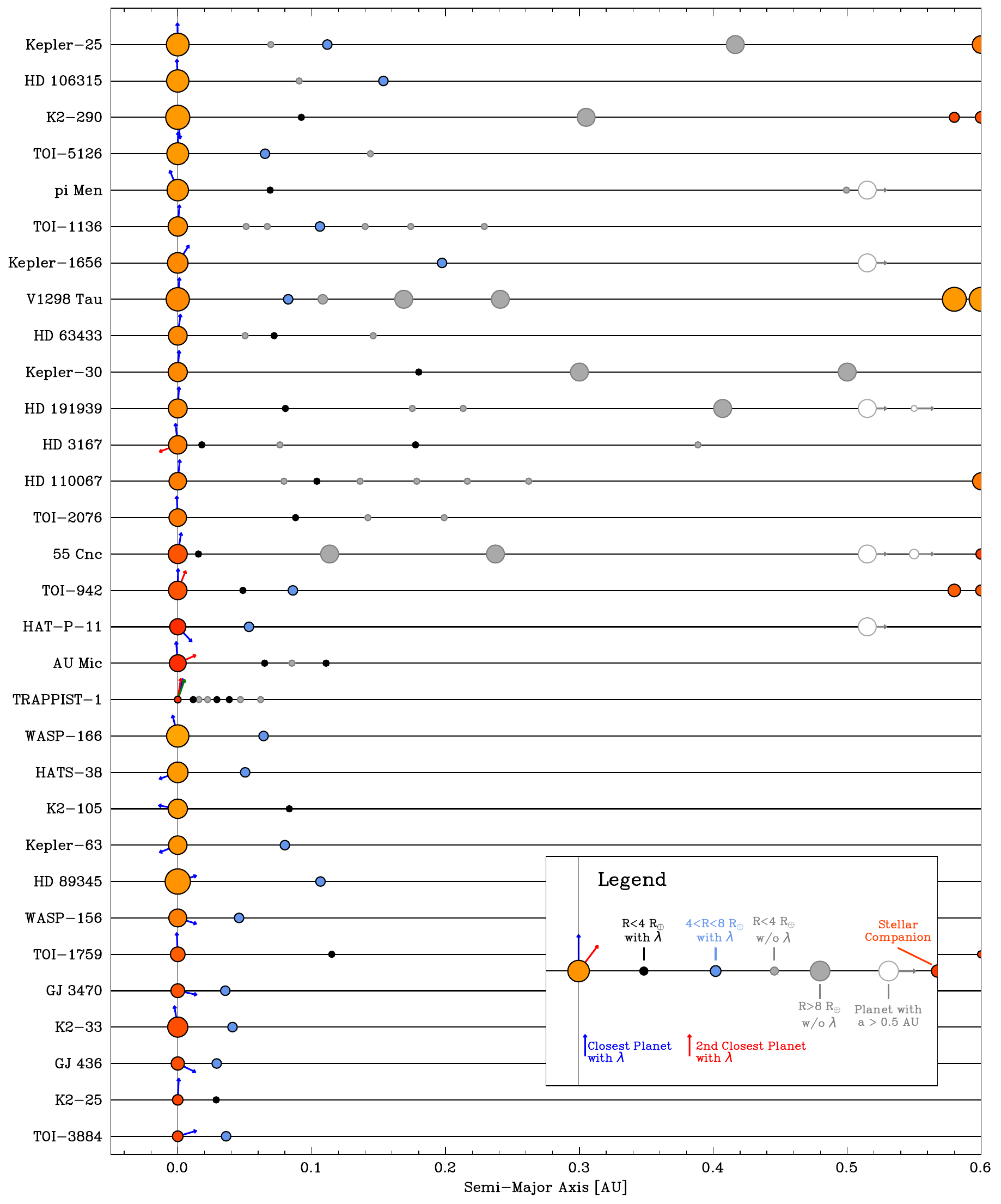}
    \caption{Abacus plot for all systems containing an R$_{P}<8$ \rearth~planet with a measured obliquity. Stars are colored and sized according to the temperature and radius. Blue arrows indicate the obliquity of the inner-most planet with $\lambda$ measured. Red arrows are for the 2nd inner-most planet with $\lambda$ measured. In the case of TRAPPIST-1, the additional green arrow indicates the obliquity for the outer-most planet with $\lambda$ measured. We also note V1298 Tau b (R$_{P}=10$ \rearth) has an obliquity measurement consistent with that of V1298 Tau c \citep{Johnson2022}. }
    \label{fig:abacus}
\end{figure*}

The preference for more polar obliquities in single planet systems (Figure \ref{fig:abacus}) in comparison to the multiplanet systems has been more challenging to explain. In our sample, these tend to be higher mass, sub-Saturn planets. \cite{EspinozaRetamal2024a} and \cite{Knudstrup2024} also report a pile-up of these types of planets at polar orbits. Resonance chain breaking can inflate the obliquity distribution of planets but not to such a high degree. Furthermore, a single planet is a rare outcome for systems that have broken their resonance chains \citep{Esteves2023,Izidoro2021}. Alternatively, these single planets may have attained their polar orientations through interactions with an outer companion. \cite{Petrovich2020} proposed a framework that shows near-polar sub-Saturn systems can be obtained through inclination excitations caused by a giant planet and was used to explain the polar orbit of HAT-P-11 b and WASP-107 b\footnote{WASP-107 b has a mass comparable to the other sub-Saturns in our sample, but it was not included due to its inflated radius.}, both of which have long-period companions \citep{Piaulet2020,Rubenzahl2021}. 

Since massive planets are readily detectable in long-term RV surveys \citep{Rosenthal2021}, we compiled archival RV data for the 13 single-planet systems in our sample and searched for long-period companions via injection and recovery testing. Using \texttt{RVSearch}, we fit for the parameters of the known planet and this signal was subtracted from the dataset. We then generated 5,000 synthetic Keplerian signals with 0.1 $< K_{\text{P}} < $ 1000 \ms and 1.2 $< P < $ 1$\times10^6$ injecting each and assessing its recovery. The results are shown in Figure \ref{fig:injection} showing that TOI-3884, HATS-38, and HD 89345 currently do not have sufficient archival RV data to have been able to detect a Jovian mass planet out to $\sim0.5$ AU. Additionally, K2-105, Kepler-63, and K2-33 only have $<10$ archival RVs. On the other hand, if there was a massive companion in the TOI-1759, WASP-166, WASP-156, GJ 3470, or GJ 436 systems, archival data should have been sufficient to find it (Figure \ref{fig:injection}) at least out to $\sim2-3$ AU. 

RV detection of giants ideally requires continuous monitoring with the same instrument, as offsets between datasets make it more difficult to track long-term trends. For the TOI-1759, WASP-166, GJ 3470, and GJ 436 systems, either one instrument covers the majority of the baseline or there is significant overlap between different datasets. WASP-156 has two epochs of data from different datasets separated by four years, making it more difficult to detect long-period companions. We also note that of these systems, GJ 3470 is the only one where a linear trend in the RVs is detected \citep{Stefansson2022}, suggesting we might be sampling just a portion of a longer-period orbit. Although it is still not definitive whether the occurrence rate of giant planets peaks or plateaus beyond the snowline \citep{Fulton2021,Wittenmyer2020}, further RV monitoring of these systems will push our sensitivity to wider orbits. For now, it is not possible to attribute the polar orbits of single Neptune systems to giant planets.

%Breaking of resonance chains may also explain the preference for more polar obliquities in single planet systems (Figure \ref{fig:abacus}) in comparison to the multiplanet systems. \cite{Izidoro2021} found that more massive planets tend to have more violent instabilities and, while the study was focused on planets R $<$ 4 \rearth, it's reasonable to assume that dynamical instability can scale with mass beyond the sub-Neptune regime. This is compatible with our current sample of single planet systems being composed mostly of planets 4 $<$ R $<$ 8 \rearth~ However, resonance breaking is thought to occur after gas disk dispersal so an outstanding question is why these planet did not undergo runaway gas accretion. A possible solution to this is late-stage accretion which \cite{Izidoro2021} also notes becomes common-place near the inner-edge of the disk resulting in more massive planets. The mean mass of single planets in our sample, 28$\pm8$ \mearth, is consistent with the tail end of the mass distribution \cite{Izidoro2021} observed for unstable scenarios. GJ 3470 b is the lowest mass among them but is currently undergoing heavy atmospheric loss and has possibly shed $\sim$40\% of its mass.

\subsection{TOI-1759A b in Context}

Where does this leave TOI-1759A b in the emerging picture of small planet obliquities? At $-4^\circ\pm18^{\circ}$, our work places this sub-Neptune among the few well-aligned single planet systems joining K2-25 b, K2-33 b, and WASP-166 b \citep{Stefansson2020,Bourrier2023,Hirano2024}. If we restrict this to sub-Neptunes and super-Earths, TOI-1759A b and K2-25 b are the only well-aligned single planet systems known to date. TOI 1759 b is also currently the longest period single sub-Neptune with a measured obliquity. Given the longer period of TOI-1759A b, no discernible eccentricity, and the lack of a (detected) companion, this sub-Neptune likely experienced a dynamically cool formation history. This is in contrast to K2-25 b, which would have required more chaotic formation mechanisms (e.g. planet-planet scattering) to explain the low obliquity but high eccentricity \citep{Stefansson2020}. A formation pathway for TOI-1759A b that is consistent with the current orbital architecture for TOI-1759A b would likely be disk-driven migration. 

Finally, TOI-1759A b is one of seven systems in this sample that have known comoving stellar objects with similar parallaxes and proper motions \citep{GonzalezPayo2024}. The TESS Input Catalog entry for TOI-1759B suggests a mid-to-late M dwarf (M= $0.112\pm0.020$ M$_{\odot}$) making the TOI-1759 system an M+M binary with a projected separation of 2770 AU. We compile astrometric data for TOI-1759 system in Table \ref{tab:companion}. While TOI-1759AB is not as widely separated as compared to the other multistar systems in this sample, the low mass of both components produces a low binding energy for their orbit. Out of the 720 M dwarf multistar systems presented by \cite{Cifuentes2025}, TOI-1759AB is among the most susceptible to having their orbit disrupted. 

Close stellar companions likely have significant effects on planet formation and evolution. Companion stars commonly exhibit low mutual inclinations with planet orbits, especially for systems of multiple small planets \citep{Behmard2022, Dupuy2022}, however recent work suggests a diminishing influence as separations increase beyond $\sim700$ AU \citep[e.g.][]{Christian2025}. In the context of planetary obliquities, \citep{Rice2024} presented evidence that multistar systems that host hot-Jupiters tend to exhibit spin-orbit alignment \textit{and} orbit-orbit alignment lending support to efficient viscous dissipation in these systems. These dissipative effects are strongly dependent on binary separation which may explain why most of their fully aligned sample had $s\sim$ 200-900 AU. Given these points, we suspect the companions in our sample, most of which have separations of order thousands of AU, played a smaller role in shaping their current alignment. Although the sample is small, we see that in all but one multistar system, the planets are well-aligned with the exception being K2-290 which is a triple system where the A and B components are separated by only $\sim100$ AU. 

\section{Summary}

In this work we measure the spin-orbit alignment of the sub-Neptune TOI-1759A b with in-transit radial velocity measurements from MAROON-X. We find that TOI-1759A b's orbit is well-aligned with the spin axis of its host star, measuring a projected obliquity of $-4^\circ\pm18^{\circ}$. This marks the largest orbital separation of a single sub-Neptune for which an obliquity measurement has been made. It joins K2-25 b as the only other single, well-aligned sub-Neptune known to date. We also conduct an early exploration of the obliquity distribution for planets with R$<$8 \rearth, and identify the following features:

\begin{itemize}
    \item Planets with size R $\leq$ 8 \rearth~are divided into aligned and misaligned populations in similar proportions found for giants planets: 2/3 aligned, 1/3 misaligned. Dividing our sample into compact and isolated planets, we find that compact systems, especially those with planets with R$<4$ \rearth, tend towards alignment suggesting a quiescent formation in a well-aligned disk. Planets with large obliquities are more likely to be misaligned at the $\sim3\sigma$ level, in agreement with results presented by \cite{Radzom2024}. 
    \item Planets at larger orbital separation, including TOI-1759A b, tend to be aligned. While the sample size is not sufficient to identify a clear cut-off, the majority of misaligned planets have orbital periods less than 15 days. An outlier is HD 3167 c, which is notable for it's near 90\degree~mutual inclination with HD 3167 b. 
    \item Only a handful of the misaligned, single planets have sufficient archival radial velocity data available to have detected a long-period companion out to $\sim$2-3 AU. The lack of a giant companion planet found in these systems suggests dynamical interactions with a giant planet may not be a universal explanation for the polar orbits observed in this population. Many of these systems do not have a long-enough baseline of observations to draw any definitive conclusions. 
\end{itemize}

The distribution of small planet obliquities is still an emerging picture. While this work adds another measurement to the sample, many more are needed to begin in-depth, population-level analyses as is currently being done with the hot- and warm-Jupiters. Additionally, many of the obliquities measured for these planets have uncertainties $>30$\degree. Improving constraints on these measurements is equally as valuable as growing the sample itself. 

\newpage

\section*{}
The authors thanks the anonymous reviewer whose comments greatly improved the quality of this work. In addition, A.S.P thanks Emma Louden, Jack Lubin, Joe Llama, and James Sikora for the insightful discussions during the preparation of this manuscript. 

A.S.P acknowledges support through the Percival Lowell Postdoctoral Fellowship which is funded in part through generous donations of the Percival Lowell Society. This organization was established to recognize individuals who create gifts to Lowell Observatory through their wills, trusts, and other estate planning vehicles. This research is partially funded through the Caltech-IPAC Visiting Graduate Research Fellowship (VGRF).

The University of Chicago group acknowledges funding for the MAROON-X project from the David and Lucile Packard Foundation, the Heising-Simons Foundation, the Gordon and Betty Moore Foundation, the Gemini Observatory, the NSF (award number 2108465), and NASA (grant number 80NSSC22K0117). This work was enabled by observations made from the Gemini North telescope, located within the Maunakea Science Reserve and adjacent to the summit of Maunakea. We are grateful for the privilege of observing the Universe from a place that is unique in both its astronomical quality and its cultural significance.

This paper is based on observations made with the MuSCAT instruments, developed by the Astrobiology Center (ABC) in Japan, the University of Tokyo, and Las Cumbres Observatory (LCOGT). MuSCAT3 was developed with financial support by JSPS KAKENHI (JP18H05439) and JST PRESTO (JPMJPR1775), and is located at the Faulkes Telescope North on Maui, HI (USA), operated by LCOGT. MuSCAT4 was developed with financial support provided by the Heising-Simons Foundation (grant 2022-3611), JST grant number JPMJCR1761, and the ABC in Japan, and is located at the Faulkes Telescope South at Siding Spring Observatory (Australia), operated by LCOGT.

This work is partly supported by JSPS KAKENHI Grant Number JP24H00017
and JSPS Bilateral Program Number JPJSBP120249910.

This research has made use of the NASA Exoplanet Archive, which is operated by the California Institute of Technology, under contract with the National Aeronautics and Space Administration under the Exoplanet Exploration Program.

\facilities{Gemini:Gillett, FTN, TESS, Exoplanet Archive}

\software{\texttt{exoplanet} \citep{Foreman-Mackey2021}, \texttt{Lightkurve} \citep{Lightkurve2018}, \texttt{corner} \citep{corner}, \texttt{emcee} \citep{emcee}, \texttt{smplotlib} \citep{smplotlib}, \texttt{lmfit} \citep{lmfit}}

\bibliography{bib.bib}
\bibliographystyle{aasjournal}

\newpage

\appendix
\restartappendixnumbering 

\section{RM Modeling for Systems with low $b$}{\label{sec:low-b-models}}

\begin{figure*}[ht!]
\gridline{
  \fig{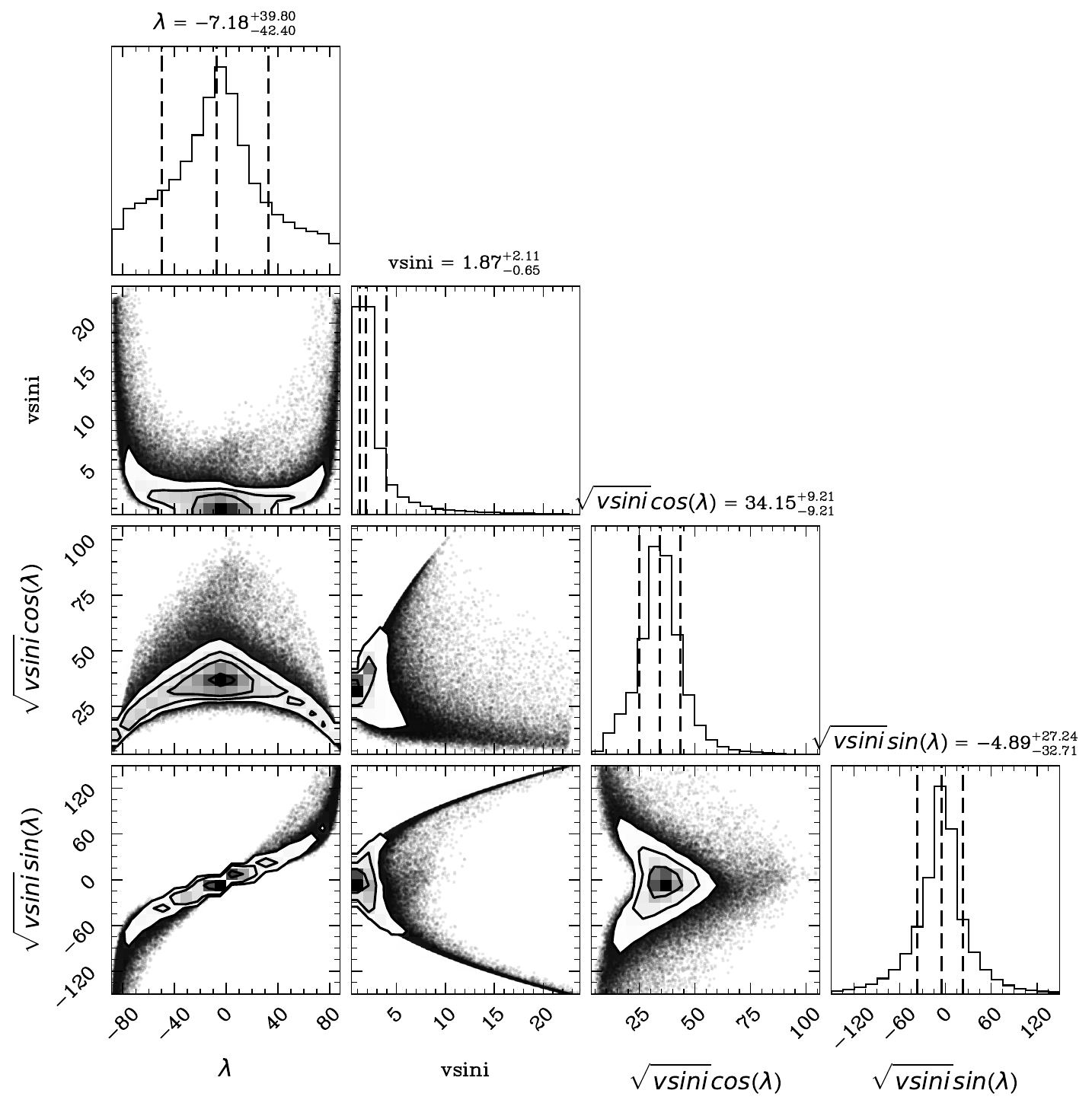}{0.5\textwidth}{}
  \fig{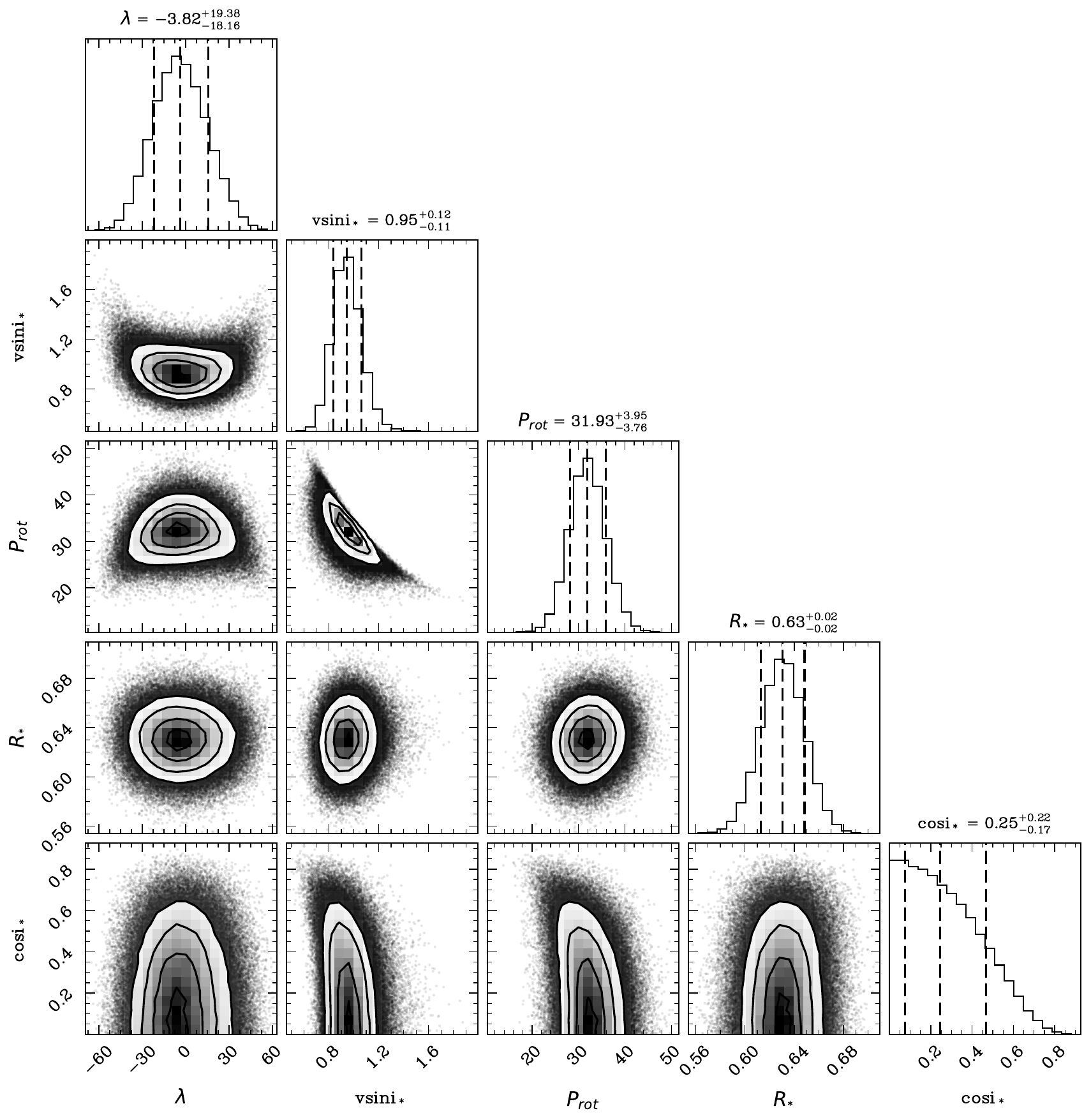}{0.5\textwidth}{}
}
\caption{\textit{Left:} Posterior distribution corner plot for a fitting basis of $\sqrt{v\sin{i}}\cos{\lambda}$ and $\sqrt{v\sin{i}}\sin{\lambda}$. No prior has been imposed on \vsini~resulting in a long velocity tail which widens the $\lambda$ distribution. \textit{Right:} Same as left but for a basis that fits $\lambda$ directly and \vsini~ indirectly via $v_{eq}\sqrt{1-\cos^2{i_{*}}}$. }
\label{fig:rm_corner}
\end{figure*}

For a system with an impact parameter (b) of zero, the shape of the RM curve will be symmetric for all values of $\lambda$, but as $b$ increases, any change in obliquity will create an increasingly asymmetric RM curve. In terms of physical parameters, $\sqrt{1-b^2}v\sin{i}\cos{\lambda}$ describes the semi-amplitude whereas $bv\sin{i}\sin{\lambda}$ describes the asymmetry \citep{Albrecht2012,Triaud2018} with the latter being poorly constrained as $b$ approaches zero. A consequence of this is that, to appropriately describe the data, the \vsini~must increase in order to match the observed amplitude for high obliquity models. This results in long posterior tails that not only inflate the \vsini~estimate but also widens the $\lambda$ distribution and can create a posterior geometry that is difficult to sample. This behavior has been noted in previous works \citep{Anderson2011a} and is especially pernicious with low-SNR detections. To address the sampling issue, the model can be reparameterized to reduce correlations between jump parameters, for example using $\sqrt{v\sin{i}}\cos{\lambda}$ and $\sqrt{v\sin{i}}\sin{\lambda}$. Using this basis, which preserves a uniform distribution on $\lambda$ and~\vsini, we ran 4 MCMC chains for 10,000 burn-in steps and an additional 50,000 sampling steps each. Inspection of the posterior distribution (Figure \ref{fig:rm_corner}) clearly exhibits the pitfalls of modeling low-$b$ systems. $\sqrt{v\sin{i}}\cos{\lambda}$, corresponding to the amplitude of the RM is much more constrained than $\sqrt{v\sin{i}}\sin{\lambda}$ which is reflective of the asymmetry. A transformation to physical parameters shows both $\lambda$ and~\vsini~to be highly correlated with a long high-velocity tail that corresponds to models with higher obliquity, reducing the significance of $\lambda$.

Breaking this correlation requires further constraints with \vsini~being the most readily informed from additional data. For many systems which have had their obliquities measured, \vsini~can be determined with spectroscopy, however, the projected rotational velocity of TOI-1759A is estimated to be below 2 \kms~ \citep{Martioli2022,Espinoza2022}. Following \cite{Albrecht2011b}  and restricting \vsini~to be between [0,2] only slightly tightens the constraints on $\lambda$ since \vsini~is still asymmetric in this range. Implementing the parametrization presented in \cite{Stefansson2022} helps address this by constraining \vsini~to physical values. .

\section{RV vs. Index Correlation Plots}

\begin{figure*}[h!]
    \centering
    \includegraphics[width=0.99\textwidth]{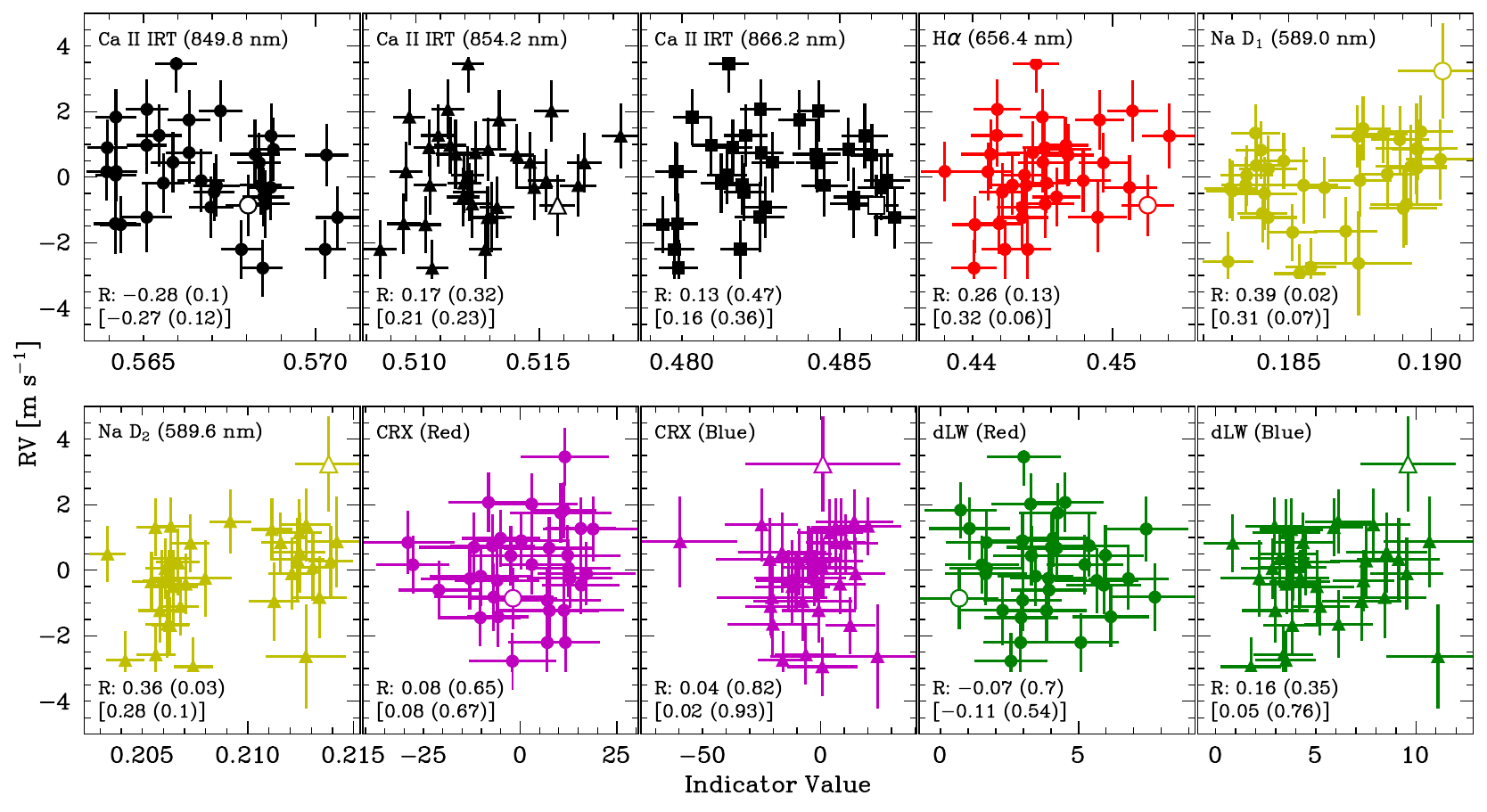}
    \caption{Radial velocity versus the various measured activity indicators from MAROON-X. Pearson R values are given with the p-value in parentheses. We observe tentative correlations in the Na D lines however the correlation is dependent on a single observation indicated by the open markers across all panels. The Pearson R and p-value for these indicators omitting this point are given in brackets. This observation also corresponds to the highest discrepancy in RVs derived between the blue and red arms.}
    \label{fig:activity}
\end{figure*}

\section{Parameters for Systems with Measured $\lambda$}
{\label{sec:lit_params_sec}}

\input{literature_obliquities}

\section{Injection \& Recovery for Single Planet Systems}{\label{sec:appen_injection}}

\begin{figure*}[h!]
    \centering
    \includegraphics[width=0.99\textwidth]{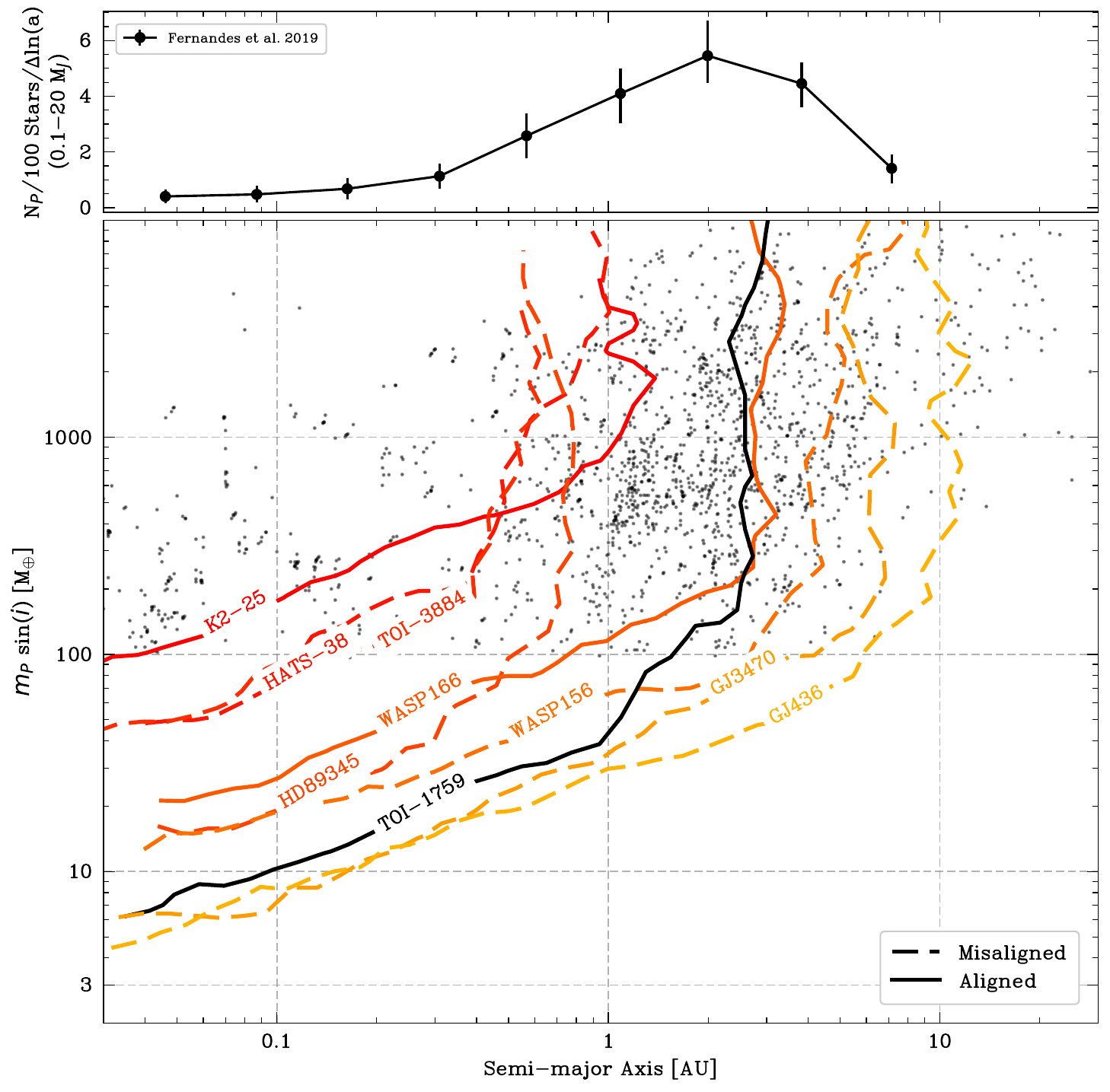}
    \caption{Injection and recovery results for nine out of twelve single planets with a measured obliquity. The remaining three (K2-105, Kepler-63 and K2-33) have too few RV measurements to be included ($<$10). Colored lines show the 50th percentile of recoveries for each archival radial velocity dataset. The black line shows the same contour but for TOI-1759. Grey points indicate RV-discovered giant planets ($m_{\text{p}}\sin{i}>0.5$ M$_{\text{J}}$) from the NASA Exoplanet Archive while the top panel shows the occurrence rate for giant planets (0.1-20 M$_{\text{Jup}}$) from \cite{Fernandes2019}. Sources for archival data are as follows: GJ 436 \citep{Rosenthal2021}, GJ 3470 \citep{Bonfils2012,Kosiarek2019}, HAT-38 \citep{Jordan2020}, HD 89345 \citep{Yu2018,vanEylen2018}, K2-25 \citep{Stefansson2020}, TOI-3884 \citep{LibbyRoberts2023}, WASP-166 \citep{Hellier2019}, WASP-156 \citep{Polanski2024,Demangeon2018}, TOI-1759A \citep{Polanski2024,Espinoza2022,Martioli2022} }
    \label{fig:injection}
\end{figure*}

\section{Radial Velocity Measurements}

\input{radial_velocities}

\end{document}

%% file: posterior_vals.tex
\renewcommand{\arraystretch}{1.2}
\begin{deluxetable*}{lccr}
\label{tab:params}
\tabletypesize{\footnotesize}
\tablecaption{Priors and Posterior Values}
\tablehead{\colhead{~~~Parameter Name [units]} &
\colhead{Posterior Value} & \colhead{Prior} & \colhead{Description}
}
\startdata
\sidehead{\textbf{Transit Parameters}}
P$_{orb}$ [days] & $18.850048\pm{0.000011}$ & $\mathcal{N}(18.850,0.001)$ & Orbital Period \\
T$_{0}$-2457000.0 [days] & $2442.91796\pm{0.00026}$ & $\mathcal{N}(2442.9,0.1)$ & Transit Midpoint \\
$b$ & $0.125^{+0.134}_{-0.087}$ & $\mathcal{U}(0.0,1.0)$ & Impact Parameter \\
$T_{\text{14}}$ [hours] & $3.491^{+0.027}_{-0.026}$ & $\mathcal{U}(0.2,9)$ & Full Transit Duration \\
\rprs & $0.04731^{+0.00078}_{-0.00076}$ & $\mathcal{U}(0.01,0.1)$ & Scaled Radius \\
$u_{0}$, TESS & $0.393^{+0.096}_{-0.094}$ & $\mathcal{N}(0.38,0.2)$ & Limb Darkening Parameter (TESS) \\
$u_{1}$, TESS & $0.49\pm{0.16}$ & $\mathcal{N}(0.37,0.2)$ & ... \\
$u_{0}$, $u_{1}$, $g$ & $\equiv$0.619, $\equiv$0.166 & - & Limb Darkening Parameters (MuSCAT3 $g$) \\
$u_{0}$, $u_{1}$, $r$ & $\equiv$0.481, $\equiv$0.269 & - & Limb Darkening Parameters (MuSCAT3 $r$) \\
$u_{0}$, $u_{1}$, $i$ & $\equiv$0.366, $\equiv$0.293 & - & Limb Darkening Parameters (MuSCAT3 $i$) \\
$u_{0}$, $u_{1}$, $z$ & $\equiv$0.253, $\equiv$0.326 & - & Limb Darkening Parameters (MuSCAT3 $z$) \\
\sidehead{\textbf{Rossiter-McLaughlin Model}}
$\lambda$ [deg] & $-4^{+19}_{-18}$ & $\mathcal{U}(-180,180)$ & Projected Obliquity \\
\vsini~ [\kms] & $0.95^{+0.12}_{-0.11}$ & - & Rotational Velocity \\
$R_{*}$ [R$_{\odot}$] & $0.631\pm{0.018}$ & $\mathcal{N}(0.628,0.018)$ & Stellar Radius \\
$P_{\text{rot}}$ [days] & $31.9^{+3.9}_{-3.8}$ & $\mathcal{N}(35,5)$ & Stellar Rotation Period \\
$\beta$ [\kms] & $3.5\pm{0.50}$ & $\mathcal{N}(3500,500)$ & Line Broadening Parameter \\
$u_{0}$, $u_{1}$, MX & $\equiv$0.497, $\equiv$0.179 & - & Limb Darkening Parameters (MAROON-X) \\
\sidehead{\textbf{Derived Parameters}}
R [R$_{\oplus}$] & $3.25\pm{0.11}$ & - & Planet Radius \\
$\psi$ [deg] & $24^{+12}_{+11}$ & - & True Obliquity \\
a/R$_*$  & $40.1\pm1.2$ & - & Scaled Semi-major Axis \\
\sidehead{\textbf{Systematics \& Offsets}}
$\rho_{\text{GP}}$ & $0.199^{+0.053}_{-0.044}$ & $\mathcal{U}(0.001,5)$ & GP Period \\
$Q_{\text{GP}}$ & $\equiv \frac{1}{3}$ & - & GP Quality Factor \\
$\sigma_{\text{GP},g}$ & $0.001075^{+0.000046}_{-0.000043}$ & $\Gamma^{-1}$($\widetilde{\sigma}_{{\text{Flux}}}$,std(Flux)) & GP Standard Deviation (MuSCAT3 $g$) \\
$\sigma_{\text{GP},r}$ & $0.000814^{+0.000047}_{-0.000044}$ & $\Gamma^{-1}$($\widetilde{\sigma}_{{\text{Flux}}}$,std(Flux)) & ... (MuSCAT3 $r$) \\
$\sigma_{\text{GP},i}$ & $0.000872^{+0.000056}_{-0.000051}$ & $\Gamma^{-1}$($\widetilde{\sigma}_{{\text{Flux}}}$,std(Flux)) & ... (MuSCAT3 $i$) \\
$\sigma_{\text{GP},z}$ & $0.00112^{+0.00011}_{-0.00010}$ & $\Gamma^{-1}$($\widetilde{\sigma}_{{\text{Flux}}}$,std(Flux)) & ... (MuSCAT3 $z$) \\
$\sigma_{\text{TESS}}$ & $0.000915^{+0.000058}_{-0.000053}$ & $\Gamma^{-1}$($\widetilde{\sigma}_{{\text{Flux}}}$,std(Flux)) & White Noise (TESS) \\
$\sigma_{g}$ & $0.00094^{+0.000033}_{-0.000032}$ & $\Gamma^{-1}$($\widetilde{\sigma}_{{\text{Flux}}}$,std(Flux)) & White Noise (MuSCAT3 $g$) \\
$\sigma_{r}$ & $0.00073\pm{0.000032}$ & $\Gamma^{-1}$($\widetilde{\sigma}_{{\text{Flux}}}$,std(Flux)) & ... (MuSCAT3 $r$) \\
$\sigma_{i}$ & $0.000756^{+0.000035}_{-0.000034}$ & $\Gamma^{-1}$($\widetilde{\sigma}_{{\text{Flux}}}$,std(Flux)) & ... (MuSCAT3 $i$) \\
$\sigma_{z}$ & $0.000833^{+0.000050}_{-0.000046}$ & $\Gamma^{-1}$($\widetilde{\sigma}_{{\text{Flux}}}$,std(Flux)) & ... (MuSCAT3 $z$) \\
$\mu_{\text{MX}}$ [\ms] & $-0.03^{+0.13}_{-0.12}$ & $\mathcal{N}(0.0,0.5)$ & Offset (MAROON-X) \\
$\mu_{\text{TESS}}$ & $0.999884^{+0.000050}_{-0.000049}$ & $\mathcal{N}(1.0,1.0)$ & Offset (TESS) \\
$\mu_{g}$ & $1.00275^{+0.00086}_{-0.00085}$ & $\mathcal{N}(1.0,1.0)$ & Offset (MuSCAT3 $g$) \\
$\mu_{r}$ & $1.00152\pm{0.00064}$ & $\mathcal{N}(1.0,1.0)$ & ... (MuSCAT3 $r$) \\
$\mu_{i}$ & $1.00154^{+0.00067}_{-0.00069}$ & $\mathcal{N}(1.0,1.0)$ & ... (MuSCAT3 $i$) \\
$\mu_{z}$ & $1.00138^{+0.00087}_{-0.00088}$ & $\mathcal{N}(1.0,1.0)$ & ... (MuSCAT3 $z$) \\
\enddata
%\tablenotetext{a}{$\mathcal{N}$ is a normal prior with $\mathcal{N}$(mean, standard deviation)}
%\tablenotetext{b}{$\mathcal{U}$ is a uniform prior with $\mathcal{U}$(lower,upper)}
\end{deluxetable*}

%% file: companion.tex
\begin{deluxetable}{lcc}[t]
\tablewidth{\textwidth} 
\tabletypesize{\normalsize}
\setlength{\tabcolsep}{0.1cm}
\tablecaption{TOI-1759AB Astrometric Properties \label{tab:companion}}
\tablehead{
  \colhead{Parameter [units]} & 
  \colhead{Value} & 
  \colhead{Reference}
}
\startdata
$\alpha_{\text{A}}$ R.A. [deg] & $326.851\pm0.009$ & \textit{Gaia} DR3$^a$ \\
$\delta_{\text{A}}$ Dec. [deg] & $62.753\pm0.009$ & ... \\
$\alpha_{\text{B}}$ R.A. [deg] & $326.85\pm0.07$ & ... \\
$\delta_{\text{B}}$ Dec. [deg] & $62.77\pm0.07$ & ... \\
$\mu_{\alpha,\text{A}}$ [mas yr$^{-1}$] & $-173.42\pm0.01$ & ...\\
$\mu_{\delta,\text{A}}$ [mas yr$^{-1}$] &  $-10.65\pm0.01$ & ...\\
$\mu_{\alpha,\text{B}}$ [mas yr$^{-1}$] & $-174.76\pm0.09$ & ...\\
$\mu_{\delta,\text{B}}$ [mas yr$^{-1}$] &  $-11.14\pm0.08$ & ...\\
$\pi_{\text{A}}$ [mas] & $24.92\pm0.01$ & ...\\
$\pi_{\text{B}}$ [mas] & $24.83\pm0.08$ & ... \\
$\rho^b$ [arcsec] & 69.04 & Calculated \\
$s^c$ [AU] & 2770 & Calculated\\
$|Q^*_g|^d$ [Joules]  & $4.3\times10^{34}$ & Calculated\\\hline
\enddata
\tablenotetext{a}{\cite{Gaia2023} }
\tablenotetext{b}{Projected angular separation. }
\tablenotetext{c}{Projected physical separation. }
\tablenotetext{d}{Reduced binding energy calculated via $\text{G}\frac{\text{M}_1 \text{M}_2}{s}$ }
\end{deluxetable}

%% file: literature_obliquities.tex
\renewcommand{\arraystretch}{1.2}
\begin{deluxetable*}{lcccccc}[h]
\label{tab:literature}
\tabletypesize{\footnotesize}
\tablecaption{Properties of Systems with Measured $\lambda$ (R$_{P}<8$\rearth )}
\tablehead{\colhead{Planet Name~$^{N_{\text{Stars}}}_{N_{\text{Planets}}}$} &
\colhead{$\lambda$ [\degree]} 
& \colhead{\teff~[K]} 
& \colhead{$R_{\text{p}}$ [R$_{\oplus}$]} 
& \colhead{Period [days]}
& \colhead{a/R$_*$}
& \colhead{Reference(s)}
}
\startdata
Kepler-25 c~$^2_3$ & $-0.9^{+7.7}_{-6.4}$ & $6354$ & $5.2$ & $12.7$ & $18.4$ & \begin{tabular}{@{}c@{}}\cite{Bourrier2023} \\ \cite{ Benomar2014} \\ \cite{ Mills2019}\end{tabular} \\
\hline
HD 106315 c~$^1_2$ & $-2.7^{+2.7}_{-2.6}$ & $6327$ & $4.2$ & $21.1$ & $25.5$ & \cite{Bourrier2023,Barros2017} \\
\hline
K2-290 b~$^3_2$ & $173.0^{+45.0}_{-53.0}$ & $6302$ & $3.1$ & $9.2$ & $13.2$ & \cite{Hjorth2021,Hjorth2019} \\
\hline
TOI-5126 b~$^1_2$ & $1.0\pm48.0$ & $6150$ & $4.7$ & $5.5$ & $11.3$ & \cite{Radzom2024,Fairnington2024} \\
\hline
pi Men c~$^1_3$ & $-24.0\pm4.1$ & $5998$ & $2.0$ & $6.3$ & $12.5$ & \cite{Kunovac2021,Damasso2020} \\
\hline
TOI-1136 d~$^1_6$ & $5.0\pm5.0$ & $5770$ & $4.6$ & $12.5$ & $23.5$ & \cite{Dai2023} \\
\hline
Kepler-1656 b~$^1_2$ & $35.0^{+15.0}_{-22.0}$ & $5731$ & $5.0$ & $31.6$ & $38.7$ & \cite{Rubenzahl2024,Brady2018} \\
\hline
V1298 Tau c~$^3_4$ & $4.9^{+15.0}_{-15.1}$ & $5700$ & $5.2$ & $8.2$ & $14.2$ & \cite{Feinstein2021,SaurezMascareno2022} \\
\hline
HD 63433 b~$^1_3$ & $8.0^{+33.0}_{-45.0}$ & $5688$ & $2.1$ & $7.1$ & $16.8$ & \cite{Mann2020,Capistrant2024} \\
\hline
Kepler-30 b~$^1_3$ & $4.0\pm10.0$ & $5498$ & $3.9$ & $29.3$ & $42.7$ & \cite{SanchisOjeda2012} \\
\hline
HD 191939 b~$^1_6$ & $3.7\pm5.0$ & $5427$ & $3.4$ & $8.9$ & $18.4$ & \begin{tabular}{@{}c@{}}\cite{Lubin2024} \\ \cite{ OrellMiqual2023} \\ \cite{ BadenasAgusti2020}\end{tabular} \\
\hline
HD 3167 b~$^1_4$ & $-6.6^{+6.6}_{-7.9}$ & $5300$ & $1.6$ & $1.0$ & $4.5$ & \cite{Bourrier2021,Bourrier2022b} \\
\hline
HD 3167 c~$^1_4$ & $-108.9^{+5.4}_{-5.5}$ & $5300$ & $2.9$ & $29.8$ & $44.0$ & \cite{Bourrier2022b,Bourrier2021} \\
\hline
HD 110067 c~$^3_6$ & $6.0^{+24.0}_{-26.0}$ & $5266$ & $2.4$ & $13.7$ & $28.4$ & \cite{Zak2024,Luque2023} \\
\hline
TOI-2076 b~$^1_3$ & $-3.0^{+15.0}_{-16.0}$ & $5201$ & $2.4$ & $10.4$ & $25.1$ & \cite{Frazier2023} \\
\hline
55 Cnc e~$^2_5$ & $10.0^{+17.0}_{-20.0}$ & $5172$ & $1.9$ & $0.7$ & $3.5$ & \cite{Zhao2023,Bourrier2018} \\
\hline
TOI-942 b~$^3_2$ & $1.0^{+41.0}_{-33.0}$ & $4969$ & $4.2$ & $4.3$ & $11.7$ & \cite{Wirth2021,Carleo2021} \\
\hline
TOI-942 c~$^3_2$ & $24.0\pm14.0$ & $4969$ & $4.8$ & $10.2$ & $20.7$ & \cite{Carleo2021,Teng2024} \\
\hline
HAT-P-11 b~$^1_2$ & $133.9^{+7.1}_{-8.3}$ & $4780$ & $4.9$ & $4.9$ & $15.1$ & \cite{Bourrier2023,Basilicata2024} \\
\hline
AU Mic b~$^1_3$ & $-4.7^{+6.8}_{-6.4}$ & $3678$ & $4.0$ & $8.5$ & $18.8$ & \cite{Wittrock2023,Hirano2020b} \\
\hline
AU Mic c~$^1_3$ & $68.0^{+32.0}_{-49.0}$ & $3678$ & $2.5$ & $18.9$ & $32.0$ & \cite{Yu2025,Hirano2020b} \\
\hline
TRAPPIST-1 b~$^1_7$ & $15.0^{+26.0}_{-30.0}$ & $2566$ & $1.1$ & $1.5$ & $20.8$ & \cite{Hirano2020a,Algol2021} \\
\hline
TRAPPIST-1 e~$^1_7$ & $9.0^{+45.0}_{-51.0}$ & $2566$ & $0.9$ & $6.1$ & $52.9$ & \cite{Hirano2020a,Algol2021} \\
\hline
TRAPPIST-1 f~$^1_7$ & $21.0\pm32.0$ & $2566$ & $1.0$ & $9.2$ & $69.5$ & \cite{Hirano2020a,Algol2021} \\
\hline
\hline
WASP-166 b~$^1_1$ & $-15.5^{+2.9}_{-2.8}$ & $6050$ & $7.1$ & $5.4$ & $11.3$ & \cite{Doyle2022,Hellier2019} \\
\hline
HATS-38 b~$^1_1$ & $-108.0^{+11.0}_{-16.0}$ & $5732$ & $6.9$ & $4.4$ & $9.8$ & \cite{EspinozaRetamal2024b,Jordan2020} \\
\hline
K2-105 b~$^1_1$ & $-81.0^{+50.0}_{-47.0}$ & $5636$ & $3.6$ & $8.3$ & $18.5$ & \cite{Bourrier2023,CastroGonzlez2022} \\
\hline
HD 89345 b~$^1_1$ & $74.0^{+34.0}_{-33.0}$ & $5576$ & $7.4$ & $11.8$ & $13.1$ & \cite{Bourrier2023,Yu2018} \\
\hline
Kepler-63 b~$^1_1$ & $-110.0^{+21.0}_{-27.0}$ & $5576$ & $6.1$ & $9.4$ & $19.1$ & \cite{Bourrier2023,SanchisOjeda2013} \\
\hline
WASP-156 b~$^1_1$ & $106.0\pm14.0$ & $4910$ & $5.7$ & $3.8$ & $12.8$ & \cite{Bourrier2023,Demangeon2018} \\
\hline
TOI-1759 b~$^2_1$ & $-4.0^{+19.0}_{-18.0}$ & $3970$ & $3.2$ & $18.9$ & $40.1$ & \cite{Martioli2022}, This Work \\
\hline
GJ 3470 b~$^1_1$ & $101.0^{+29.0}_{-14.0}$ & $3652$ & $3.9$ & $3.3$ & $12.9$ & \cite{Stefansson2022,Kosiarek2019} \\
\hline
K2-33 b~$^1_1$ & $-10.0^{+22.0}_{-24.0}$ & $3540$ & $5.0$ & $5.4$ & $10.4$ & \cite{Hirano2024,Mann2016} \\
\hline
GJ 436 b~$^1_1$ & $114.0^{+23.0}_{-17.0}$ & $3479$ & $4.2$ & $2.6$ & $13.7$ & \cite{Bourrier2022a,Maciejewski2014} \\
\hline
K2-25 b~$^1_1$ & $3.0\pm16.0$ & $3207$ & $3.4$ & $3.5$ & $21.1$ & \cite{Stefansson2020} \\
\hline
TOI-3884 b~$^1_1$ & $75.0\pm10.0$ & $3180$ & $6.4$ & $4.5$ & $25.9$ & \cite{LibbyRoberts2023} \\
\enddata
\tablecomments{Uncertainties for \teff, period, R$_{P}$, and a/R$_*$ have been omitted and period values are truncated. Uncertainties and full-precision values are available in an online machine-readable form. }
\end{deluxetable*}

%% file: radial_velocities.tex
\renewcommand{\arraystretch}{1.2}
\begin{deluxetable*}{lcccc}[h]
\label{tab:priors}
\tabletypesize{\footnotesize}
\tablecaption{MAROON-X Radial Velocities}
\tablehead{\colhead{Time [BJD-2457000.0]} &
\colhead{RV$_{\text{Red}}$ [\ms]} 
& \colhead{$\sigma_{\text{Red}}$ [\ms]} 
& \colhead{RV$_{\text{Blue}}$ [\ms]} 
& \colhead{$\sigma_{\text{Blue}}$ [\ms]}
}
\startdata
2442.76290 & 0.06 & 0.96 & -2.31 & 1.58 \\
2442.77096 & -0.54 & 0.94 & 3.56 & 1.45 \\
2442.77917 & 0.21 & 0.97 & 1.19 & 1.38 \\
2442.78725 & 0.77 & 0.91 & 0.65 & 1.25 \\
2442.79539 & -0.60 & 0.93 & 0.86 & 1.22 \\
2442.80351 & 1.07 & 0.95 & -0.63 & 1.20 \\
2442.81154 & -0.90 & 1.07 & -0.51 & 1.25 \\
2442.81985 & 0.75 & 0.91 & 0.60 & 1.08 \\
2442.82809 & 1.57 & 1.00 & 1.70 & 1.11 \\
2442.83618 & -0.00 & 1.00 & 0.21 & 1.10 \\
2442.84440 & 2.33 & 0.94 & 0.40 & 1.02 \\
2442.85250 & 2.06 & 0.91 & 1.80 & 0.98 \\
2442.86054 & 1.58 & 0.96 & 1.46 & 1.02 \\
2442.86874 & 0.12 & 0.91 & 1.54 & 0.94 \\
2442.87691 & 2.37 & 0.91 & 1.14 & 0.94 \\
2442.88512 & 1.27 & 0.87 & 1.61 & 0.89 \\
2442.89319 & 3.76 & 0.88 & 1.12 & 0.89 \\
2442.90137 & 2.14 & 0.86 & 0.06 & 0.86 \\
2442.90944 & 0.36 & 0.88 & -0.81 & 0.88 \\
2442.91767 & 1.20 & 0.87 & 0.79 & 0.87 \\
2442.92578 & -1.16 & 0.88 & 1.64 & 0.88 \\
2442.93395 & -1.13 & 0.91 & -0.13 & 0.92 \\
2442.94208 & 0.46 & 0.91 & -0.06 & 0.91 \\
2442.95025 & 0.45 & 0.88 & 0.65 & 0.89 \\
2442.95846 & -1.92 & 0.89 & -2.29 & 0.90 \\
2442.96659 & -2.49 & 0.87 & -1.39 & 0.87 \\
2442.97469 & -0.18 & 0.88 & -2.47 & 0.91 \\
2442.98290 & -1.92 & 0.89 & -2.66 & 0.91 \\
2442.99100 & 0.03 & 0.91 & 0.33 & 0.94 \\
2442.99923 & -0.33 & 0.90 & -0.04 & 0.93 \\
2443.00723 & 0.95 & 0.94 & -0.95 & 0.99 \\
2443.01567 & 1.13 & 0.96 & 0.53 & 1.02 \\
2443.02355 & -0.95 & 0.95 & -1.37 & 1.03 \\
2443.03165 & -0.55 & 1.04 & -0.23 & 1.13 \\
2443.03998 & 0.97 & 1.06 & 0.02 & 1.18 \\
\enddata
\tablecomments{This table is available in its entirety in a machine-readable form. The online table also contains additional information pertaining to observing conditions (airmass, SNR, etc.) and the stellar activity indicators discussed in \S\ref{sec:rv_activity_corr}.} 
\end{deluxetable*}

%% file: main.bbl
\begin{thebibliography}{}
\expandafter\ifx\csname natexlab\endcsname\relax\def\natexlab#1{#1}\fi
\providecommand{\url}[1]{\href{#1}{#1}}
\providecommand{\dodoi}[1]{doi:~\href{http://doi.org/#1}{\nolinkurl{#1}}}
\providecommand{\doeprint}[1]{\href{http://ascl.net/#1}{\nolinkurl{http://ascl.net/#1}}}
\providecommand{\doarXiv}[1]{\href{https://arxiv.org/abs/#1}{\nolinkurl{https://arxiv.org/abs/#1}}}

\bibitem[{{Agol} {et~al.}(2021){Agol}, {Dorn}, {Grimm}, {Turbet}, {Ducrot}, {Delrez}, {Gillon}, {Demory}, {Burdanov}, {Barkaoui}, {Benkhaldoun}, {Bolmont}, {Burgasser}, {Carey}, {de Wit}, {Fabrycky}, {Foreman-Mackey}, {Haldemann}, {Hernandez}, {Ingalls}, {Jehin}, {Langford}, {Leconte}, {Lederer}, {Luger}, {Malhotra}, {Meadows}, {Morris}, {Pozuelos}, {Queloz}, {Raymond}, {Selsis}, {Sestovic}, {Triaud}, \& {Van Grootel}}]{Algol2021}
{Agol}, E., {Dorn}, C., {Grimm}, S.~L., {et~al.} 2021, \psj, 2, 1, \dodoi{10.3847/PSJ/abd022}

\bibitem[{{Akeson} {et~al.}(2013){Akeson}, {Chen}, {Ciardi}, {Crane}, {Good}, {Harbut}, {Jackson}, {Kane}, {Laity}, {Leifer}, {Lynn}, {McElroy}, {Papin}, {Plavchan}, {Ram{\'\i}rez}, {Rey}, {von Braun}, {Wittman}, {Abajian}, {Ali}, {Beichman}, {Beekley}, {Berriman}, {Berukoff}, {Bryden}, {Chan}, {Groom}, {Lau}, {Payne}, {Regelson}, {Saucedo}, {Schmitz}, {Stauffer}, {Wyatt}, \& {Zhang}}]{NEA}
{Akeson}, R.~L., {Chen}, X., {Ciardi}, D., {et~al.} 2013, \pasp, 125, 989, \dodoi{10.1086/672273}

\bibitem[{{Albrecht} {et~al.}(2012){Albrecht}, {Winn}, {Butler}, {Crane}, {Shectman}, {Thompson}, {Hirano}, \& {Wittenmyer}}]{Albrecht2012}
{Albrecht}, S., {Winn}, J.~N., {Butler}, R.~P., {et~al.} 2012, \apj, 744, 189, \dodoi{10.1088/0004-637X/744/2/189}

\bibitem[{{Albrecht} {et~al.}(2011){Albrecht}, {Winn}, {Johnson}, {Butler}, {Crane}, {Shectman}, {Thompson}, {Narita}, {Sato}, {Hirano}, {Enya}, \& {Fischer}}]{Albrecht2011b}
{Albrecht}, S., {Winn}, J.~N., {Johnson}, J.~A., {et~al.} 2011, \apj, 738, 50, \dodoi{10.1088/0004-637X/738/1/50}

\bibitem[{{Anderson} {et~al.}(2011){Anderson}, {Collier Cameron}, {Gillon}, {Hellier}, {Jehin}, {Lendl}, {Queloz}, {Smalley}, {Triaud}, \& {Vanhuysse}}]{Anderson2011a}
{Anderson}, D.~R., {Collier Cameron}, A., {Gillon}, M., {et~al.} 2011, \aap, 534, A16, \dodoi{10.1051/0004-6361/201117597}

\bibitem[{{Badenas-Agusti} {et~al.}(2020){Badenas-Agusti}, {G{\"u}nther}, {Daylan}, {Mikal-Evans}, {Vanderburg}, {Huang}, {Matthews}, {Rackham}, {Bieryla}, {Stassun}, {Kane}, {Shporer}, {Fulton}, {Hill}, {Nowak}, {Ribas}, {Pall{\'e}}, {Jenkins}, {Latham}, {Seager}, {Ricker}, {Vanderspek}, {Winn}, {Abril-Pla}, {Collins}, {Serra}, {Niraula}, {Rustamkulov}, {Barclay}, {Crossfield}, {Howell}, {Ciardi}, {Gonzales}, {Schlieder}, {Caldwell}, {Fausnaugh}, {McDermott}, {Paegert}, {Pepper}, {Rose}, \& {Twicken}}]{BadenasAgusti2020}
{Badenas-Agusti}, M., {G{\"u}nther}, M.~N., {Daylan}, T., {et~al.} 2020, \aj, 160, 113, \dodoi{10.3847/1538-3881/aba0b5}

\bibitem[{{Barros} {et~al.}(2017){Barros}, {Gosselin}, {Lillo-Box}, {Bayliss}, {Delgado Mena}, {Brugger}, {Santerne}, {Armstrong}, {Adibekyan}, {Armstrong}, {Barrado}, {Bento}, {Boisse}, {Bonomo}, {Bouchy}, {Brown}, {Cochran}, {Collier Cameron}, {Deleuil}, {Demangeon}, {D{\'\i}az}, {Doyle}, {Dumusque}, {Ehrenreich}, {Espinoza}, {Faedi}, {Faria}, {Figueira}, {Foxell}, {H{\'e}brard}, {Hojjatpanah}, {Jackman}, {Lendl}, {Ligi}, {Lovis}, {Melo}, {Mousis}, {Neal}, {Osborn}, {Pollacco}, {Santos}, {Sefako}, {Shporer}, {Sousa}, {Triaud}, {Udry}, {Vigan}, \& {Wyttenbach}}]{Barros2017}
{Barros}, S.~C.~C., {Gosselin}, H., {Lillo-Box}, J., {et~al.} 2017, \aap, 608, A25, \dodoi{10.1051/0004-6361/201731276}

\bibitem[{{Basilicata} {et~al.}(2024){Basilicata}, {Giacobbe}, {Bonomo}, {Scandariato}, {Brogi}, {Singh}, {Di Paola}, {Mancini}, {Sozzetti}, {Lanza}, {Cubillos}, {Damasso}, {Desidera}, {Biazzo}, {Bignamini}, {Borsa}, {Cabona}, {Carleo}, {Ghedina}, {Guilluy}, {Maggio}, {Mainella}, {Micela}, {Molinari}, {Molinaro}, {Nardiello}, {Pedani}, {Pino}, {Poretti}, {Southworth}, {Stangret}, \& {Turrini}}]{Basilicata2024}
{Basilicata}, M., {Giacobbe}, P., {Bonomo}, A.~S., {et~al.} 2024, \aap, 686, A127, \dodoi{10.1051/0004-6361/202347659}

\bibitem[{{Beard} {et~al.}(2024){Beard}, {Robertson}, {Dai}, {Holcomb}, {Lubin}, {Akana Murphy}, {Batalha}, {Blunt}, {Crossfield}, {Dressing}, {Fulton}, {Howard}, {Huber}, {Isaacson}, {Kane}, {Nowak}, {Petigura}, {Roy}, {Rubenzahl}, {Weiss}, {Barrena}, {Behmard}, {Brinkman}, {Carleo}, {Chontos}, {Dalba}, {Fetherolf}, {Giacalone}, {Hill}, {Kawauchi}, {Korth}, {Luque}, {MacDougall}, {Mayo}, {Mo{\v{c}}nik}, {Morello}, {Murgas}, {Orell-Miquel}, {Palle}, {Polanski}, {Rice}, {Scarsdale}, {Tyler}, \& {Van Zandt}}]{Beard2024a}
{Beard}, C., {Robertson}, P., {Dai}, F., {et~al.} 2024, \aj, 167, 70, \dodoi{10.3847/1538-3881/ad1330}

\bibitem[{{Behmard} {et~al.}(2022){Behmard}, {Dai}, \& {Howard}}]{Behmard2022}
{Behmard}, A., {Dai}, F., \& {Howard}, A.~W. 2022, \aj, 163, 160, \dodoi{10.3847/1538-3881/ac53a7}

\bibitem[{{Benomar} {et~al.}(2014){Benomar}, {Masuda}, {Shibahashi}, \& {Suto}}]{Benomar2014}
{Benomar}, O., {Masuda}, K., {Shibahashi}, H., \& {Suto}, Y. 2014, \pasj, 66, 94, \dodoi{10.1093/pasj/psu069}

\bibitem[{{Bonfils} {et~al.}(2012){Bonfils}, {Gillon}, {Udry}, {Armstrong}, {Bouchy}, {Delfosse}, {Forveille}, {Fumel}, {Jehin}, {Lendl}, {Lovis}, {Mayor}, {McCormac}, {Neves}, {Pepe}, {Perrier}, {Pollaco}, {Queloz}, \& {Santos}}]{Bonfils2012}
{Bonfils}, X., {Gillon}, M., {Udry}, S., {et~al.} 2012, \aap, 546, A27, \dodoi{10.1051/0004-6361/201219623}

\bibitem[{Bourque {et~al.}(2021)Bourque, Espinoza, Filippazzo, Fix, King, Martlin, Medina, Batalha, Fox, Fowler, Fraine, Hill, Lewis, Stevenson, Valenti, \& Wakeford}]{exoctk}
Bourque, M., Espinoza, N., Filippazzo, J., {et~al.} 2021, The Exoplanet Characterization Toolkit (ExoCTK), 1.0.0,  Zenodo, \dodoi{10.5281/zenodo.4556063}

\bibitem[{{Bourrier} {et~al.}(2018){Bourrier}, {Dumusque}, {Dorn}, {Henry}, {Astudillo-Defru}, {Rey}, {Benneke}, {H{\'e}brard}, {Lovis}, {Demory}, {Moutou}, \& {Ehrenreich}}]{Bourrier2018}
{Bourrier}, V., {Dumusque}, X., {Dorn}, C., {et~al.} 2018, \aap, 619, A1, \dodoi{10.1051/0004-6361/201833154}

\bibitem[{{Bourrier} {et~al.}(2021){Bourrier}, {Lovis}, {Cretignier}, {Allart}, {Dumusque}, {Delisle}, {Deline}, {Sousa}, {Adibekyan}, {Alibert}, {Barros}, {Borsa}, {Cristiani}, {Demangeon}, {Ehrenreich}, {Figueira}, {Gonz{\'a}lez Hern{\'a}ndez}, {Lendl}, {Lillo-Box}, {Lo Curto}, {Di Marcantonio}, {Martins}, {M{\'e}gevand}, {Mehner}, {Micela}, {Molaro}, {Oshagh}, {Palle}, {Pepe}, {Poretti}, {Rebolo}, {Santos}, {Scandariato}, {Seidel}, {Sozzetti}, {Su{\'a}rez Mascare{\~n}o}, \& {Zapatero Osorio}}]{Bourrier2021}
{Bourrier}, V., {Lovis}, C., {Cretignier}, M., {et~al.} 2021, \aap, 654, A152, \dodoi{10.1051/0004-6361/202141527}

\bibitem[{{Bourrier} {et~al.}(2022{\natexlab{a}}){Bourrier}, {Deline}, {Krenn}, {Egger}, {Petit}, {Malavolta}, {Cretignier}, {Billot}, {Broeg}, {Flor{\'e}n}, {Queloz}, {Alibert}, {Bonfanti}, {Bonomo}, {Delisle}, {Demangeon}, {Demory}, {Dumusque}, {Ehrenreich}, {Haywood}, {Howell}, {Lendl}, {Mortier}, {Nigro}, {Salmon}, {Sousa}, {Wilson}, {Adibekyan}, {Alonso}, {Anglada}, {B{\'a}rczy}, {Barrado y Navascues}, {Barros}, {Baumjohann}, {Beck}, {Benz}, {Biondi}, {Bonfils}, {Brandeker}, {Cabrera}, {Charnoz}, {Csizmadia}, {Collier Cameron}, {Damasso}, {Davies}, {Deleuil}, {Delrez}, {Di Fabrizio}, {Erikson}, {Fortier}, {Fossati}, {Fridlund}, {Gandolfi}, {Gillon}, {G{\"u}del}, {Heng}, {Hoyer}, {Isaak}, {Kiss}, {Laskar}, {Lecavelier des Etangs}, {Lorenzi}, {Lovis}, {Magrin}, {Massa}, {Maxted}, {Nascimbeni}, {Olofsson}, {Ottensamer}, {Pagano}, {Pall{\'e}}, {Peter}, {Piotto}, {Pollacco}, {Ragazzoni}, {Rando}, {Rauer}, {Ribas}, {Santos}, {Scandariato}, {S{\'e}gransan}, {Simon}, {Smith}, {Steller}, {Szab{\'o}}, {Thomas},
  {Udry}, {Van Grootel}, {Verrecchia}, {Walton}, {Beck}, {Buder}, {Ratti}, {Ulmer}, \& {Viotto}}]{Bourrier2022b}
{Bourrier}, V., {Deline}, A., {Krenn}, A., {et~al.} 2022{\natexlab{a}}, \aap, 668, A31, \dodoi{10.1051/0004-6361/202243778}

\bibitem[{{Bourrier} {et~al.}(2022{\natexlab{b}}){Bourrier}, {Zapatero Osorio}, {Allart}, {Attia}, {Cretignier}, {Dumusque}, {Lovis}, {Adibekyan}, {Borsa}, {Figueira}, {Gonz{\'a}lez Hern{\'a}ndez}, {Mehner}, {Santos}, {Schmidt}, {Seidel}, {Sozzetti}, {Alibert}, {Casasayas-Barris}, {Ehrenreich}, {Lo Curto}, {Martins}, {Di Marcantonio}, {M{\'e}gevand}, {Nunes}, {Palle}, {Poretti}, \& {Sousa}}]{Bourrier2022a}
{Bourrier}, V., {Zapatero Osorio}, M.~R., {Allart}, R., {et~al.} 2022{\natexlab{b}}, \aap, 663, A160, \dodoi{10.1051/0004-6361/202142559}

\bibitem[{{Bourrier} {et~al.}(2023){Bourrier}, {Attia}, {Mallonn}, {Marret}, {Lendl}, {Konig}, {Krenn}, {Cretignier}, {Allart}, {Henry}, {Bryant}, {Leleu}, {Nielsen}, {Hebrard}, {Hara}, {Ehrenreich}, {Seidel}, {dos Santos}, {Lovis}, {Bayliss}, {Cegla}, {Dumusque}, {Boisse}, {Boucher}, {Bouchy}, {Pepe}, {Lavie}, {Rey Cerda}, {S{\'e}gransan}, {Udry}, \& {Vrignaud}}]{Bourrier2023}
{Bourrier}, V., {Attia}, M., {Mallonn}, M., {et~al.} 2023, \aap, 669, A63, \dodoi{10.1051/0004-6361/202245004}

\bibitem[{{Brady} {et~al.}(2023){Brady}, {Bean}, {Seifahrt}, {Kasper}, {Luque}, {Reiners}, {Benneke}, {Stef{\'a}nsson}, \& {St{\"u}rmer}}]{Brady2023}
{Brady}, M., {Bean}, J.~L., {Seifahrt}, A., {et~al.} 2023, \aj, 165, 129, \dodoi{10.3847/1538-3881/acb5f7}

\bibitem[{{Brady} {et~al.}(2018){Brady}, {Petigura}, {Knutson}, {Sinukoff}, {Isaacson}, {Hirsch}, {Fulton}, {Kosiarek}, \& {Howard}}]{Brady2018}
{Brady}, M.~T., {Petigura}, E.~A., {Knutson}, H.~A., {et~al.} 2018, \aj, 156, 147, \dodoi{10.3847/1538-3881/aad773}

\bibitem[{{Brinkman} {et~al.}(2023){Brinkman}, {Weiss}, {Dai}, {Huber}, {Kite}, {Valencia}, {Bean}, {Beard}, {Behmard}, {Blunt}, {Brady}, {Fulton}, {Giacalone}, {Howard}, {Isaacson}, {Kasper}, {Lubin}, {MacDougall}, {Akana Murphy}, {Plotnykov}, {Polanski}, {Rice}, {Seifahrt}, {Stef{\'a}nsson}, \& {St{\"u}rmer}}]{Brinkman2023}
{Brinkman}, C.~L., {Weiss}, L.~M., {Dai}, F., {et~al.} 2023, \aj, 165, 88, \dodoi{10.3847/1538-3881/acad83}

\bibitem[{{Capistrant} {et~al.}(2024){Capistrant}, {Soares-Furtado}, {Vanderburg}, {Jankowski}, {Mann}, {Ross}, {Srdoc}, {Hinkel}, {Becker}, {Magliano}, {Limbach}, {Stephan}, {Nine}, {Tofflemire}, {Kraus}, {Giacalone}, {Winn}, {Bieryla}, {Bouma}, {Ciardi}, {Collins}, {Covone}, {de Beurs}, {Huang}, {Jenkins}, {Kreidberg}, {Latham}, {Quinn}, {Seager}, {Shporer}, {Twicken}, {Wohler}, {Vanderspek}, {Yarza}, \& {Ziegler}}]{Capistrant2024}
{Capistrant}, B.~K., {Soares-Furtado}, M., {Vanderburg}, A., {et~al.} 2024, \aj, 167, 54, \dodoi{10.3847/1538-3881/ad1039}

\bibitem[{{Carleo} {et~al.}(2021){Carleo}, {Desidera}, {Nardiello}, {Malavolta}, {Lanza}, {Livingston}, {Locci}, {Marzari}, {Messina}, {Turrini}, {Baratella}, {Borsa}, {D'Orazi}, {Nascimbeni}, {Pinamonti}, {Rainer}, {Alei}, {Bignamini}, {Gratton}, {Micela}, {Montalto}, {Sozzetti}, {Squicciarini}, {Affer}, {Benatti}, {Biazzo}, {Bonomo}, {Claudi}, {Cosentino}, {Covino}, {Damasso}, {Esposito}, {Fiorenzano}, {Frustagli}, {Giacobbe}, {Harutyunyan}, {Leto}, {Magazz{\`u}}, {Maggio}, {Mainella}, {Maldonado}, {Mallonn}, {Mancini}, {Molinari}, {Molinaro}, {Pagano}, {Pedani}, {Piotto}, {Poretti}, {Redfield}, \& {Scandariato}}]{Carleo2021}
{Carleo}, I., {Desidera}, S., {Nardiello}, D., {et~al.} 2021, \aap, 645, A71, \dodoi{10.1051/0004-6361/202039042}

\bibitem[{{Castro-Gonz{\'a}lez} {et~al.}(2022){Castro-Gonz{\'a}lez}, {D{\'\i}ez Alonso}, {Men{\'e}ndez Blanco}, {Livingston}, {de Leon}, {Lillo-Box}, {Korth}, {Fern{\'a}ndez Men{\'e}ndez}, {Recio}, {Izquierdo-Ruiz}, {Coya Lozano}, {Garc{\'\i}a de la Cuesta}, {G{\'o}mez Hern{\'a}ndez}, {Vidal Blanco}, {Hevia D{\'\i}az}, {Pardo Silva}, {P{\'e}rez Acevedo}, {Polancos Ruiz}, {Padilla Tijer{\'\i}n}, {V{\'a}zquez Garc{\'\i}a}, {Su{\'a}rez G{\'o}mez}, {Garc{\'\i}a Riesgo}, {Gonz{\'a}lez Guti{\'e}rrez}, {Bonavera}, {Gonz{\'a}lez-Nuevo}, {Rodr{\'\i}guez Pereira}, {S{\'a}nchez Lasheras}, {S{\'a}nchez Rodr{\'\i}guez}, {Mu{\~n}iz}, {Santos Rodr{\'\i}guez}, \& {de Cos Juez}}]{CastroGonzlez2022}
{Castro-Gonz{\'a}lez}, A., {D{\'\i}ez Alonso}, E., {Men{\'e}ndez Blanco}, J., {et~al.} 2022, \mnras, 509, 1075, \dodoi{10.1093/mnras/stab2669}

\bibitem[{{Christian} {et~al.}(2025){Christian}, {Vanderburg}, {Becker}, {Kraus}, {Pearce}, {Collins}, {Rice}, {Jensen}, {Baker}, {Bozza}, {Benni}, {Bieryla}, {Binnenfeld}, {Collins}, {Conti}, {Crossfield}, {Evans}, {Johnson}, {Girardin}, {Gregorio}, {Lewin}, {Mazeh}, {Murgas}, {Panahi}, {Pozuelos}, {Radford}, {Relles}, {Rodriguez Frustaglia}, {Schwarz}, {Srdoc}, {Stockdale}, {Tan}, {Waalkes}, {Wang}, {Wittrock}, \& {Zucker}}]{Christian2025}
{Christian}, S., {Vanderburg}, A., {Becker}, J., {et~al.} 2025, \aj, 169, 308, \dodoi{10.3847/1538-3881/adc933}

\bibitem[{{Christiansen} {et~al.}(2025){Christiansen}, {McElroy}, {Harbut}, {Ciardi}, {Crane}, {Good}, {Hardegree-Ullman}, {Kesseli}, {Lund}, {Lynn}, {Muthiar}, {Nilsson}, {Oluyide}, {Papin}, {Rivera}, {Swain}, {Susemiehl}, {Tam}, {van Eyken}, \& {Beichman}}]{nea_new}
{Christiansen}, J.~L., {McElroy}, D.~L., {Harbut}, M., {et~al.} 2025, arXiv e-prints, arXiv:2506.03299, \dodoi{10.48550/arXiv.2506.03299}

\bibitem[{{Cifuentes} {et~al.}(2025){Cifuentes}, {Caballero}, {Gonz{\'a}lez-Payo}, {Amado}, {B{\'e}jar}, {Burgasser}, {Cort{\'e}s-Contreras}, {Lodieu}, {Montes}, {Quirrenbach}, {Reiners}, {Ribas}, {Sanz-Forcada}, {Seifert}, \& {Zapatero Osorio}}]{Cifuentes2025}
{Cifuentes}, C., {Caballero}, J.~A., {Gonz{\'a}lez-Payo}, J., {et~al.} 2025, \aap, 693, A228, \dodoi{10.1051/0004-6361/202452527}

\bibitem[{{Claret} \& {Bloemen}(2011)}]{Claret2011}
{Claret}, A., \& {Bloemen}, S. 2011, \aap, 529, A75, \dodoi{10.1051/0004-6361/201116451}

\bibitem[{{Dai} {et~al.}(2023){Dai}, {Masuda}, {Beard}, {Robertson}, {Goldberg}, {Batygin}, {Bouma}, {Lissauer}, {Knudstrup}, {Albrecht}, {Howard}, {Knutson}, {Petigura}, {Weiss}, {Isaacson}, {Kristiansen}, {Osborn}, {Wang}, {Wang}, {Behmard}, {Greklek-McKeon}, {Vissapragada}, {Batalha}, {Brinkman}, {Chontos}, {Crossfield}, {Dressing}, {Fetherolf}, {Fulton}, {Hill}, {Huber}, {Kane}, {Lubin}, {MacDougall}, {Mayo}, {Mo{\v{c}}nik}, {Akana Murphy}, {Rubenzahl}, {Scarsdale}, {Tyler}, {Zandt}, {Polanski}, {Schwengeler}, {Terentev}, {Benni}, {Bieryla}, {Ciardi}, {Falk}, {Furlan}, {Girardin}, {Guerra}, {Hesse}, {Howell}, {Lillo-Box}, {Matthews}, {Twicken}, {Villase{\~n}or}, {Latham}, {Jenkins}, {Ricker}, {Seager}, {Vanderspek}, \& {Winn}}]{Dai2023}
{Dai}, F., {Masuda}, K., {Beard}, C., {et~al.} 2023, \aj, 165, 33, \dodoi{10.3847/1538-3881/aca327}

\bibitem[{{Damasso} {et~al.}(2020){Damasso}, {Sozzetti}, {Lovis}, {Barros}, {Sousa}, {Demangeon}, {Faria}, {Lillo-Box}, {Cristiani}, {Pepe}, {Rebolo}, {Santos}, {Zapatero Osorio}, {Gonz{\'a}lez Hern{\'a}ndez}, {Amate}, {Pasquini}, {Zerbi}, {Adibekyan}, {Abreu}, {Affolter}, {Alibert}, {Aliverti}, {Allart}, {Allende Prieto}, {{\'A}lvarez}, {Alves}, {Avila}, {Baldini}, {Bandy}, {Benz}, {Bianco}, {Borsa}, {Bossini}, {Bourrier}, {Bouchy}, {Broeg}, {Cabral}, {Calderone}, {Cirami}, {Coelho}, {Conconi}, {Coretti}, {Cumani}, {Cupani}, {D'Odorico}, {Deiries}, {Dekker}, {Delabre}, {Di Marcantonio}, {Dumusque}, {Ehrenreich}, {Figueira}, {Fragoso}, {Genolet}, {Genoni}, {G{\'e}nova Santos}, {Hughes}, {Iwert}, {Kerber}, {Knudstrup}, {Landoni}, {Lavie}, {Lizon}, {Lo Curto}, {Maire}, {Martins}, {M{\'e}gevand}, {Mehner}, {Micela}, {Modigliani}, {Molaro}, {Monteiro}, {Monteiro}, {Moschetti}, {Mueller}, {Murphy}, {Nunes}, {Oggioni}, {Oliveira}, {Oshagh}, {Pall{\'e}}, {Pariani}, {Poretti}, {Rasilla}, {Rebord{\~a}o}, {Redaelli},
  {Riva}, {Santana Tschudi}, {Santin}, {Santos}, {S{\'e}gransan}, {Schmidt}, {Segovia}, {Sosnowska}, {Span{\`o}}, {Su{\'a}rez Mascare{\~n}o}, {Tabernero}, {Tenegi}, {Udry}, \& {Zanutta}}]{Damasso2020}
{Damasso}, M., {Sozzetti}, A., {Lovis}, C., {et~al.} 2020, \aap, 642, A31, \dodoi{10.1051/0004-6361/202038416}

\bibitem[{{Demangeon} {et~al.}(2018){Demangeon}, {Faedi}, {H{\'e}brard}, {Brown}, {Barros}, {Doyle}, {Maxted}, {Collier Cameron}, {Hay}, {Alikakos}, {Anderson}, {Armstrong}, {Boumis}, {Bonomo}, {Bouchy}, {Delrez}, {Gillon}, {Haswell}, {Hellier}, {Jehin}, {Kiefer}, {Lam}, {Lendl}, {Mancini}, {McCormac}, {Norton}, {Osborn}, {Palle}, {Pepe}, {Pollacco}, {Prieto-Arranz}, {Queloz}, {S{\'e}gransan}, {Smalley}, {Triaud}, {Udry}, {West}, \& {Wheatley}}]{Demangeon2018}
{Demangeon}, O.~D.~S., {Faedi}, F., {H{\'e}brard}, G., {et~al.} 2018, \aap, 610, A63, \dodoi{10.1051/0004-6361/201731735}

\bibitem[{{Doyle} {et~al.}(2022){Doyle}, {Cegla}, {Bryant}, {Bayliss}, {Lafarga}, {Anderson}, {Allart}, {Bourrier}, {Brogi}, {Buchschacher}, {Kunovac}, {Lendl}, {Lovis}, {Moyano}, {Roguet-Kern}, {Seidel}, {Sosnowska}, {Wheatley}, {Acton}, {Burleigh}, {Casewell}, {Gill}, {Goad}, {Henderson}, {Jenkins}, {Tilbrook}, \& {West}}]{Doyle2022}
{Doyle}, L., {Cegla}, H.~M., {Bryant}, E., {et~al.} 2022, \mnras, 516, 298, \dodoi{10.1093/mnras/stac2178}

\bibitem[{{Dupuy} {et~al.}(2022){Dupuy}, {Kraus}, {Kratter}, {Rizzuto}, {Mann}, {Huber}, \& {Ireland}}]{Dupuy2022}
{Dupuy}, T.~J., {Kraus}, A.~L., {Kratter}, K.~M., {et~al.} 2022, \mnras, 512, 648, \dodoi{10.1093/mnras/stac306}

\bibitem[{{Eastman} {et~al.}(2013){Eastman}, {Gaudi}, \& {Agol}}]{Eastman2013}
{Eastman}, J., {Gaudi}, B.~S., \& {Agol}, E. 2013, \pasp, 125, 83, \dodoi{10.1086/669497}

\bibitem[{{Espinoza} {et~al.}(2022){Espinoza}, {Pall{\'e}}, {Kemmer}, {Luque}, {Caballero}, {Cifuentes}, {Herrero}, {S{\'a}nchez B{\'e}jar}, {Stock}, {Molaverdikhani}, {Morello}, {Kossakowski}, {Schlecker}, {Amado}, {Bluhm}, {Cort{\'e}s-Contreras}, {Henning}, {Kreidberg}, {K{\"u}rster}, {Lafarga}, {Lodieu}, {Morales}, {Oshagh}, {Passegger}, {Pavlov}, {Quirrenbach}, {Reffert}, {Reiners}, {Ribas}, {Rodr{\'\i}guez}, {Rodr{\'\i}guez L{\'o}pez}, {Schweitzer}, {Trifonov}, {Chaturvedi}, {Dreizler}, {Jeffers}, {Kaminski}, {L{\'o}pez-Gonz{\'a}lez}, {Lillo-Box}, {Montes}, {Nowak}, {Pedraz}, {Vanaverbeke}, {Zapatero Osorio}, {Zechmeister}, {Collins}, {Girardin}, {Guerra}, {Naves}, {Crossfield}, {Matthews}, {Howell}, {Ciardi}, {Gonzales}, {Matson}, {Beichman}, {Schlieder}, {Barclay}, {Vezie}, {Villase{\~n}or}, {Daylan}, {Mireies}, {Dragomir}, {Twicken}, {Jenkins}, {Winn}, {Latham}, {Ricker}, \& {Seager}}]{Espinoza2022}
{Espinoza}, N., {Pall{\'e}}, E., {Kemmer}, J., {et~al.} 2022, \aj, 163, 133, \dodoi{10.3847/1538-3881/ac4af0}

\bibitem[{{Espinoza-Retamal} {et~al.}(2024{\natexlab{a}}){Espinoza-Retamal}, {Jord{\'a}n}, {Brahm}, {Petrovich}, {Sedaghati}, {Stef{\'a}nsson}, {Hobson}, {Tala Pinto}, {Mu{\~n}oz}, {Boyle}, {Leiva}, \& {Suc}}]{EspinozaRetamal2024b}
{Espinoza-Retamal}, J.~I., {Jord{\'a}n}, A., {Brahm}, R., {et~al.} 2024{\natexlab{a}}, arXiv e-prints, arXiv:2412.08692, \dodoi{10.48550/arXiv.2412.08692}

\bibitem[{{Espinoza-Retamal} {et~al.}(2024{\natexlab{b}}){Espinoza-Retamal}, {Stef{\'a}nsson}, {Petrovich}, {Brahm}, {Jord{\'a}n}, {Sedaghati}, {Lucero}, {Tala Pinto}, {Mu{\~n}oz}, {Boyle}, {Leiva}, \& {Suc}}]{EspinozaRetamal2024a}
{Espinoza-Retamal}, J.~I., {Stef{\'a}nsson}, G., {Petrovich}, C., {et~al.} 2024{\natexlab{b}}, \aj, 168, 185, \dodoi{10.3847/1538-3881/ad70b8}

\bibitem[{{Esteves} {et~al.}(2023){Esteves}, {Izidoro}, {Winter}, {Bitsch}, \& {Isella}}]{Esteves2023}
{Esteves}, L., {Izidoro}, A., {Winter}, O.~C., {Bitsch}, B., \& {Isella}, A. 2023, \mnras, 521, 5776, \dodoi{10.1093/mnras/stad756}

\bibitem[{{Fabrycky} \& {Winn}(2009)}]{Fabrycky2009}
{Fabrycky}, D.~C., \& {Winn}, J.~N. 2009, \apj, 696, 1230, \dodoi{10.1088/0004-637X/696/2/1230}

\bibitem[{{Fairnington} {et~al.}(2024){Fairnington}, {Nabbie}, {Huang}, {Zhou}, {Foo}, {Millholland}, {Wright}, {Belinski}, {Bieryla}, {Ciardi}, {Collins}, {Collins}, {Everett}, {Howell}, {Lissauer}, {Lund}, {Murgas}, {Palle}, {Quinn}, {Relles}, {Safonov}, {Schwarz}, {Scott}, {Srdoc}, {Ricker}, {Vanderspek}, {Seager}, {Latham}, {Winn}, {Jenkins}, {Bouma}, {Shporer}, {Ting}, {Dragomir}, {Kunimoto}, \& {Eisner}}]{Fairnington2024}
{Fairnington}, T.~R., {Nabbie}, E., {Huang}, C.~X., {et~al.} 2024, \mnras, 527, 8768, \dodoi{10.1093/mnras/stad3036}

\bibitem[{{Feinstein} {et~al.}(2021){Feinstein}, {Montet}, {Johnson}, {Bean}, {David}, {Gully-Santiago}, {Livingston}, \& {Luger}}]{Feinstein2021}
{Feinstein}, A.~D., {Montet}, B.~T., {Johnson}, M.~C., {et~al.} 2021, \aj, 162, 213, \dodoi{10.3847/1538-3881/ac1f24}

\bibitem[{{Fernandes} {et~al.}(2019){Fernandes}, {Mulders}, {Pascucci}, {Mordasini}, \& {Emsenhuber}}]{Fernandes2019}
{Fernandes}, R.~B., {Mulders}, G.~D., {Pascucci}, I., {Mordasini}, C., \& {Emsenhuber}, A. 2019, \apj, 874, 81, \dodoi{10.3847/1538-4357/ab0300}

\bibitem[{Foreman-Mackey(2016)}]{corner}
Foreman-Mackey, D. 2016, The Journal of Open Source Software, 1, 24, \dodoi{10.21105/joss.00024}

\bibitem[{{Foreman-Mackey}(2018)}]{celerite2}
{Foreman-Mackey}, D. 2018, Research Notes of the American Astronomical Society, 2, 31, \dodoi{10.3847/2515-5172/aaaf6c}

\bibitem[{{Foreman-Mackey} {et~al.}(2013){Foreman-Mackey}, {Hogg}, {Lang}, \& {Goodman}}]{emcee}
{Foreman-Mackey}, D., {Hogg}, D.~W., {Lang}, D., \& {Goodman}, J. 2013, \pasp, 125, 306, \dodoi{10.1086/670067}

\bibitem[{{Foreman-Mackey} {et~al.}(2021){Foreman-Mackey}, {Luger}, {Agol}, {Barclay}, {Bouma}, {Brandt}, {Czekala}, {David}, {Dong}, {Gilbert}, {Gordon}, {Hedges}, {Hey}, {Morris}, {Price-Whelan}, \& {Savel}}]{Foreman-Mackey2021}
{Foreman-Mackey}, D., {Luger}, R., {Agol}, E., {et~al.} 2021, arXiv e-prints, arXiv:2105.01994.
\newblock \doarXiv{2105.01994}

\bibitem[{{Frazier} {et~al.}(2023){Frazier}, {Stef{\'a}nsson}, {Mahadevan}, {Yee}, {Ca{\~n}as}, {Winn}, {Luhn}, {Dai}, {Doyle}, {Cegla}, {Kanodia}, {Robertson}, {Wisniewski}, {Bender}, {Dong}, {Gupta}, {Halverson}, {Hawley}, {Hebb}, {Holcomb}, {Kowalski}, {Libby-Roberts}, {Lin}, {McElwain}, {Ninan}, {Petrovich}, {Roy}, {Schwab}, {Terrien}, \& {Wright}}]{Frazier2023}
{Frazier}, R.~C., {Stef{\'a}nsson}, G., {Mahadevan}, S., {et~al.} 2023, \apjl, 944, L41, \dodoi{10.3847/2041-8213/acba18}

\bibitem[{{Fukui} {et~al.}(2011){Fukui}, {Narita}, {Tristram}, {Sumi}, {Abe}, {Itow}, {Sullivan}, {Bond}, {Hirano}, {Tamura}, {Bennett}, {Furusawa}, {Hayashi}, {Hearnshaw}, {Hosaka}, {Kamiya}, {Kobara}, {Korpela}, {Kilmartin}, {Lin}, {Ling}, {Makita}, {Masuda}, {Matsubara}, {Miyake}, {Muraki}, {Nagaya}, {Nishimoto}, {Ohnishi}, {Omori}, {Perrott}, {Rattenbury}, {Saito}, {Skuljan}, {Suzuki}, {Sweatman}, \& {Wada}}]{2011PASJ...63..287F}
{Fukui}, A., {Narita}, N., {Tristram}, P.~J., {et~al.} 2011, \pasj, 63, 287, \dodoi{10.1093/pasj/63.1.287}

\bibitem[{{Fulton} {et~al.}(2021){Fulton}, {Rosenthal}, {Hirsch}, {Isaacson}, {Howard}, {Dedrick}, {Sherstyuk}, {Blunt}, {Petigura}, {Knutson}, {Behmard}, {Chontos}, {Crepp}, {Crossfield}, {Dalba}, {Fischer}, {Henry}, {Kane}, {Kosiarek}, {Marcy}, {Rubenzahl}, {Weiss}, \& {Wright}}]{Fulton2021}
{Fulton}, B.~J., {Rosenthal}, L.~J., {Hirsch}, L.~A., {et~al.} 2021, \apjs, 255, 14, \dodoi{10.3847/1538-4365/abfcc1}

\bibitem[{{Gaia Collaboration} {et~al.}(2023){Gaia Collaboration}, {Vallenari}, {Brown}, {Prusti}, {de Bruijne}, {Arenou}, {Babusiaux}, {Biermann}, {Creevey}, {Ducourant}, {Evans}, {Eyer}, {Guerra}, {Hutton}, {Jordi}, {Klioner}, {Lammers}, {Lindegren}, {Luri}, {Mignard}, {Panem}, {Pourbaix}, {Randich}, {Sartoretti}, {Soubiran}, {Tanga}, {Walton}, {Bailer-Jones}, {Bastian}, {Drimmel}, {Jansen}, {Katz}, {Lattanzi}, {van Leeuwen}, {Bakker}, {Cacciari}, {Casta{\~n}eda}, {De Angeli}, {Fabricius}, {Fouesneau}, {Fr{\'e}mat}, {Galluccio}, {Guerrier}, {Heiter}, {Masana}, {Messineo}, {Mowlavi}, {Nicolas}, {Nienartowicz}, {Pailler}, {Panuzzo}, {Riclet}, {Roux}, {Seabroke}, {Sordo}, {Th{\'e}venin}, {Gracia-Abril}, {Portell}, {Teyssier}, {Altmann}, {Andrae}, {Audard}, {Bellas-Velidis}, {Benson}, {Berthier}, {Blomme}, {Burgess}, {Busonero}, {Busso}, {C{\'a}novas}, {Carry}, {Cellino}, {Cheek}, {Clementini}, {Damerdji}, {Davidson}, {de Teodoro}, {Nu{\~n}ez Campos}, {Delchambre}, {Dell'Oro}, {Esquej},
  {Fern{\'a}ndez-Hern{\'a}ndez}, {Fraile}, {Garabato}, {Garc{\'\i}a-Lario}, {Gosset}, {Haigron}, {Halbwachs}, {Hambly}, {Harrison}, {Hern{\'a}ndez}, {Hestroffer}, {Hodgkin}, {Holl}, {Jan{\ss}en}, {Jevardat de Fombelle}, {Jordan}, {Krone-Martins}, {Lanzafame}, {L{\"o}ffler}, {Marchal}, {Marrese}, {Moitinho}, {Muinonen}, {Osborne}, {Pancino}, {Pauwels}, {Recio-Blanco}, {Reyl{\'e}}, {Riello}, {Rimoldini}, {Roegiers}, {Rybizki}, {Sarro}, {Siopis}, {Smith}, {Sozzetti}, {Utrilla}, {van Leeuwen}, {Abbas}, {{\'A}brah{\'a}m}, {Abreu Aramburu}, {Aerts}, {Aguado}, {Ajaj}, {Aldea-Montero}, {Altavilla}, {{\'A}lvarez}, {Alves}, {Anders}, {Anderson}, {Anglada Varela}, {Antoja}, {Baines}, {Baker}, {Balaguer-N{\'u}{\~n}ez}, {Balbinot}, {Balog}, {Barache}, {Barbato}, {Barros}, {Barstow}, {Bartolom{\'e}}, {Bassilana}, {Bauchet}, {Becciani}, {Bellazzini}, {Berihuete}, {Bernet}, {Bertone}, {Bianchi}, {Binnenfeld}, {Blanco-Cuaresma}, {Blazere}, {Boch}, {Bombrun}, {Bossini}, {Bouquillon}, {Bragaglia}, {Bramante}, {Breedt},
  {Bressan}, {Brouillet}, {Brugaletta}, {Bucciarelli}, {Burlacu}, {Butkevich}, {Buzzi}, {Caffau}, {Cancelliere}, {Cantat-Gaudin}, {Carballo}, {Carlucci}, {Carnerero}, {Carrasco}, {Casamiquela}, {Castellani}, {Castro-Ginard}, {Chaoul}, {Charlot}, {Chemin}, {Chiaramida}, {Chiavassa}, {Chornay}, {Comoretto}, {Contursi}, {Cooper}, {Cornez}, {Cowell}, {Crifo}, {Cropper}, {Crosta}, {Crowley}, {Dafonte}, {Dapergolas}, {David}, {David}, {de Laverny}, {De Luise}, \& {De March}}]{Gaia2023}
{Gaia Collaboration}, {Vallenari}, A., {Brown}, A.~G.~A., {et~al.} 2023, \aap, 674, A1, \dodoi{10.1051/0004-6361/202243940}

\bibitem[{{Gillon} {et~al.}(2016){Gillon}, {Jehin}, {Lederer}, {Delrez}, {de Wit}, {Burdanov}, {Van Grootel}, {Burgasser}, {Triaud}, {Opitom}, {Demory}, {Sahu}, {Bardalez Gagliuffi}, {Magain}, \& {Queloz}}]{Gillon2016}
{Gillon}, M., {Jehin}, E., {Lederer}, S.~M., {et~al.} 2016, \nat, 533, 221, \dodoi{10.1038/nature17448}

\bibitem[{{Gonz{\'a}lez-Payo} {et~al.}(2024){Gonz{\'a}lez-Payo}, {Caballero}, {Gorgas}, {Cort{\'e}s-Contreras}, {G{\'a}lvez-Ortiz}, \& {Cifuentes}}]{GonzalezPayo2024}
{Gonz{\'a}lez-Payo}, J., {Caballero}, J.~A., {Gorgas}, J., {et~al.} 2024, \aap, 689, A302, \dodoi{10.1051/0004-6361/202450048}

\bibitem[{{H{\'e}brard} {et~al.}(2011){H{\'e}brard}, {Ehrenreich}, {Bouchy}, {Delfosse}, {Moutou}, {Arnold}, {Boisse}, {Bonfils}, {D{\'\i}az}, {Eggenberger}, {Forveille}, {Lagrange}, {Lovis}, {Pepe}, {Perrier}, {Queloz}, {Santerne}, {Santos}, {S{\'e}gransan}, {Udry}, \& {Vidal-Madjar}}]{Hebrard2011a}
{H{\'e}brard}, G., {Ehrenreich}, D., {Bouchy}, F., {et~al.} 2011, \aap, 527, L11, \dodoi{10.1051/0004-6361/201016331}

\bibitem[{{Hellier} {et~al.}(2019){Hellier}, {Anderson}, {Triaud}, {Bouchy}, {Burdanov}, {Collier Cameron}, {Delrez}, {Ehrenreich}, {Gillon}, {Jehin}, {Lendl}, {Linder}, {Nielsen}, {Maxted}, {Pepe}, {Pollacco}, {Queloz}, {S{\'e}gransan}, {Smalley}, {Spake}, {Temple}, {Udry}, {West}, \& {Wyttenbach}}]{Hellier2019}
{Hellier}, C., {Anderson}, D.~R., {Triaud}, A.~H.~M.~J., {et~al.} 2019, \mnras, 488, 3067, \dodoi{10.1093/mnras/stz1903}

\bibitem[{{Hippke} {et~al.}(2019){Hippke}, {David}, {Mulders}, \& {Heller}}]{Hippke}
{Hippke}, M., {David}, T.~J., {Mulders}, G.~D., \& {Heller}, R. 2019, \aj, 158, 143, \dodoi{10.3847/1538-3881/ab3984}

\bibitem[{{Hirano} {et~al.}(2010){Hirano}, {Suto}, {Taruya}, {Narita}, {Sato}, {Johnson}, \& {Winn}}]{Hirano2010}
{Hirano}, T., {Suto}, Y., {Taruya}, A., {et~al.} 2010, \apj, 709, 458, \dodoi{10.1088/0004-637X/709/1/458}

\bibitem[{{Hirano} {et~al.}(2011){Hirano}, {Suto}, {Winn}, {Taruya}, {Narita}, {Albrecht}, \& {Sato}}]{Hirano2011}
{Hirano}, T., {Suto}, Y., {Winn}, J.~N., {et~al.} 2011, \apj, 742, 69, \dodoi{10.1088/0004-637X/742/2/69}

\bibitem[{{Hirano} {et~al.}(2020{\natexlab{a}}){Hirano}, {Gaidos}, {Winn}, {Dai}, {Fukui}, {Kuzuhara}, {Kotani}, {Tamura}, {Hjorth}, {Albrecht}, {Huber}, {Bolmont}, {Harakawa}, {Hodapp}, {Ishizuka}, {Jacobson}, {Konishi}, {Kudo}, {Kurokawa}, {Nishikawa}, {Omiya}, {Serizawa}, {Ueda}, \& {Weiss}}]{Hirano2020a}
{Hirano}, T., {Gaidos}, E., {Winn}, J.~N., {et~al.} 2020{\natexlab{a}}, \apjl, 890, L27, \dodoi{10.3847/2041-8213/ab74dc}

\bibitem[{{Hirano} {et~al.}(2020{\natexlab{b}}){Hirano}, {Krishnamurthy}, {Gaidos}, {Flewelling}, {Mann}, {Narita}, {Plavchan}, {Kotani}, {Tamura}, {Harakawa}, {Hodapp}, {Ishizuka}, {Jacobson}, {Konishi}, {Kudo}, {Kurokawa}, {Kuzuhara}, {Nishikawa}, {Omiya}, {Serizawa}, {Ueda}, \& {Vievard}}]{Hirano2020b}
{Hirano}, T., {Krishnamurthy}, V., {Gaidos}, E., {et~al.} 2020{\natexlab{b}}, \apjl, 899, L13, \dodoi{10.3847/2041-8213/aba6eb}

\bibitem[{{Hirano} {et~al.}(2024){Hirano}, {Gaidos}, {Harakawa}, {Hodapp}, {Kotani}, {Kudo}, {Kurokawa}, {Kuzuhara}, {Mann}, {Nishikawa}, {Omiya}, {Serizawa}, {Tamura}, {Thao}, {Ueda}, \& {Vievard}}]{Hirano2024}
{Hirano}, T., {Gaidos}, E., {Harakawa}, H., {et~al.} 2024, \mnras, 530, 3117, \dodoi{10.1093/mnras/stae998}

\bibitem[{{Hjorth} {et~al.}(2021){Hjorth}, {Albrecht}, {Hirano}, {Winn}, {Dawson}, {Zanazzi}, {Knudstrup}, \& {Sato}}]{Hjorth2021}
{Hjorth}, M., {Albrecht}, S., {Hirano}, T., {et~al.} 2021, Proceedings of the National Academy of Science, 118, e2017418118, \dodoi{10.1073/pnas.2017418118}

\bibitem[{{Hjorth} {et~al.}(2019){Hjorth}, {Justesen}, {Hirano}, {Albrecht}, {Gandolfi}, {Dai}, {Alonso}, {Barrag{\'a}n}, {Esposito}, {Kuzuhara}, {Lam}, {Livingston}, {Montanes-Rodriguez}, {Narita}, {Nowak}, {Prieto-Arranz}, {Redfield}, {Rodler}, {Van Eylen}, {Winn}, {Antoniciello}, {Cabrera}, {Cochran}, {Csizmadia}, {de Leon}, {Deeg}, {Eigm{\"u}ller}, {Endl}, {Erikson}, {Fridlund}, {Grziwa}, {Guenther}, {Hatzes}, {Heeren}, {Hidalgo}, {Korth}, {Luque}, {Nespral}, {Palle}, {P{\"a}tzold}, {Persson}, {Rauer}, {Smith}, \& {Trifonov}}]{Hjorth2019}
{Hjorth}, M., {Justesen}, A.~B., {Hirano}, T., {et~al.} 2019, \mnras, 484, 3522, \dodoi{10.1093/mnras/stz139}

\bibitem[{{Izidoro} {et~al.}(2021){Izidoro}, {Bitsch}, {Raymond}, {Johansen}, {Morbidelli}, {Lambrechts}, \& {Jacobson}}]{Izidoro2021}
{Izidoro}, A., {Bitsch}, B., {Raymond}, S.~N., {et~al.} 2021, \aap, 650, A152, \dodoi{10.1051/0004-6361/201935336}

\bibitem[{{Jenkins} {et~al.}(2016){Jenkins}, {Twicken}, {McCauliff}, {Campbell}, {Sanderfer}, {Lung}, {Mansouri-Samani}, {Girouard}, {Tenenbaum}, {Klaus}, {Smith}, {Caldwell}, {Chacon}, {Henze}, {Heiges}, {Latham}, {Morgan}, {Swade}, {Rinehart}, \& {Vanderspek}}]{Jenkins2016}
{Jenkins}, J.~M., {Twicken}, J.~D., {McCauliff}, S., {et~al.} 2016, in Society of Photo-Optical Instrumentation Engineers (SPIE) Conference Series, Vol. 9913, Software and Cyberinfrastructure for Astronomy IV, ed. G.~{Chiozzi} \& J.~C. {Guzman}, 99133E, \dodoi{10.1117/12.2233418}

\bibitem[{{Johnson} {et~al.}(2022){Johnson}, {David}, {Petigura}, {Isaacson}, {Van Zandt}, {Ilyin}, {Strassmeier}, {Mallonn}, {Zhou}, {Mann}, {Livingston}, {Luger}, {Dai}, {Weiss}, {Mo{\v{c}}nik}, {Giacalone}, {Hill}, {Rice}, {Blunt}, {Rubenzahl}, {Dalba}, {Esquerdo}, {Berlind}, {Calkins}, \& {Foreman-Mackey}}]{Johnson2022}
{Johnson}, M.~C., {David}, T.~J., {Petigura}, E.~A., {et~al.} 2022, \aj, 163, 247, \dodoi{10.3847/1538-3881/ac6271}

\bibitem[{{Jord{\'a}n} {et~al.}(2020){Jord{\'a}n}, {Bakos}, {Bayliss}, {Bento}, {Bhatti}, {Brahm}, {Csubry}, {Espinoza}, {Hartman}, {Henning}, {Mancini}, {Penev}, {Rabus}, {Sarkis}, {Suc}, {de Val-Borro}, {Zhou}, {Butler}, {Teske}, {Crane}, {Shectman}, {Tan}, {Thompson}, {Wallace}, {L{\'a}z{\'a}r}, {Papp}, \& {S{\'a}ri}}]{Jordan2020}
{Jord{\'a}n}, A., {Bakos}, G.~{\'A}., {Bayliss}, D., {et~al.} 2020, \aj, 160, 222, \dodoi{10.3847/1538-3881/aba530}

\bibitem[{{Knudstrup} {et~al.}(2024){Knudstrup}, {Albrecht}, {Winn}, {Gandolfi}, {Zanazzi}, {Persson}, {Fridlund}, {Marcussen}, {Chontos}, {Keniger}, {Eisner}, {Bieryla}, {Isaacson}, {Howard}, {Hirsch}, {Murgas}, {Narita}, {Palle}, {Kawai}, \& {Baker}}]{Knudstrup2024}
{Knudstrup}, E., {Albrecht}, S.~H., {Winn}, J.~N., {et~al.} 2024, \aap, 690, A379, \dodoi{10.1051/0004-6361/202450627}

\bibitem[{{Kosiarek} {et~al.}(2019){Kosiarek}, {Crossfield}, {Hardegree-Ullman}, {Livingston}, {Benneke}, {Henry}, {Howard}, {Berardo}, {Blunt}, {Fulton}, {Hirsch}, {Howard}, {Isaacson}, {Petigura}, {Sinukoff}, {Weiss}, {Bonfils}, {Dressing}, {Knutson}, {Schlieder}, {Werner}, {Gorjian}, {Krick}, {Morales}, {Astudillo-Defru}, {Almenara}, {Delfosse}, {Forveille}, {Lovis}, {Mayor}, {Murgas}, {Pepe}, {Santos}, {Udry}, {Corbett}, {Fors}, {Law}, {Ratzloff}, \& {del Ser}}]{Kosiarek2019}
{Kosiarek}, M.~R., {Crossfield}, I. J.~M., {Hardegree-Ullman}, K.~K., {et~al.} 2019, \aj, 157, 97, \dodoi{10.3847/1538-3881/aaf79c}

\bibitem[{{Kraft}(1967)}]{Kraft1967}
{Kraft}, R.~P. 1967, \apj, 150, 551, \dodoi{10.1086/149359}

\bibitem[{{Kreidberg}(2015)}]{batman}
{Kreidberg}, L. 2015, \pasp, 127, 1161, \dodoi{10.1086/683602}

\bibitem[{{Kunovac Hod{\v{z}}i{\'c}} {et~al.}(2021){Kunovac Hod{\v{z}}i{\'c}}, {Triaud}, {Cegla}, {Chaplin}, \& {Davies}}]{Kunovac2021}
{Kunovac Hod{\v{z}}i{\'c}}, V., {Triaud}, A. H.~M.~J., {Cegla}, H.~M., {Chaplin}, W.~J., \& {Davies}, G.~R. 2021, \mnras, 502, 2893, \dodoi{10.1093/mnras/stab237}

\bibitem[{Li(2023)}]{smplotlib}
Li, J. 2023, AstroJacobLi/smplotlib: v0.0.9, v0.0.9,  Zenodo, \dodoi{10.5281/zenodo.8126529}

\bibitem[{{Libby-Roberts} {et~al.}(2023){Libby-Roberts}, {Schutte}, {Hebb}, {Kanodia}, {Ca{\~n}as}, {Stef{\'a}nsson}, {Lin}, {Mahadevan}, {Parts}, {Powers}, {Wisniewski}, {Bender}, {Cochran}, {Diddams}, {Everett}, {Gupta}, {Halverson}, {Kobulnicky}, {Kowalski}, {Larsen}, {Monson}, {Ninan}, {Parker}, {Ramsey}, {Robertson}, {Schwab}, {Swaby}, \& {Terrien}}]{LibbyRoberts2023}
{Libby-Roberts}, J.~E., {Schutte}, M., {Hebb}, L., {et~al.} 2023, \aj, 165, 249, \dodoi{10.3847/1538-3881/accc2f}

\bibitem[{{Lightkurve Collaboration} {et~al.}(2018){Lightkurve Collaboration}, {Cardoso}, {Hedges}, {Gully-Santiago}, {Saunders}, {Cody}, {Barclay}, {Hall}, {Sagear}, {Turtelboom}, {Zhang}, {Tzanidakis}, {Mighell}, {Coughlin}, {Bell}, {Berta-Thompson}, {Williams}, {Dotson}, \& {Barentsen}}]{Lightkurve2018}
{Lightkurve Collaboration}, {Cardoso}, J. V. d.~M., {Hedges}, C., {et~al.} 2018, {Lightkurve: Kepler and TESS time series analysis in Python}.
\newblock \doeprint{1812.013}

\bibitem[{{Louden} {et~al.}(2021){Louden}, {Winn}, {Petigura}, {Isaacson}, {Howard}, {Masuda}, {Albrecht}, \& {Kosiarek}}]{Louden2021}
{Louden}, E.~M., {Winn}, J.~N., {Petigura}, E.~A., {et~al.} 2021, \aj, 161, 68, \dodoi{10.3847/1538-3881/abcebd}

\bibitem[{{Louden} {et~al.}(2024){Louden}, {Wang}, {Winn}, {Petigura}, {Isaacson}, {Handley}, {Yee}, {Beard}, {Murphy}, \& {Laughlin}}]{Louden2024}
{Louden}, E.~M., {Wang}, S., {Winn}, J.~N., {et~al.} 2024, \apjl, 968, L2, \dodoi{10.3847/2041-8213/ad4b1b}

\bibitem[{{Lubin} {et~al.}(2022){Lubin}, {Van Zandt}, {Holcomb}, {Weiss}, {Petigura}, {Robertson}, {Akana Murphy}, {Scarsdale}, {Batygin}, {Polanski}, {Batalha}, {Crossfield}, {Dressing}, {Fulton}, {Howard}, {Huber}, {Isaacson}, {Kane}, {Roy}, {Beard}, {Blunt}, {Chontos}, {Dai}, {Dalba}, {Gary}, {Giacalone}, {Hill}, {Mayo}, {Mo{\v{c}}nik}, {Kosiarek}, {Rice}, {Rubenzahl}, {Latham}, {Seager}, {Winn}, \& {Gary}}]{Lubin2022}
{Lubin}, J., {Van Zandt}, J., {Holcomb}, R., {et~al.} 2022, \aj, 163, 101, \dodoi{10.3847/1538-3881/ac3d38}

\bibitem[{{Lubin} {et~al.}(2023){Lubin}, {Wang}, {Rice}, {Dong}, {Wang}, {Radzom}, {Robertson}, {Stefansson}, {Alvarado-Montes}, {Beard}, {Bender}, {Gupta}, {Halverson}, {Kanodia}, {Li}, {Lin}, {Logsdon}, {Lubar}, {Mahadevan}, {Ninan}, {Rajagopal}, {Roy}, {Schwab}, \& {Wright}}]{Lubin2023}
{Lubin}, J., {Wang}, X.-Y., {Rice}, M., {et~al.} 2023, \apjl, 959, L5, \dodoi{10.3847/2041-8213/ad0fea}

\bibitem[{{Lubin} {et~al.}(2024){Lubin}, {Petigura}, {Van Zandt}, {Beard}, {Dai}, {Halverson}, {Holcomb}, {Howard}, {Isaacson}, {Luhn}, {Robertson}, {Rubenzahl}, {Stef{\'a}nsson}, {Winn}, {Brodheim}, {Deich}, {Hill}, {Gibson}, {Holden}, {Householder}, {Laher}, {Lanclos}, {Payne}, {Roy}, {Smith}, {Shaum}, {Schwab}, \& {Walawender}}]{Lubin2024}
{Lubin}, J., {Petigura}, E.~A., {Van Zandt}, J., {et~al.} 2024, \aj, 168, 196, \dodoi{10.3847/1538-3881/ad79ed}

\bibitem[{{Luque} {et~al.}(2023){Luque}, {Osborn}, {Leleu}, {Pall{\'e}}, {Bonfanti}, {Barrag{\'a}n}, {Wilson}, {Broeg}, {Cameron}, {Lendl}, {Maxted}, {Alibert}, {Gandolfi}, {Delisle}, {Hooton}, {Egger}, {Nowak}, {Lafarga}, {Rapetti}, {Twicken}, {Morales}, {Carleo}, {Orell-Miquel}, {Adibekyan}, {Alonso}, {Alqasim}, {Amado}, {Anderson}, {Anglada-Escud{\'e}}, {Bandy}, {B{\'a}rczy}, {Barrado Navascues}, {Barros}, {Baumjohann}, {Bayliss}, {Bean}, {Beck}, {Beck}, {Benz}, {Billot}, {Bonfils}, {Borsato}, {Boyle}, {Brandeker}, {Bryant}, {Cabrera}, {Carrazco-Gaxiola}, {Charbonneau}, {Charnoz}, {Ciardi}, {Cochran}, {Collins}, {Crossfield}, {Csizmadia}, {Cubillos}, {Dai}, {Davies}, {Deeg}, {Deleuil}, {Deline}, {Delrez}, {Demangeon}, {Demory}, {Ehrenreich}, {Erikson}, {Esparza-Borges}, {Falk}, {Fortier}, {Fossati}, {Fridlund}, {Fukui}, {Garcia-Mejia}, {Gill}, {Gillon}, {Goffo}, {G{\'o}mez Maqueo Chew}, {G{\"u}del}, {Guenther}, {G{\"u}nther}, {Hatzes}, {Helling}, {Hesse}, {Howell}, {Hoyer}, {Ikuta}, {Isaak}, {Jenkins},
  {Kagetani}, {Kiss}, {Kodama}, {Korth}, {Lam}, {Laskar}, {Latham}, {Lecavelier des Etangs}, {Leon}, {Livingston}, {Magrin}, {Matson}, {Matthews}, {Mordasini}, {Mori}, {Moyano}, {Munari}, {Murgas}, {Narita}, {Nascimbeni}, {Olofsson}, {Osborne}, {Ottensamer}, {Pagano}, {Parviainen}, {Peter}, {Piotto}, {Pollacco}, {Queloz}, {Quinn}, {Quirrenbach}, {Ragazzoni}, {Rando}, {Ratti}, {Rauer}, {Redfield}, {Ribas}, {Ricker}, {Rudat}, {Sabin}, {Salmon}, {Santos}, {Scandariato}, {Schanche}, {Schlieder}, {Seager}, {S{\'e}gransan}, {Shporer}, {Simon}, {Smith}, {Sousa}, {Stalport}, {Szab{\'o}}, {Thomas}, {Tuson}, {Udry}, {Vanderburg}, {Van Eylen}, {Van Grootel}, {Venturini}, {Walter}, {Walton}, {Watanabe}, {Winn}, \& {Zingales}}]{Luque2023}
{Luque}, R., {Osborn}, H.~P., {Leleu}, A., {et~al.} 2023, \nat, 623, 932, \dodoi{10.1038/s41586-023-06692-3}

\bibitem[{{Maciejewski} {et~al.}(2014){Maciejewski}, {Niedzielski}, {Nowak}, {Pall{\'e}}, {Tingley}, {Errmann}, \& {Neuh{\"a}user}}]{Maciejewski2014}
{Maciejewski}, G., {Niedzielski}, A., {Nowak}, G., {et~al.} 2014, \actaa, 64, 323, \dodoi{10.48550/arXiv.1501.02711}

\bibitem[{{Mann} {et~al.}(2016){Mann}, {Newton}, {Rizzuto}, {Irwin}, {Feiden}, {Gaidos}, {Mace}, {Kraus}, {James}, {Ansdell}, {Charbonneau}, {Covey}, {Ireland}, {Jaffe}, {Johnson}, {Kidder}, \& {Vanderburg}}]{Mann2016}
{Mann}, A.~W., {Newton}, E.~R., {Rizzuto}, A.~C., {et~al.} 2016, \aj, 152, 61, \dodoi{10.3847/0004-6256/152/3/61}

\bibitem[{{Mann} {et~al.}(2020){Mann}, {Johnson}, {Vanderburg}, {Kraus}, {Rizzuto}, {Wood}, {Bush}, {Rockcliffe}, {Newton}, {Latham}, {Mamajek}, {Zhou}, {Quinn}, {Thao}, {Benatti}, {Cosentino}, {Desidera}, {Harutyunyan}, {Lovis}, {Mortier}, {Pepe}, {Poretti}, {Wilson}, {Kristiansen}, {Gagliano}, {Jacobs}, {LaCourse}, {Omohundro}, {Schwengeler}, {Terentev}, {Kane}, {Hill}, {Rabus}, {Esquerdo}, {Berlind}, {Collins}, {Murawski}, {Sallam}, {Aitken}, {Massey}, {Ricker}, {Vanderspek}, {Seager}, {Winn}, {Jenkins}, {Barclay}, {Caldwell}, {Dragomir}, {Doty}, {Glidden}, {Tenenbaum}, {Torres}, {Twicken}, \& {Villanueva}}]{Mann2020}
{Mann}, A.~W., {Johnson}, M.~C., {Vanderburg}, A., {et~al.} 2020, \aj, 160, 179, \dodoi{10.3847/1538-3881/abae64}

\bibitem[{{Martioli} {et~al.}(2022){Martioli}, {H{\'e}brard}, {Fouqu{\'e}}, {Artigau}, {Donati}, {Cadieux}, {Bellotti}, {Lecavelier des Etangs}, {Doyon}, {do Nascimento}, {Arnold}, {Carmona}, {Cook}, {Cortes-Zuleta}, {de Almeida}, {Delfosse}, {Folsom}, {K{\"o}nig}, {Moutou}, {Ould-Elhkim}, {Petit}, {Stassun}, {Vidotto}, {Vandal}, {Benneke}, {Boisse}, {Bonfils}, {Boyd}, {Brasseur}, {Charbonneau}, {Cloutier}, {Collins}, {Cristofari}, {Crossfield}, {D{\'\i}az}, {Fausnaugh}, {Figueira}, {Forveille}, {Furlan}, {Girardin}, {Gnilka}, {Gomes da Silva}, {Gu}, {Guerra}, {Howell}, {Hussain}, {Jenkins}, {Kiefer}, {Latham}, {Matson}, {Matthews}, {Morin}, {Naves}, {Ricker}, {Seager}, {Takami}, {Twicken}, {Vanderburg}, {Vanderspek}, \& {Winn}}]{Martioli2022}
{Martioli}, E., {H{\'e}brard}, G., {Fouqu{\'e}}, P., {et~al.} 2022, \aap, 660, A86, \dodoi{10.1051/0004-6361/202142540}

\bibitem[{{Masuda} \& {Winn}(2020)}]{Masuda2020}
{Masuda}, K., \& {Winn}, J.~N. 2020, \aj, 159, 81, \dodoi{10.3847/1538-3881/ab65be}

\bibitem[{{Mazeh} {et~al.}(2015){Mazeh}, {Perets}, {McQuillan}, \& {Goldstein}}]{Mazeh2015}
{Mazeh}, T., {Perets}, H.~B., {McQuillan}, A., \& {Goldstein}, E.~S. 2015, \apj, 801, 3, \dodoi{10.1088/0004-637X/801/1/3}

\bibitem[{{McLaughlin}(1924)}]{McLaughlin1924}
{McLaughlin}, D.~B. 1924, \apj, 60, 22, \dodoi{10.1086/142826}

\bibitem[{{Millholland} \& {Spalding}(2020)}]{MillholandSpalding2020}
{Millholland}, S.~C., \& {Spalding}, C. 2020, \apj, 905, 71, \dodoi{10.3847/1538-4357/abc4e5}

\bibitem[{{Mills} {et~al.}(2019){Mills}, {Howard}, {Weiss}, {Steffen}, {Isaacson}, {Fulton}, {Petigura}, {Kosiarek}, {Hirsch}, \& {Boisvert}}]{Mills2019}
{Mills}, S.~M., {Howard}, A.~W., {Weiss}, L.~M., {et~al.} 2019, \aj, 157, 145, \dodoi{10.3847/1538-3881/ab0899}

\bibitem[{{Narita} {et~al.}(2020){Narita}, {Fukui}, {Yamamuro}, {Harbeck}, {Bowman}, {Elphick}, {Nation}, {Armstrong}, {Han}, {Abe}, {Ikoma}, {Isogai}, {Kawauchi}, {Kurita}, {Kusakabe}, {de Leon}, {Livingston}, {Mori}, {Nishiumi}, {Tamura}, {Watanabe}, {Volgenau}, {Heinrich-Josties}, {Foale}, {Daily}, {McCully}, {Kirby}, {Smith}, {Haworth}, {Conway}, {Storrie-Lombardi}, {Rosing}, {Chatelain}, {Bachelet}, {Johnson}, \& {Rabus}}]{2020SPIE11447E..5KN}
{Narita}, N., {Fukui}, A., {Yamamuro}, T., {et~al.} 2020, in Society of Photo-Optical Instrumentation Engineers (SPIE) Conference Series, Vol. 11447, Society of Photo-Optical Instrumentation Engineers (SPIE) Conference Series, 114475K, \dodoi{10.1117/12.2559947}

\bibitem[{Newville {et~al.}(2025)Newville, Otten, Nelson, Stensitzki, Ingargiola, Allan, Fox, Carter, \& Rawlik}]{lmfit}
Newville, M., Otten, R., Nelson, A., {et~al.} 2025, LMFIT: Non-Linear Least-Squares Minimization and Curve-Fitting for Python, 1.3.3,  Zenodo, \dodoi{10.5281/zenodo.15014437}

\bibitem[{{Orell-Miquel} {et~al.}(2023){Orell-Miquel}, {Nowak}, {Murgas}, {Palle}, {Morello}, {Luque}, {Badenas-Agusti}, {Ribas}, {Lafarga}, {Espinoza}, {Morales}, {Zechmeister}, {Alqasim}, {Cochran}, {Gandolfi}, {Goffo}, {Kab{\'a}th}, {Korth}, {Lam}, {Livingston}, {Muresan}, {Persson}, \& {Van Eylen}}]{OrellMiqual2023}
{Orell-Miquel}, J., {Nowak}, G., {Murgas}, F., {et~al.} 2023, \aap, 669, A40, \dodoi{10.1051/0004-6361/202244120}

\bibitem[{{Petigura} {et~al.}(2017){Petigura}, {Howard}, {Marcy}, {Johnson}, {Isaacson}, {Cargile}, {Hebb}, {Fulton}, {Weiss}, {Morton}, {Winn}, {Rogers}, {Sinukoff}, {Hirsch}, \& {Crossfield}}]{Petigura2017}
{Petigura}, E.~A., {Howard}, A.~W., {Marcy}, G.~W., {et~al.} 2017, \aj, 154, 107, \dodoi{10.3847/1538-3881/aa80de}

\bibitem[{{Petigura} {et~al.}(2018){Petigura}, {Marcy}, {Winn}, {Weiss}, {Fulton}, {Howard}, {Sinukoff}, {Isaacson}, {Morton}, \& {Johnson}}]{Petigura2018}
{Petigura}, E.~A., {Marcy}, G.~W., {Winn}, J.~N., {et~al.} 2018, \aj, 155, 89, \dodoi{10.3847/1538-3881/aaa54c}

\bibitem[{{Petrovich} {et~al.}(2020){Petrovich}, {Mu{\~n}oz}, {Kratter}, \& {Malhotra}}]{Petrovich2020}
{Petrovich}, C., {Mu{\~n}oz}, D.~J., {Kratter}, K.~M., \& {Malhotra}, R. 2020, \apjl, 902, L5, \dodoi{10.3847/2041-8213/abb952}

\bibitem[{{Piaulet} {et~al.}(2020){Piaulet}, {Benneke}, {Rubenzahl}, {Howard}, {Lee}, {Thorngren}, {Angus}, {Peterson}, {Schlieder}, {Werner}, {Kreidberg}, {Jaouni}, {Crossfield}, {Ciardi}, {Petigura}, {Livingston}, {Dressing}, {Fulton}, {Beichman}, {Christiansen}, {Gorjian}, {Hardegree-Ullman}, {Krick}, \& {Sinukoff}}]{Piaulet2020}
{Piaulet}, C., {Benneke}, B., {Rubenzahl}, R.~A., {et~al.} 2020, arXiv e-prints, arXiv:2011.13444.
\newblock \doarXiv{2011.13444}

\bibitem[{{Polanski} {et~al.}(2024){Polanski}, {Lubin}, {Beard}, {Akana Murphy}, {Rubenzahl}, {Hill}, {Crossfield}, {Chontos}, {Robertson}, {Isaacson}, {Kane}, {Ciardi}, {Batalha}, {Dressing}, {Fulton}, {Howard}, {Huber}, {Petigura}, {Weiss}, {Angelo}, {Behmard}, {Blunt}, {Brinkman}, {Dai}, {Dalba}, {Fetherolf}, {Giacalone}, {Hirsch}, {Holcomb}, {Kosiarek}, {Mayo}, {MacDougall}, {Mo{\v{c}}nik}, {Pidhorodetska}, {Rice}, {Rosenthal}, {Scarsdale}, {Turtelboom}, {Tyler}, {Van Zandt}, {Yee}, {Coria}, {Dulz}, {Hartman}, {Householder}, {Lange}, {Langford}, {Louden}, {Siegel}, {Gilbert}, {Gonzales}, {Schlieder}, {Boyle}, {Christiansen}, {Clark}, {Fernandes}, {Lund}, {Savel}, {Gill}, {Beichman}, {Matson}, {Matthews}, {Furlan}, {Howell}, {Scott}, {Everett}, {Livingston}, {Ershova}, {Cheryasov}, {Safonov}, {Lillo-Box}, {Barrado}, \& {Morales-Calder{\'o}n}}]{Polanski2024}
{Polanski}, A.~S., {Lubin}, J., {Beard}, C., {et~al.} 2024, \apjs, 272, 32, \dodoi{10.3847/1538-4365/ad4484}

\bibitem[{{Pu} \& {Lai}(2019)}]{PuLai2019}
{Pu}, B., \& {Lai}, D. 2019, \mnras, 488, 3568, \dodoi{10.1093/mnras/stz1817}

\bibitem[{{Radzom} {et~al.}(2024){Radzom}, {Dong}, {Rice}, {Wang}, {Yee}, {Fairnington}, {Petrovich}, \& {Wang}}]{Radzom2024}
{Radzom}, B.~T., {Dong}, J., {Rice}, M., {et~al.} 2024, \aj, 168, 116, \dodoi{10.3847/1538-3881/ad61d8}

\bibitem[{{Rice} {et~al.}(2024){Rice}, {Gerbig}, \& {Vanderburg}}]{Rice2024}
{Rice}, M., {Gerbig}, K., \& {Vanderburg}, A. 2024, \aj, 167, 126, \dodoi{10.3847/1538-3881/ad1bed}

\bibitem[{{Rice} {et~al.}(2022{\natexlab{a}}){Rice}, {Wang}, \& {Laughlin}}]{Rice2022a}
{Rice}, M., {Wang}, S., \& {Laughlin}, G. 2022{\natexlab{a}}, \apjl, 926, L17, \dodoi{10.3847/2041-8213/ac502d}

\bibitem[{{Rice} {et~al.}(2022{\natexlab{b}}){Rice}, {Wang}, {Wang}, {Stef{\'a}nsson}, {Isaacson}, {Howard}, {Logsdon}, {Schweiker}, {Dai}, {Brinkman}, {Giacalone}, \& {Holcomb}}]{Rice2022b}
{Rice}, M., {Wang}, S., {Wang}, X.-Y., {et~al.} 2022{\natexlab{b}}, \aj, 164, 104, \dodoi{10.3847/1538-3881/ac8153}

\bibitem[{{Rosenthal} {et~al.}(2021){Rosenthal}, {Fulton}, {Hirsch}, {Isaacson}, {Howard}, {Dedrick}, {Sherstyuk}, {Blunt}, {Petigura}, {Knutson}, {Behmard}, {Chontos}, {Crepp}, {Crossfield}, {Dalba}, {Fischer}, {Henry}, {Kane}, {Kosiarek}, {Marcy}, {Rubenzahl}, {Weiss}, \& {Wright}}]{Rosenthal2021}
{Rosenthal}, L.~J., {Fulton}, B.~J., {Hirsch}, L.~A., {et~al.} 2021, \apjs, 255, 8, \dodoi{10.3847/1538-4365/abe23c}

\bibitem[{{Rossiter}(1924)}]{Rossiter1924}
{Rossiter}, R.~A. 1924, \apj, 60, 15, \dodoi{10.1086/142825}

\bibitem[{{Rubenzahl} {et~al.}(2021){Rubenzahl}, {Dai}, {Howard}, {Chontos}, {Giacalone}, {Lubin}, {Rosenthal}, {Isaacson}, {Batalha}, {Crossfield}, {Dressing}, {Fulton}, {Huber}, {Kane}, {Petigura}, {Robertson}, {Roy}, {Weiss}, {Beard}, {Hill}, {Mayo}, {Mocnik}, {Murphy}, \& {Scarsdale}}]{Rubenzahl2021}
{Rubenzahl}, R.~A., {Dai}, F., {Howard}, A.~W., {et~al.} 2021, \aj, 161, 119, \dodoi{10.3847/1538-3881/abd177}

\bibitem[{{Rubenzahl} {et~al.}(2024){Rubenzahl}, {Howard}, {Halverson}, {Petrovich}, {Angelo}, {Stef{\'a}nsson}, {Dai}, {Householder}, {Fulton}, {Gibson}, {Roy}, {Shaum}, {Isaacson}, {Brodheim}, {Deich}, {Hill}, {Holden}, {Huber}, {Laher}, {Lanclos}, {Payne}, {Petigura}, {Schwab}, {Walawender}, {Wang}, {Weiss}, {Winn}, \& {Wright}}]{Rubenzahl2024}
{Rubenzahl}, R.~A., {Howard}, A.~W., {Halverson}, S., {et~al.} 2024, \apjl, 971, L40, \dodoi{10.3847/2041-8213/ad6985}

\bibitem[{{Sanchis-Ojeda} {et~al.}(2012){Sanchis-Ojeda}, {Fabrycky}, {Winn}, {Barclay}, {Clarke}, {Ford}, {Fortney}, {Geary}, {Holman}, {Howard}, {Jenkins}, {Koch}, {Lissauer}, {Marcy}, {Mullally}, {Ragozzine}, {Seader}, {Still}, \& {Thompson}}]{SanchisOjeda2012}
{Sanchis-Ojeda}, R., {Fabrycky}, D.~C., {Winn}, J.~N., {et~al.} 2012, \nat, 487, 449, \dodoi{10.1038/nature11301}

\bibitem[{{Sanchis-Ojeda} {et~al.}(2013){Sanchis-Ojeda}, {Winn}, {Marcy}, {Howard}, {Isaacson}, {Johnson}, {Torres}, {Albrecht}, {Campante}, {Chaplin}, {Davies}, {Lund}, {Carter}, {Dawson}, {Buchhave}, {Everett}, {Fischer}, {Geary}, {Gilliland}, {Horch}, {Howell}, \& {Latham}}]{SanchisOjeda2013}
{Sanchis-Ojeda}, R., {Winn}, J.~N., {Marcy}, G.~W., {et~al.} 2013, \apj, 775, 54, \dodoi{10.1088/0004-637X/775/1/54}

\bibitem[{{Schatzman}(1962)}]{Schatzman1962}
{Schatzman}, E. 1962, Annales d'Astrophysique, 25, 18

\bibitem[{{Seifahrt} {et~al.}(2018){Seifahrt}, {St{\"u}rmer}, {Bean}, \& {Schwab}}]{Seifahrt2018}
{Seifahrt}, A., {St{\"u}rmer}, J., {Bean}, J.~L., \& {Schwab}, C. 2018, in Society of Photo-Optical Instrumentation Engineers (SPIE) Conference Series, Vol. 10702, Ground-based and Airborne Instrumentation for Astronomy VII, ed. C.~J. {Evans}, L.~{Simard}, \& H.~{Takami}, 107026D, \dodoi{10.1117/12.2312936}

\bibitem[{{Seifahrt} {et~al.}(2020){Seifahrt}, {Bean}, {St{\"u}rmer}, {Kasper}, {Gers}, {Schwab}, {Zechmeister}, {Stef{\'a}nsson}, {Montet}, {Dos Santos}, {Peck}, {White}, \& {Tapia}}]{Seifahrt2020}
{Seifahrt}, A., {Bean}, J.~L., {St{\"u}rmer}, J., {et~al.} 2020, in Society of Photo-Optical Instrumentation Engineers (SPIE) Conference Series, Vol. 11447, Ground-based and Airborne Instrumentation for Astronomy VIII, ed. C.~J. {Evans}, J.~J. {Bryant}, \& K.~{Motohara}, 114471F, \dodoi{10.1117/12.2561564}

\bibitem[{{Southworth}(2011)}]{Southworth2011}
{Southworth}, J. 2011, \mnras, 417, 2166, \dodoi{10.1111/j.1365-2966.2011.19399.x}

\bibitem[{{Stefansson} {et~al.}(2020){Stefansson}, {Mahadevan}, {Maney}, {Ninan}, {Robertson}, {Rajagopal}, {Haase}, {Allen}, {Ford}, {Winn}, {Wolfgang}, {Dawson}, {Wisniewski}, {Bender}, {Ca{\~n}as}, {Cochran}, {Diddams}, {Fredrick}, {Halverson}, {Hearty}, {Hebb}, {Kanodia}, {Levi}, {Metcalf}, {Monson}, {Ramsey}, {Roy}, {Schwab}, {Terrien}, \& {Wright}}]{Stefansson2020}
{Stefansson}, G., {Mahadevan}, S., {Maney}, M., {et~al.} 2020, \aj, 160, 192, \dodoi{10.3847/1538-3881/abb13a}

\bibitem[{{Stef{\`a}nsson} {et~al.}(2022){Stef{\`a}nsson}, {Mahadevan}, {Petrovich}, {Winn}, {Kanodia}, {Millholland}, {Maney}, {Ca{\~n}as}, {Wisniewski}, {Robertson}, {Ninan}, {Ford}, {Bender}, {Blake}, {Cegla}, {Cochran}, {Diddams}, {Dong}, {Endl}, {Fredrick}, {Halverson}, {Hearty}, {Hebb}, {Hirano}, {Lin}, {Logsdon}, {Lubar}, {McElwain}, {Metcalf}, {Monson}, {Rajagopal}, {Ramsey}, {Roy}, {Schwab}, {Schweiker}, {Terrien}, \& {Wright}}]{Stefansson2022}
{Stef{\`a}nsson}, G., {Mahadevan}, S., {Petrovich}, C., {et~al.} 2022, \apjl, 931, L15, \dodoi{10.3847/2041-8213/ac6e3c}

\bibitem[{{St{\"u}rmer} {et~al.}(2017){St{\"u}rmer}, {Seifahrt}, {Schwab}, \& {Bean}}]{Sturmer2017}
{St{\"u}rmer}, J., {Seifahrt}, A., {Schwab}, C., \& {Bean}, J.~L. 2017, Journal of Astronomical Telescopes, Instruments, and Systems, 3, 025003, \dodoi{10.1117/1.JATIS.3.2.025003}

\bibitem[{{Su{\'a}rez Mascare{\~n}o} {et~al.}(2021){Su{\'a}rez Mascare{\~n}o}, {Damasso}, {Lodieu}, {Sozzetti}, {B{\'e}jar}, {Benatti}, {Zapatero Osorio}, {Micela}, {Rebolo}, {Desidera}, {Murgas}, {Claudi}, {Gonz{\'a}lez Hern{\'a}ndez}, {Malavolta}, {del Burgo}, {D'Orazi}, {Amado}, {Locci}, {Tabernero}, {Marzari}, {Aguado}, {Turrini}, {Cardona Guill{\'e}n}, {Toledo-Padr{\'o}n}, {Maggio}, {Aceituno}, {Bauer}, {Caballero}, {Chinchilla}, {Esparza-Borges}, {Gonz{\'a}lez-{\'A}lvarez}, {Granzer}, {Luque}, {Mart{\'\i}n}, {Nowak}, {Oshagh}, {Pall{\'e}}, {Parviainen}, {Quirrenbach}, {Reiners}, {Ribas}, {Strassmeier}, {Weber}, \& {Mallonn}}]{SaurezMascareno2022}
{Su{\'a}rez Mascare{\~n}o}, A., {Damasso}, M., {Lodieu}, N., {et~al.} 2021, Nature Astronomy, 6, 232, \dodoi{10.1038/s41550-021-01533-7}

\bibitem[{{Teng} {et~al.}(2024){Teng}, {Dai}, {Howard}, {Isaacson}, {Rubenzahl}, {Angelo}, \& {Polanski}}]{Teng2024}
{Teng}, H.-Y., {Dai}, F., {Howard}, A.~W., {et~al.} 2024, \aj, 168, 194, \dodoi{10.3847/1538-3881/ad7022}

\bibitem[{{Triaud}(2018)}]{Triaud2018}
{Triaud}, A. H.~M.~J. 2018, in Handbook of Exoplanets, ed. H.~J. {Deeg} \& J.~A. {Belmonte}, 2, \dodoi{10.1007/978-3-319-55333-7_2}

\bibitem[{{Triaud} {et~al.}(2010){Triaud}, {Collier Cameron}, {Queloz}, {Anderson}, {Gillon}, {Hebb}, {Hellier}, {Loeillet}, {Maxted}, {Mayor}, {Pepe}, {Pollacco}, {S{\'e}gransan}, {Smalley}, {Udry}, {West}, \& {Wheatley}}]{Triaud2010}
{Triaud}, A.~H.~M.~J., {Collier Cameron}, A., {Queloz}, D., {et~al.} 2010, \aap, 524, A25, \dodoi{10.1051/0004-6361/201014525}

\bibitem[{{Tyler} {et~al.}(2025){Tyler}, {Petigura}, {Rogers}, {Lubin}, {Seifhart}, {Bean}, {Brady}, \& {Luque}}]{Tyler2025}
{Tyler}, D., {Petigura}, E.~A., {Rogers}, J., {et~al.} 2025, \aj, 169, 109, \dodoi{10.3847/1538-3881/ada121}

\bibitem[{{Van Eylen} {et~al.}(2018){Van Eylen}, {Dai}, {Mathur}, {Gandolfi}, {Albrecht}, {Fridlund}, {Garc{\'\i}a}, {Guenther}, {Hjorth}, {Justesen}, {Livingston}, {Lund}, {P{\'e}rez Hern{\'a}ndez}, {Prieto-Arranz}, {Regulo}, {Bugnet}, {Everett}, {Hirano}, {Nespral}, {Nowak}, {Palle}, {Silva Aguirre}, {Trifonov}, {Winn}, {Barrag{\'a}n}, {Beck}, {Chaplin}, {Cochran}, {Csizmadia}, {Deeg}, {Endl}, {Heeren}, {Grziwa}, {Hatzes}, {Hidalgo}, {Korth}, {Mathis}, {Monta{\~n}es Rodriguez}, {Narita}, {Patzold}, {Persson}, {Rodler}, \& {Smith}}]{vanEylen2018}
{Van Eylen}, V., {Dai}, F., {Mathur}, S., {et~al.} 2018, \mnras, 478, 4866, \dodoi{10.1093/mnras/sty1390}

\bibitem[{{Vanderburg} {et~al.}(2016){Vanderburg}, {Plavchan}, {Johnson}, {Ciardi}, {Swift}, \& {Kane}}]{Vanderburg2016}
{Vanderburg}, A., {Plavchan}, P., {Johnson}, J.~A., {et~al.} 2016, \mnras, 459, 3565, \dodoi{10.1093/mnras/stw863}

\bibitem[{{Wang} {et~al.}(2024){Wang}, {Rice}, {Wang}, {Kanodia}, {Dai}, {Logsdon}, {Schweiker}, {Teske}, {Butler}, {Crane}, {Shectman}, {Quinn}, {Kostov}, {Osborn}, {Goeke}, {Eastman}, {Shporer}, {Rapetti}, {Collins}, {Watkins}, {Relles}, {Ricker}, {Seager}, {Winn}, \& {Jenkins}}]{Wang2024}
{Wang}, X.-Y., {Rice}, M., {Wang}, S., {et~al.} 2024, \apjl, 973, L21, \dodoi{10.3847/2041-8213/ad7469}

\bibitem[{{Winn} {et~al.}(2010){Winn}, {Fabrycky}, {Albrecht}, \& {Johnson}}]{Winn2010}
{Winn}, J.~N., {Fabrycky}, D., {Albrecht}, S., \& {Johnson}, J.~A. 2010, \apjl, 718, L145, \dodoi{10.1088/2041-8205/718/2/L145}

\bibitem[{{Winn} {et~al.}(2017){Winn}, {Petigura}, {Morton}, {Weiss}, {Dai}, {Schlaufman}, {Howard}, {Isaacson}, {Marcy}, {Justesen}, \& {Albrecht}}]{Winn2017}
{Winn}, J.~N., {Petigura}, E.~A., {Morton}, T.~D., {et~al.} 2017, \aj, 154, 270, \dodoi{10.3847/1538-3881/aa93e3}

\bibitem[{{Wirth} {et~al.}(2021){Wirth}, {Zhou}, {Quinn}, {Mann}, {Bouma}, {Latham}, {Teske}, {Wang}, {Shectman}, {Butler}, \& {Crane}}]{Wirth2021}
{Wirth}, C.~P., {Zhou}, G., {Quinn}, S.~N., {et~al.} 2021, \apjl, 917, L34, \dodoi{10.3847/2041-8213/ac13a9}

\bibitem[{{Wittenmyer} {et~al.}(2020){Wittenmyer}, {Butler}, {Horner}, {Clark}, {Tinney}, {Carter}, {Wang}, {Johnson}, \& {Collins}}]{Wittenmyer2020}
{Wittenmyer}, R.~A., {Butler}, R.~P., {Horner}, J., {et~al.} 2020, \mnras, 491, 5248, \dodoi{10.1093/mnras/stz3378}

\bibitem[{{Wittrock} {et~al.}(2023){Wittrock}, {Plavchan}, {Cale}, {Barclay}, {Ludwig}, {Schwarz}, {M{\'e}karnia}, {Triaud}, {Abe}, {Suarez}, {Guillot}, {Conti}, {Collins}, {Waite}, {Kielkopf}, {Collins}, {Dreizler}, {El Mufti}, {Feliz}, {Gaidos}, {Geneser}, {Horne}, {Kane}, {Lowrance}, {Martioli}, {Radford}, {Reefe}, {Roccatagliata}, {Shporer}, {Stassun}, {Stockdale}, {Tan}, {Tanner}, \& {Vega}}]{Wittrock2023}
{Wittrock}, J.~M., {Plavchan}, P.~P., {Cale}, B.~L., {et~al.} 2023, \aj, 166, 232, \dodoi{10.3847/1538-3881/acfda8}

\bibitem[{{Xu} \& {Lai}(2017)}]{Xu2017}
{Xu}, W., \& {Lai}, D. 2017, \mnras, 468, 3223, \dodoi{10.1093/mnras/stx668}

\bibitem[{{Yu} {et~al.}(2025){Yu}, {Garai}, {Cretignier}, {Szab{\'o}}, {Aigrain}, {Gandolfi}, {Bryant}, {Correia}, {Klein}, {Brandeker}, {Owen}, {G{\"u}nther}, {Winn}, {Heitzmann}, {Cegla}, {Wilson}, {Gill}, {Kriskovics}, {Barrag{\'a}n}, {Boldog}, {Nielsen}, {Billot}, {Lafarga}, {Meech}, {Alibert}, {Alonso}, {B{\'a}rczy}, {Barrado}, {Barros}, {Baumjohann}, {Bayliss}, {Benz}, {Bergomi}, {Borsato}, {Broeg}, {Cameron}, {Csizmadia}, {Cubillos}, {Davies}, {Deleuil}, {Deline}, {Demangeon}, {Demory}, {Derekas}, {Doyle}, {Edwards}, {Egger}, {Ehrenreich}, {Erikson}, {Fortier}, {Fossati}, {Fridlund}, {Gazeas}, {Gillon}, {G{\"u}del}, {Helling}, {Isaak}, {Kiss}, {Korth}, {Lam}, {Laskar}, {Lecavelier des Etangs}, {Lendl}, {Magrin}, {Maxted}, {McCormac}, {Mer{\'\i}n}, {Mordasini}, {Nascimbeni}, {O'Brien}, {Olofsson}, {Ottensamer}, {Pagano}, {Pall{\'e}}, {Peter}, {Piazza}, {Piotto}, {Pollacco}, {Queloz}, {Ragazzoni}, {Rando}, {Rauer}, {Ribas}, {Santos}, {Scandariato}, {S{\'e}gransan}, {Simon}, {Smith}, {Sousa},
  {Southworth}, {Stalport}, {Steinberger}, {Sulis}, {Udry}, {Ulmer}, {Ulmer-Moll}, {Van Grootel}, {Venturini}, {Villaver}, {Walton}, \& {Wheatley}}]{Yu2025}
{Yu}, H., {Garai}, Z., {Cretignier}, M., {et~al.} 2025, \mnras, 536, 2046, \dodoi{10.1093/mnras/stae2655}

\bibitem[{{Yu} {et~al.}(2018){Yu}, {Rodriguez}, {Eastman}, {Crossfield}, {Shporer}, {Gaudi}, {Burt}, {Fulton}, {Sinukoff}, {Howard}, {Isaacson}, {Kosiarek}, {Ciardi}, {Schlieder}, {Penev}, {Vanderburg}, {Stassun}, {Bieryla}, {Butler}, {Berlind}, {Calkins}, {Esquerdo}, {Latham}, {Murawski}, {Stevens}, {Petigura}, {Kreidberg}, \& {Bristow}}]{Yu2018}
{Yu}, L., {Rodriguez}, J.~E., {Eastman}, J.~D., {et~al.} 2018, \aj, 156, 127, \dodoi{10.3847/1538-3881/aad6e7}

\bibitem[{{Zak} {et~al.}(2024){Zak}, {Boffin}, {Sedaghati}, {Bocchieri}, {Changeat}, {Fukui}, {Hatzes}, {Hillwig}, {Hornoch}, {Itrich}, {Ivanov}, {Jones}, {Kabath}, {Kawai}, {Mugnai}, {Murgas}, {Narita}, {Palle}, {Pascale}, {Pravec}, {Redfield}, {Roccetti}, {Roth}, {Srba}, {Tian}, {Tsiaras}, {Turrini}, \& {Vignes}}]{Zak2024}
{Zak}, J., {Boffin}, H.~M.~J., {Sedaghati}, E., {et~al.} 2024, \aap, 687, L2, \dodoi{10.1051/0004-6361/202450570}

\bibitem[{{Zhao} {et~al.}(2023){Zhao}, {Kunovac}, {Brewer}, {Llama}, {Millholland}, {Hedges}, {Szymkowiak}, {Roettenbacher}, {Cabot}, {Weiss}, \& {Fischer}}]{Zhao2023}
{Zhao}, L.~L., {Kunovac}, V., {Brewer}, J.~M., {et~al.} 2023, Nature Astronomy, 7, 198, \dodoi{10.1038/s41550-022-01837-2}

\end{thebibliography}
